\documentclass[amssymb,amsmath,prd,twocolumn,superscriptaddress,nofootinbib]{revtex4-1}
\usepackage[10pt]{type1ec}  
\usepackage[T1]{fontenc}
\usepackage{CJKutf8}
\usepackage[overlap, CJK]{ruby}
\usepackage{CJKulem}
\newenvironment{SChinese}{%
  \CJKfamily{gbsn}%
  \CJKtilde
  \CJKnospace}{}

\usepackage{graphicx}
\usepackage{bm}
\usepackage{amssymb,amsmath}
\usepackage{mathrsfs}
\usepackage{latexsym}
\usepackage{color}
\usepackage{dcolumn}
\usepackage[colorlinks=true,citecolor=blue,urlcolor=blue]{hyperref}
\usepackage[usenames,dvipsnames]{xcolor}
\usepackage[caption=false]{subfig}

\newcommand{\beq}{\begin{equation}}
\newcommand{\eeq}{\end{equation}}
\newcommand{\ba}{\begin{array}}
\newcommand{\ea}{\end{array}}
\newcommand{\bea}{\begin{eqnarray}}
\newcommand{\eea}{\end{eqnarray}}
\newcommand{\bc}{\begin{center}}
\newcommand{\ec}{\end{center}}

\newcommand{\De}{\Delta}
\newcommand{\ep}{\varepsilon}

\newcommand{\La}{\Lambda}

\newcommand{\ptrans}{p_{\rm trans}}
\newcommand{\ntrans}{n_{\rm trans}}

\newcommand{\etrans}{\varepsilon_{\rm trans}}
\newcommand{\Mtrans}{M_{\rm t}}
\newcommand{\Rtrans}{R_{\rm t}}
\newcommand{\Msolar}{{\rm M}_{\odot}}

\newcommand{\cQMsq}{c^2_{\rm QM}}

\newcommand{\Mmax}{M_{\rm max}}
\newcommand{\Rtyp}{R_{1.4}}

\newcommand{\Mchirp}{{\mathcal M}}
\newcommand{\Rchirp}{{\mathcal R}}

\newcommand\Eqn[1]{Eq.~(\ref{#1})}  

\allowdisplaybreaks

\newcommand{\CCA}{\affiliation{Center for Computational Astrophysics, Flatiron Institute, 162 5th Ave, New York, NY 10010}}

\definecolor{azgreen}{rgb}{0.03,0.47,0.19}
\definecolor{kcmagenta}{rgb}{0.54, 0.17, 0.88}
\definecolor{chorange}{rgb}{0.851, 0.372, 0.007}

\newcommand{\nn}{\nonumber}

\begin{document}

\title{Studying strong phase transitions in neutron stars with gravitational waves}

\author{Katerina Chatziioannou}\email{kchatziioannou@flatironinstitute.org}
\CCA
\author{Sophia Han 
(\begin{CJK}{UTF8}{}\begin{SChinese}韩 君\end{SChinese}\end{CJK})
}
\email{sjhan@berkeley.edu}
\affiliation{Department of Physics and Astronomy, Ohio University,
Athens, OH 45701, USA}
\affiliation{Department of Physics, University of California Berkeley, Berkeley, CA~94720, USA}

\date{January 23, 2020}

\begin{abstract}
The composition of neutron stars at the extreme densities reached in their cores is currently unknown. Besides nuclear matter of normal neutrons and protons, the cores of neutron stars might harbor exotic matter such as deconfined quarks. In this paper we study strong hadron-quark phase transitions in the context of gravitational wave observations of inspiraling neutron stars. We consider upcoming detections of neutron star coalescences and model the neutron star equations of state with phase transitions through the Constant-Speed-of-Sound parametrization. 
We use the fact that neutron star binaries with one or more hadron-quark hybrid stars can exhibit qualitatively different tidal properties than binaries with hadronic stars of the same mass, and hierarchically model the masses and tidal properties of simulated populations of binary neutron star inspiral signals. We explore the parameter space of phase transitions and discuss under which conditions future observations of binary neutron star inspirals can identify this effect and constrain its properties, in particular the threshold 
density at which the transition happens and the strength of the transition.
We find that if the detected population of binary neutron stars contains both hadronic and hybrid stars, the onset mass and strength of a sufficiently strong phase transition can be constrained with 50--100 detections. If the detected neutron stars are exclusively hadronic or hybrid, then it is possible to place lower or upper limits on the transition density and strength.
\end{abstract}

\maketitle
  
\section{Introduction}
\label{sec:intro}

The detection of the gravitational wave (GW) signal GW170817~\cite{TheLIGOScientific:2017qsa} by the LIGO~\cite{TheLIGOScientific:2014jea} and Virgo~\cite{TheVirgo:2014hva} detectors, followed by electromagnetic (EM) observations~\cite{GBM:2017lvd}, opens up the possibility of multimessenger astronomy as a probe of ultra-dense matter on the quantum chromodynamics (QCD) phase diagram. 
Estimates of the tidal properties of the binary neutron star (BNS) source of GW170817~\cite{Abbott:2018wiz,Abbott:2018exr} are inconsistent with too stiff hadronic matter, and lead to NS radii that are consistent with previous X-ray binary observations~\cite{Steiner:2015aea,Lattimer:2015nhk,Ozel:2016oaf,Steiner:2017vmg}. 
With further future BNS detections and improved detector sensitivity~\cite{Aasi:2013wya}, GW observations will yield stringent constraints on the NS interior properties~\cite{Hinderer:2009ca,Read:2009yp,DelPozzo:2013ala,Lackey:2014fwa,Agathos:2015uaa,Chatziioannou:2015uea,Vivanco:2019qnt} that are expected to encounter smaller systematic uncertainties than those from X-ray binaries~\cite{Degenaar:2018lle}. 

An interesting feature of the QCD phase diagram that can be tested with future BNS observations is the potential for phase transitions, drastic transitions from normal hadronic matter to exotic matter such as deconfined quarks that may appear in the inner cores of NSs~\cite{Lattimer:2015nhk}.
Possible effects from nonhadronic degrees of freedom complicate the analyses of GW data through LIGO detection of BNSs, as they can alter assumptions under which certain observational constraints are derived; a fully comprehensive framework with proper treatment of phase transitions is required as we observe more BNS signals. 
Recently there have been emerging investigations on the role of a possible hadron-quark phase transition in extracting tidal parameters and constraining the equation of state (EoS) of NS, by utilizing either physical models or generic parametrizations; see e.g.~\cite{Paschalidis:2017qmb,Nandi:2017rhy,Baym:2017whm,Li:2018ayl,Alvarez-Castillo:2018pve,Burgio:2018yix,Gomes:2018eiv,Han:2019bub,Annala:2019puf,Christian:2018jyd,Sieniawska:2018zzj,Han:2018mtj,Montana:2018bkb,Chen:2019rja,Essick:2019ldf}. The primary quantity that characterizes the transition is the onset density for deconfined quarks, as it determines whether the densities encountered within stable NSs or transient supernova and merger remnant stars can possibly access the phase of quark matter. 
A low onset density implies that most NSs in the Universe could harbor quark cores, while a high onset density would mean that most NSs remain hadronic throughout.

The unknown nature of the hadron-quark transition, being of first-order (either with a sharp transition or in mixed phases depending on the surface tension~\cite{Alford:2001zr,Mintz:2009ay,Lugones:2013ema,Fraga:2018cvr}) or a smooth crossover~\cite{Kojo:2014rca,Fukushima:2015bda,Baym:2017whm,McLerran:2018hbz,Jeong:2019lhv}, further complicates its study. Formally, quarks --if they exist-- are expected to be strongly interacting at the typical NS densities, so the EoS of quark matter could resemble that of very dense hadronic matter; this scenario would lead to almost negligible differences between hadronic stars and hybrid stars with a quark core in the mass-radius diagram~\cite{Alford:2004pf}. 
In this case the hybrid branch looks roughly like a continuation of the purely hadronic branch, making the experimental verification of the existence of hybrid stars challenging. 
One exception would be the emergence of a disconnected branch, the ``third-family'' of hybrid stars with significantly smaller radii, which is characteristic of sharp, sufficiently strong first-order phase transition and could be distinguishable from normal hadronic stars~\cite{Glendenning:2000,Schertler:2000xq}. 
As an example of how this effect complicates the interpretation of astrophysical data,
it has been argued that allowing for strong first-order phase transitions in the EoS decreases the inferred radii from observations of eight quiescent low-mass X-ray binaries in globular clusters, and in that case, the radius is likely smaller than 12 km~\cite{Steiner:2017vmg}.

In this paper, we study the effect of a sharp phase transition on static NS observables, in particular the measurement of tidal properties from the inspiral stage of BNS coalescences, and examine how well future BNS observations can possibly constrain the properties of the phase transition\footnote{It is worth mentioning that dynamical NS properties such as NS cooling and spin-down~\cite{Yakovlev:2004iq,Page:2009fu,Haskell:2015jra}, global oscillation modes~\cite{Lindblom:1998wf,Andersson:2000mf}, and the GW evolution of merger products~\cite{Most:2018eaw,Bauswein:2018bma} that are sensitive to transport properties would potentially provide more distinct signatures of possible phase transition in NSs~\cite{Alford:2019oge}. However, some of these signals are more challenging to observe than BNS inspirals, see e.g.~\cite{Torres-Rivas:2018svp}.}. 
We employ hybrid EoSs constructed with the generic ``Constant-Speed-of-Sound'' (CSS) parametrization~\cite{Alford:2013aca}, which allows us to systematically consider the detectability of sharp phase transitions with different onset density and of different strength.
Our analysis hinges on the \textit{chirp radius} parameter ${\cal{R}}$~\cite{Wade:2014vqa} which is a combination of the best measured tidal parameter, the weighted-average binary tidal deformability $\tilde{\Lambda}$, and the best measured mass parameter, the chirp mass ${\cal{M}}$, with GW signals. We show that the chirp radius for binaries involving hybrid stars can exhibit a qualitatively different behavior than binaries of hadronic stars only. The latter follows a unique ${\cal{R}}({\cal{M}},R)$ curve, where $R$ is the typical hadronic radius, while the former deviates from this curve.

We simulate populations of potential future BNS detections with different mass distributions and employ a hierarchical model that takes advantage of the dependence of the chirp radius ${\cal{R}}$ on the nature of the binary components. 
The GW signal from each BNS results in an estimate of its chirp mass, mass ratio, and chirp radius to within some statistical measurement uncertainty.  
Our hierarchical model then assumes that binaries of NSs below a certain transition mass $\Mtrans$ approximately obey a theoretical relation ${\cal{R}}({\cal{M}},R)$, while heavier binaries exhibit smaller (chirp) radii and tidal effects. 
This allows us to obtain an estimate of the population properties of the simulated BNSs, namely the NS mass $\Mtrans$ at which the phase transition occurs and the typical radius $R$ of the hadronic stars. 

We find that both the NS mass distribution and the details of the phase transition play an important role in whether the phase transition is detectable. If the detected population of BNSs contains both hybrid and hadronic stars and the transition is sufficiently strong, the onset mass and strength of the transition can be determined with $\sim50-100$ detections, while the hadronic radius can simultaneously be reliably measured. Such strong phase transitions (that are compatible with $\Mmax\geq2\,\Msolar$) can be realized when the energy density jump is large enough and the subsequent quark phase is stiff.
If, on the other hand, the transition happens at high masses, or it happens at masses that are higher than the detected population, then we will be able to place a lower limit on the onset mass as well as an upper limit on the transition strength. 
On the opposite extreme, if the transition happens at densities lower than the central NS densities, all detected NSs are hadron-quark hybrids and the underlying EoS is degenerate with a softer hadronic EoS. 
In that case it might still be possible to determine that a phase transition has occurred if we consider the recent heavy pulsar detections, which rule out very soft hadronic EoSs and break the degeneracy~\cite{Demorest:2010bx,Antoniadis:2013pzd,Fonseca:2016tux,Cromartie:2019kug}. 
This leads to an upper limit on the transition mass, while the hadronic radius is unmeasurable since our detected population contains no purely hadronic NSs.

The rest of the paper presents the details of our study and it is organized as follows. 
Section~\ref{sec:CSS} describes the basics of the CSS parametrization and its applicability to identify the observable features of sharp first-order phase transitions. 
In Sec.~\ref{sec:estimate}, we illustrate EoS models with a strong transition that induce potential bias in the radius estimates inferred from BNS observations. 
In Sec.~\ref{sec:popModel} we present a mapping onto chirp mass-chirp radius relation for selected EoSs with and without a strong phase transition, and provide a fit using EoS-insensitive relations that assume no phase transitions.  
We then describe a method to identify hadronic and hybrid EoSs within a mixed population of BNS signals detected. 
In Sec.~\ref{sec:results} we apply this method to simulated BNS observations and discuss the prospects of inferring phase transition parameters. 
In Sec.~\ref{sec:con} we summarize our main conclusions.

We work in units where $\hbar=G=c=1$.

\section{The constant-sound-speed parametrization}
\label{sec:CSS}

To conduct a model-independent survey of the observable signatures of a first-order phase transition, we combine a plausible hadronic matter EoS at low densities with a generic parametrization of such a transition into a higher-density phase. 
One example of this is the CSS parametrization \cite{Alford:2013aca} which contains only three parameters to relate the NS pressure $p$ and energy density $\ep$. 
The three CSS parameters specify the critical pressure at which the transition occurs $\ptrans$ (or equivalently the critical baryon number density $\ntrans$), the strength of the transition $\De\ep/\etrans$, and the ``stiffness'' of the high-density phase, where the speed of sound $\cQMsq$ (that characterizes how rapidly pressure rises with energy density) is assumed independent of density.  
The CSS form can be viewed as the lowest-order terms of a Taylor expansion of the high-density EoS about the transition pressure
\beq \ep(p) = \left\{\!
\begin{array}{ll}
\ep_{\rm HM}(p), & p<\ptrans \\
\ep_{\rm HM}(\ptrans)+\De\ep+c_{\rm QM}^{-2} (p-\ptrans), & p>\ptrans
\end{array}
\right.\ 
\label{eqn:CSS_EoS}
\eeq
where $\ep_{\rm HM}(p)$ is the hadronic matter EoS. If the surface tension at the hadron-quark interface were low enough, such a transition would be smoothed out by the appearance of charge-separated mixed phases (Gibbs construction).
However, given the uncertainty in the estimates of surface tension \cite{Alford:2001zr,Mintz:2009ay,Lugones:2013ema,Fraga:2018cvr} we will assume a sharp interface (Maxwell construction).

The CSS parametrization can be used as a generic language for relating different models to each other, and for expressing experimental and observational constraints. 
There are two restrictive and commonly used observational constraints on the stiffness of EoS 
(or, equivalently, the speed of sound). First, EoSs that are too soft at high densities are being eliminated by the progressively more stringent discovery of massive pulsars~\cite{Demorest:2010bx,Antoniadis:2013pzd,Fonseca:2016tux,Cromartie:2019kug}. 
Secondly, EoSs that are very stiff at the relevant mass range are inconsistent with the premerger GW signal from the BNS merger GW170817, which disfavored a large tidal deformation (and large radii)~\cite{TheLIGOScientific:2017qsa,Abbott:2018wiz}. Both constraints lead to constraints on the CSS parameter space~\cite{Alford:2015gna,Han:2018mtj} which are obeyed by the example EoSs considered in this study.
Besides, the CSS parametrization can be naturally combined with other parametrizations for the hadronic matter EoS $\ep_{\rm HM}(p)$, such as the spectral one~\cite{Carney:2018sdv}, to enable a more complete analysis in the future.

\begin{figure*}[]
\parbox{0.48\hsize}{
\includegraphics[width=\hsize]{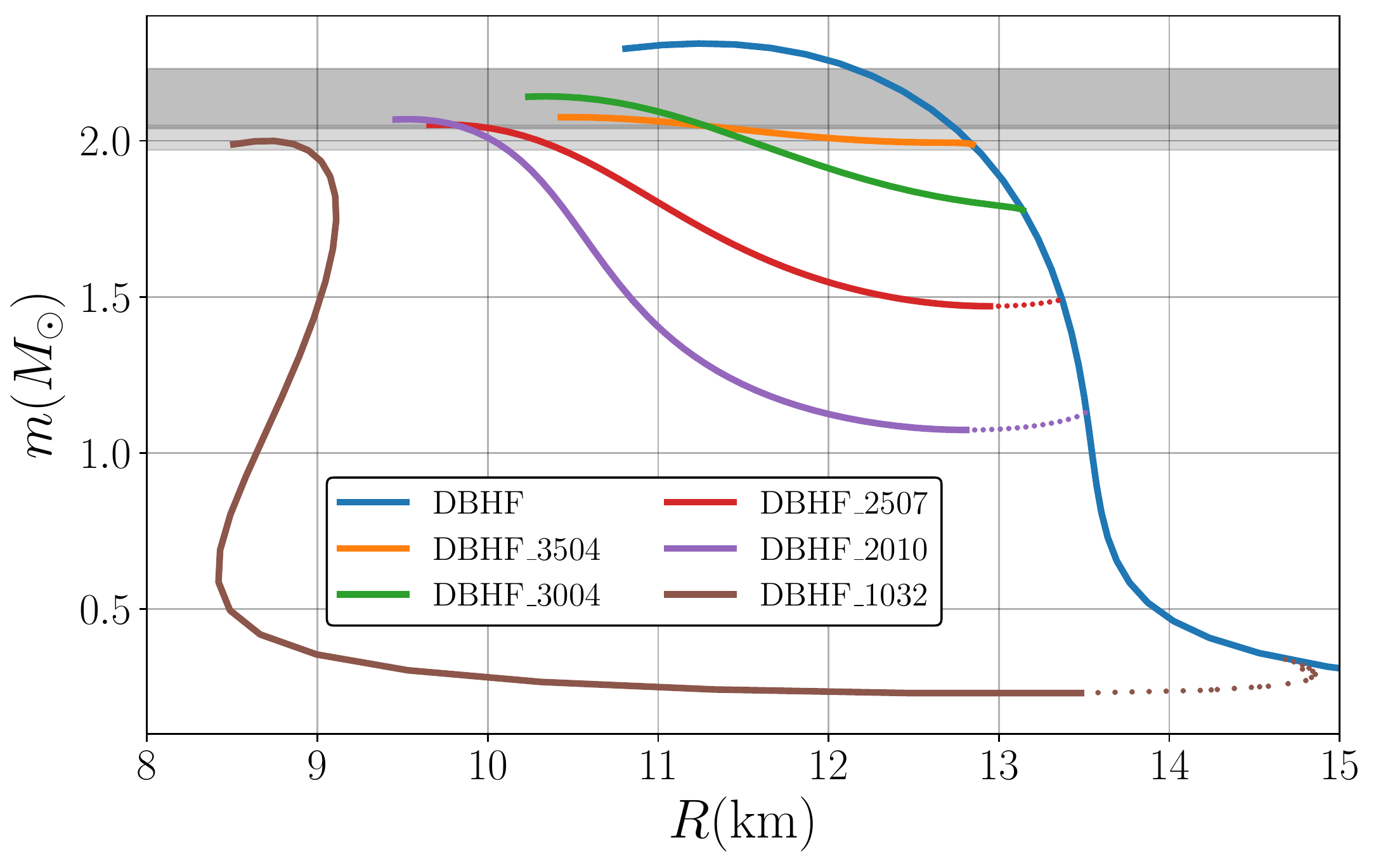}\\[-2ex]
}\parbox{0.48\hsize}{
\includegraphics[width=\hsize]{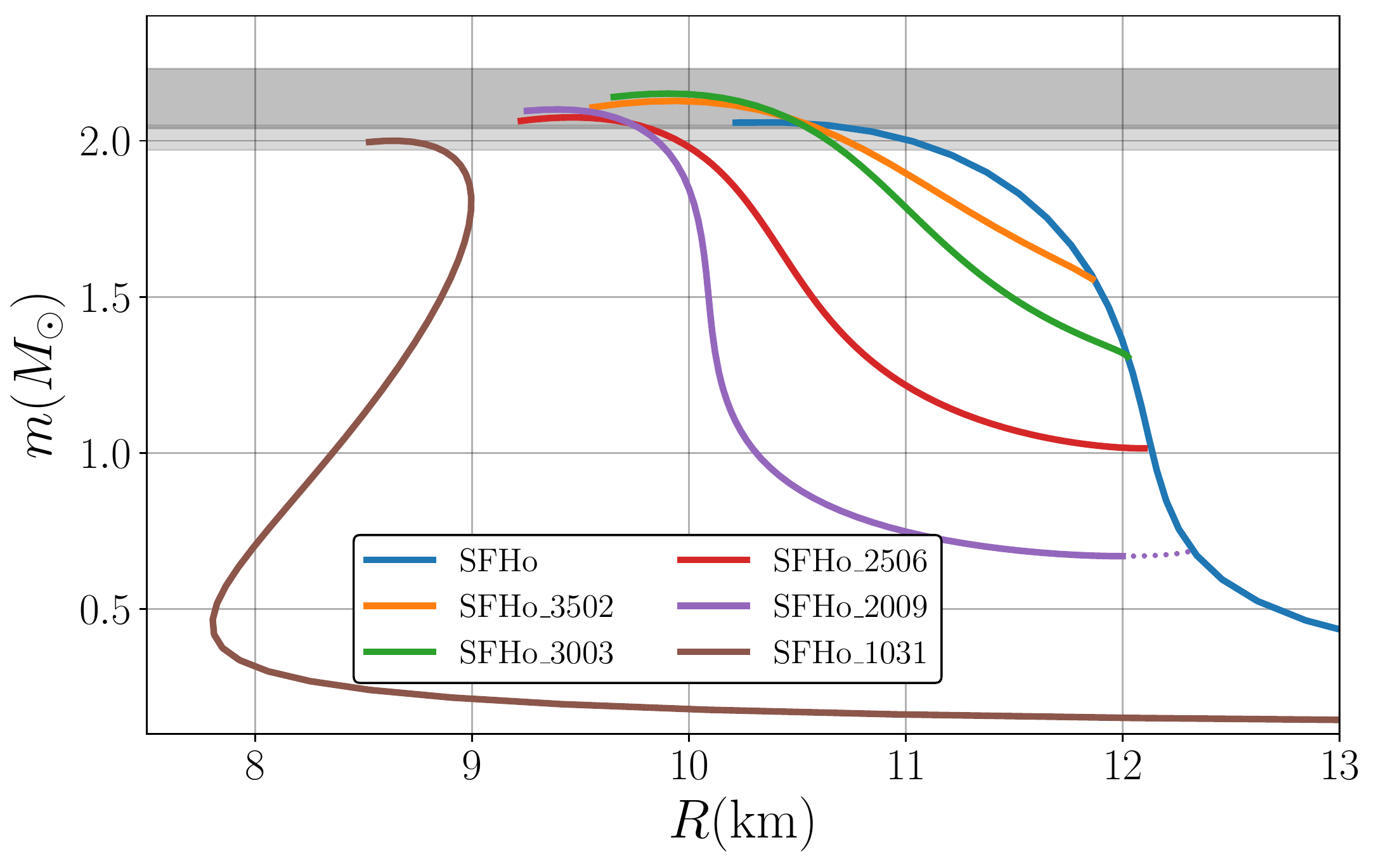}\\[-2ex]
}
\caption{Mass-radius relation for the EoSs considered in this study. The EoSs are based on the hadronic baseline EoSs DBHF (left) and SFHo (right). They are constructed by choosing different values of the CSS parameters ($\ntrans,\De \ep/\etrans$) and setting maximal stiffness in quark matter ($\cQMsq=1$). Solid lines (dotted lines) correspond to the stable (unstable) part of each EoS. Grey bands denote the 
measured masses of heavy pulsars from~\cite{Antoniadis:2013pzd} (light grey) and~\cite{Cromartie:2019kug} (dark grey) at the $68\%$ level.
}
\label{fig:MR}
\end{figure*}

First-order phase transition(s) can produce qualitative and quantitative features in the mass-radius relation for NSs. Qualitatively, a sharp transition to quark matter leads to a branch of hybrid stars with central pressures exceeding $\ptrans$~\cite{Seidov:1971,Haensel:1983,Lindblom:1998dp}. 
Depending on the energy density discontinuity and the speed of sound, there may or may not be a stable branch of hybrid stars, which may or may not be disconnected from the hadronic branch \cite{Alford:2013aca}.
If the phase transition is sufficiently strongly first-order (with large enough $\De \ep$) and occurs at low enough pressure $\ptrans$, then there will be a disconnected hybrid branch. This ``third family'' of stars with significantly smaller radii (and hence smaller tidal deformability) is separated by an unstable branch where no NSs with these radii should be detected; see, e.g., the dotted part of the brown, purple, and red curves in the left panel of Fig.~\ref{fig:MR}. If there is a second first-order phase transition, then there might be a fourth-family of hybrid stars with even smaller radius and tidal deformability~\cite{Alford:2017qgh,Han:2018mtj}.

We choose two representative hadronic matter models, DBHF~\cite{GrossBoelting:1998jg} and SFHo~\cite{Steiner:2012rk}, with large $L=69.4$ MeV and small $L=45.7$ MeV values, respectively, for the symmetry energy slope parameter $L$ at saturation, to illustrate the dependence of our results on the choice of low-density EoS. 
Both EoSs satisfy current constraints from the nuclear experiments, although DBHF lies on the upper side of stiffness consistent with GW170817. We assume the NS crust is as described in~\cite{Baym71tg,Negele73ns}. 
In addition, to fulfill the observational constraint on the maximum NS mass $\Mmax\gtrsim 2\,\Msolar$~\cite{Demorest:2010bx,Antoniadis:2013pzd,Fonseca:2016tux,Cromartie:2019kug}, strong first-order phase transitions 
that substantially soften the transition region require a relatively large speed of sound elsewhere in the star. 
Preferably, $\cQMsq \gtrsim 0.5$ is needed for softer hadronic EoSs and $\cQMsq \gtrsim 0.4$ is needed for stiffer hadronic EoSs~\cite{Alford:2013aca}; if $\cQMsq \approx 1/3$ which is characteristic of weakly interacting massless quarks, even with sufficiently stiff hadronic EoSs nearly no \textit{detectable} hybrid configurations could exist (except for ultra-low ($\lesssim1.5\,n_0$) transition densities)~\cite{Alford:2015gna,Chamel:2012ea}. 
We therefore assume that quark matter is maximally stiff ($c_s^2 =1$) throughout this paper to explore the largest parameter space available.

For each hadronic matter EoS, we vary the other two CSS parameters ($\ntrans,\De \ep/\etrans$) while fixing $\cQMsq=1$ 
to specify phase transitions of different onset density and strength. Table~\ref{tab:pt-MR} lists some properties of the obtained hybrid EoSs, such as their transition mass $\Mtrans$ (the maximum hadronic NS mass), the transition radius $\Rtrans$ (the radius of the heaviest hadronic NS), a typical radius $\Rtyp$ (the radius of a $1.4\,\Msolar$ star), and the maximum mass supported $\Mmax$. Mass-radius plots for all the EoSs considered are presented in Fig.~\ref{fig:MR}. 

\begin{table*}[]
\begin{center}
\begin{tabular}{c|@{\quad}c|@{\quad}c|@{\quad}c@{\quad}|@{\quad}c@{\quad}|@{\quad}c@{\quad}|@{\quad}c@{\quad}|@{\quad}c}
\hline 
&&&&&&\\[-2ex]
 EoS&$\ntrans/n_0$&$\De \ep/\etrans$& $\Mtrans/\Msolar$  & $\Rtrans$/${\rm km}$  & $\Rtyp$/${\rm km}$ &$\Mmax/\Msolar$& Case   \\[0.5ex]
  \hline 
&&&&&&\\[-2ex]
 DBHF&N/A&N/A& N/A &N/A & 13.41 &2.31& A   \\[0.5ex]
\hline  
&&&&&&\\[-2ex]
  DBHF\_1032 &1.0&3.2& 0.34 & 14.66 & 8.96 &2.0&  B  \\[0.5ex]
 \hline 
&&&&&&\\[-2ex]
 DBHF\_2010 &2.0&1.0& 1.13  & 13.51 & 11.02 & 2.07 & C \\[0.5ex]
 \hline
 &&&&&&\\[-2ex]
 DBHF\_2507 &2.5&0.7& 1.49  & 13.37 & 13.41 & 2.05 & C \\[0.5ex]
 \hline 
&&&&&&\\[-2ex]
 DBHF\_3004 &3.0&0.4& 1.78 & 13.14 & 13.41 & 2.14 &  C \\[0.5ex]
  \hline 
  &&&&&&\\[-2ex]
 DBHF\_3504 &3.5&0.4& 1.99 & 12.85 & 13.41 & 2.08 &  D  \\[0.5ex]
  \hline 
&&&&&&\\[-2ex]
 SFHo&N/A&N/A& N/A & N/A & 11.97 &2.06& A    \\[0.5ex]
 \hline 
&&&&&&\\[-2ex]
 SFHo\_1031 &1.0&3.1& 0.20 & 17.81 & 8.78  & 2.0 &   B \\[0.5ex]
\hline
&&&&&&\\[-2ex]
 SFHo\_2009 &2.0&0.9& 0.69 & 12.32 & 10.09 & 2.10 &  B \\[0.5ex]
\hline
&&&&&&\\[-2ex]
 SFHo\_2506 &2.5&0.6& 1.01 & 12.13 & 10.67 & 2.07 &  B, C\\[0.5ex]
\hline
&&&&&&\\[-2ex]
 SFHo\_3003 &3.0&0.3& 1.31 & 12.02 & 11.73  & 2.15 &  C\\[0.5ex]
\hline
&&&&&&\\[-2ex]
 SFHo\_3502 &3.5&0.2& 1.55 & 11.88 & 11.97  & 2.13 &  C\\[0.5ex]
\hline
\end{tabular}
\end{center}
\caption{Properties of the EoSs we consider, including the transition density $\ntrans/n_0$, the transition strength $\De \ep/\etrans$, the transition mass $\Mtrans$, 
the transition radius $\Rtrans$, the radius of a $1.4\,\Msolar$ star $\Rtyp$, and the maximum mass supported $\Mmax$. 
The last column denotes which of the characteristic cases discussed in Sec.~\ref{sec:results} each EoS belongs in for a BNS population with masses uniformly distributed in $[1\,\Msolar,\Mmax]$.
}
\label{tab:pt-MR}
\end{table*} 

\section{Detectability estimates}
\label{sec:estimate}

The presence of deconfined quarks in a NS core causes its pressure to decrease, leading to a reduction in the radius of the star compared to that of a purely hadronic NS of the same mass. GW inference of the properties of NSs could then lead to a systematic error in the radius extraction if a phase transition is not taken into consideration. In this section we introduce the basic elements of inferring NS properties from GWs, and present estimates for the potential systematic errors. The magnitude of these errors also offers some indication of whether the effect of phase transitions is measurable: a negligible systematic error would suggest that the effect is weak enough that extracting it would be challenging.

The main finite-size effect on GW signals comes from the tidal deformation the stars experience during the late stages of the coalescence~\cite{Read:2013zra}. The dominant tidal contribution to the phase of the GW enters at the fifth post-Newtonian (5PN) order\footnote{The post-Newtonian series is an expansion in terms of the characteristic velocity of the system $u$ compared to the speed of light. A term proportional to $u^{N/2}$ compared to the leading order term, is referred to as an $N$PN effect.} and it is proportional to the effective tidal deformability parameter~\cite{Flanagan:2007ix,Favata:2013rwa}
\begin{align}
\tilde{\Lambda} &\equiv \frac{16}{13}\frac{(m_1+12m_2)m_1^4 \Lambda_1 +(m_2+12m_1)m_2^4 \Lambda_2}{(m_1+m_2)^5},\label{def:lambdaT}
\end{align}
where $m_i$ are the component masses, $\Lambda_i$ the component tidal deformabilities, and $i\in \{1,2\}$ enumerates the two binary components.
The component tidal deformabilities are defined as $\Lambda_i\equiv 2/3 \,k_{2i} \,R_i^5/m_i^5$, where $k_{2i}$ and $R_i$ are each star's tidal Love number and radius, respectively. The tidal deformability quantifies how easily the NS is deformed under the influence of its companion and is, thus, related to its size and internal structure.
The effective tidal parameter $\tilde{\Lambda}$ is the best measured tidal parameter, and perhaps the only measurable tidal parameter given the detector sensitivities expected in the coming years~\cite{Wade:2014vqa}. It is therefore reasonable to expect that a GW detection of a BNS will result in a posterior estimate of $\tilde{\Lambda}$, and that this estimate is independent of whether the EoS contains a phase transition or not.

Once an unbiased measurement of $\tilde{\Lambda}$ and the individual masses has been achieved from the GW data, further assumptions about the EoS need to be employed in order to arrive at an estimate of the NS radii. Common methods to achieve this involve the use of EoS-insensitive relations~\cite{Maselli:2013mva,Yagi:2015pkc,Chatziioannou:2018vzf,Zhao:2018nyf}, or representations of the EoS directly~\cite{Lackey:2014fwa,Carney:2018sdv,Landry:2018prl}. 
The EoS-insensitive relations, specifically, are certain relations between macroscopic NS properties that do not depend strongly on the EoS assuming the latter is hadronic~\cite{Carson:2019rjx}. They can be used to translate an estimate of the BNS masses and tidal deformabilities into estimates of the NS radii that are valid under the assumption of a hadronic EoS. If this assumption is violated, then the resulting radius estimate will suffer from a systematic bias which we explore below.

For this study we follow the approach of~\cite{Chatziioannou:2018vzf,TheLIGOScientific:2017qsa} and use the two relations proposed in~\cite{Yagi:2015pkc,Maselli:2013mva}. The first relation connects the tidal deformability of one star $\Lambda_2$ to the tidal deformability of the other star $\Lambda_1$ and the ratio of their masses $q=m_2/m_1$~\cite{Yagi:2015pkc,Chatziioannou:2018vzf}. This allows us to express the effective tidal deformability $\tilde{\Lambda}(\Lambda_1,\Lambda_2,q)$ as a function of the mass ratio and one component tidal deformability only $\tilde{\Lambda}(\Lambda_1(\Lambda_2,q),\Lambda_2,q)$. The second relation expresses the compactness of a star as a function of its tidal deformability $C_i(\Lambda_i)$~\cite{Maselli:2013mva} and can be used to compute the radius through $R_i=m_i/C_i(\Lambda_i)$. An improved fit with more data has been presented in~\cite{Yagi:2016bkt}, the results of which are utilized in the following. In summary, we use the GW data to measure the component masses and the effective tidal deformability. We then restrict to a purely hadronic EoS and use the relation $\Lambda_1(\Lambda_2,q)$ to translate the measurement of $\tilde{\Lambda}$ into a measurement of $\Lambda_2$. The relation $C_2(\Lambda_2)$ is then used to obtain an estimate of $R_2$, and a similar procedure can be used to obtain $R_1$.

The above procedure applied to GW170817 results in a radius measurement of $R_1 \sim R_2 = 10.7^{+2.0}_{-1.7}$~km at the $90\%$ credible level, for a radius uncertainty of about $3.7$~km~\cite{TheLIGOScientific:2017qsa}. This uncertainty can be reduced if one takes into account the existence of a $1.97\,\Msolar$ star~\cite{Antoniadis:2013pzd} through the spectral EoS parametrization of~\cite{2012PhRvD..86h4003L,Lindblom:2013kra,Carney:2018sdv}, reducing the radius uncertainty to $2.8$~km at the $90\%$ level. The uncertainty can be further reduced by incorporating information from the EM emission from GW170817~\cite{Radice:2018ozg,Coughlin:2018fis} or working within a nuclear physics framework~\cite{Capano:2019eae}.

For this study we are interested in the effect of a sharp phase transition in the EoS on the radius estimate. For each CSS model of Fig.~\ref{fig:MR}, we consider BNS systems with different values of the component masses. Then for each of these systems, we assume that analysis of the GW data can return an unbiased measurement of $\tilde{\Lambda}$, and follow the steps outlined above and compute $R_1$ and $R_2$ assuming the EoS is hadronic through the relations of~\cite{Yagi:2015pkc,Yagi:2016bkt}. The assumption of a purely hadronic EoS is obviously wrong here, so we expect a biased estimate of the radii. We plot the difference between the inferred and the true radius as a function of the systems' chirp mass ${\cal{M}}$ and mass ratio $q$.

\begin{figure*}[]
\includegraphics[width=.65\columnwidth,clip=true]{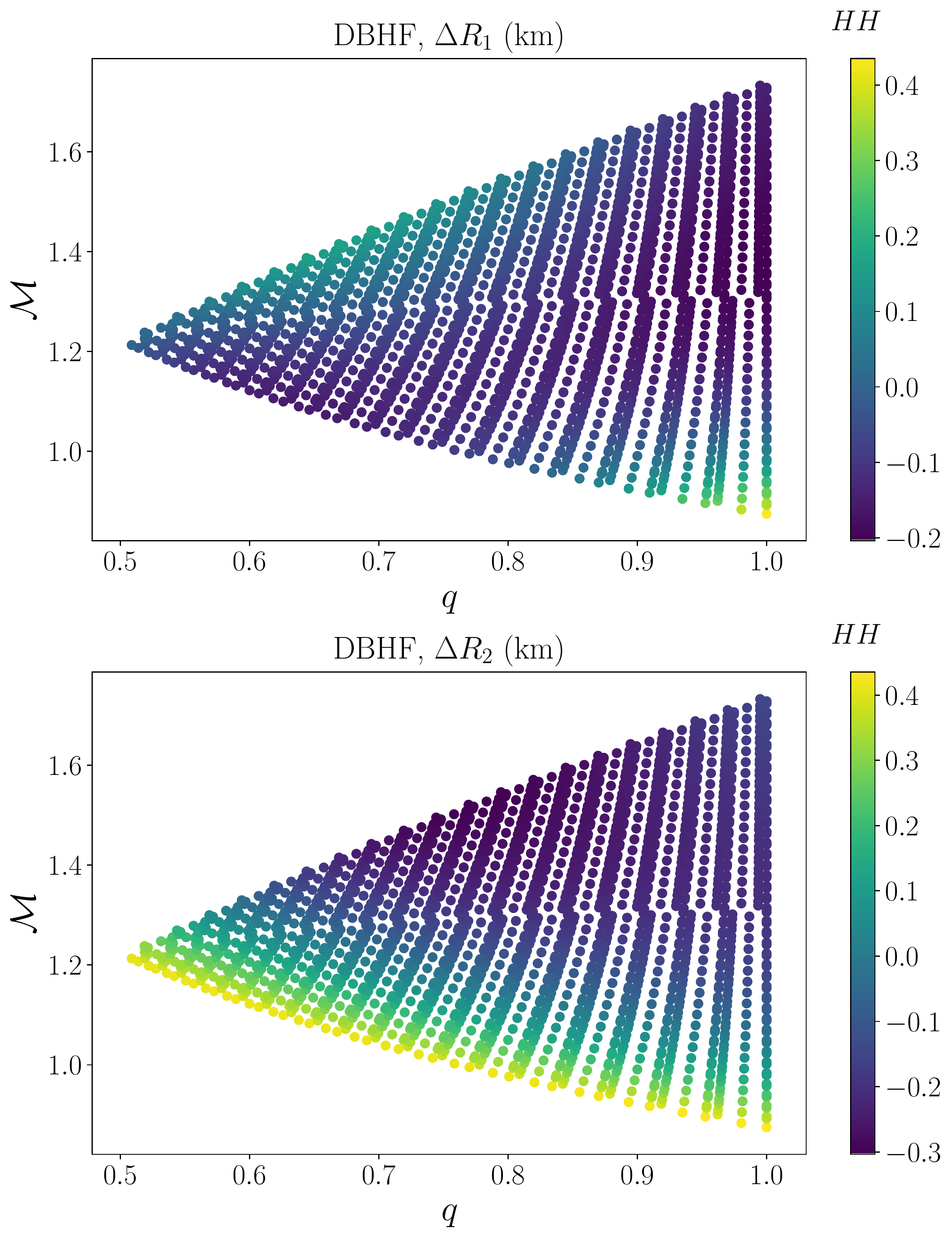}
\includegraphics[width=.65\columnwidth,clip=true]{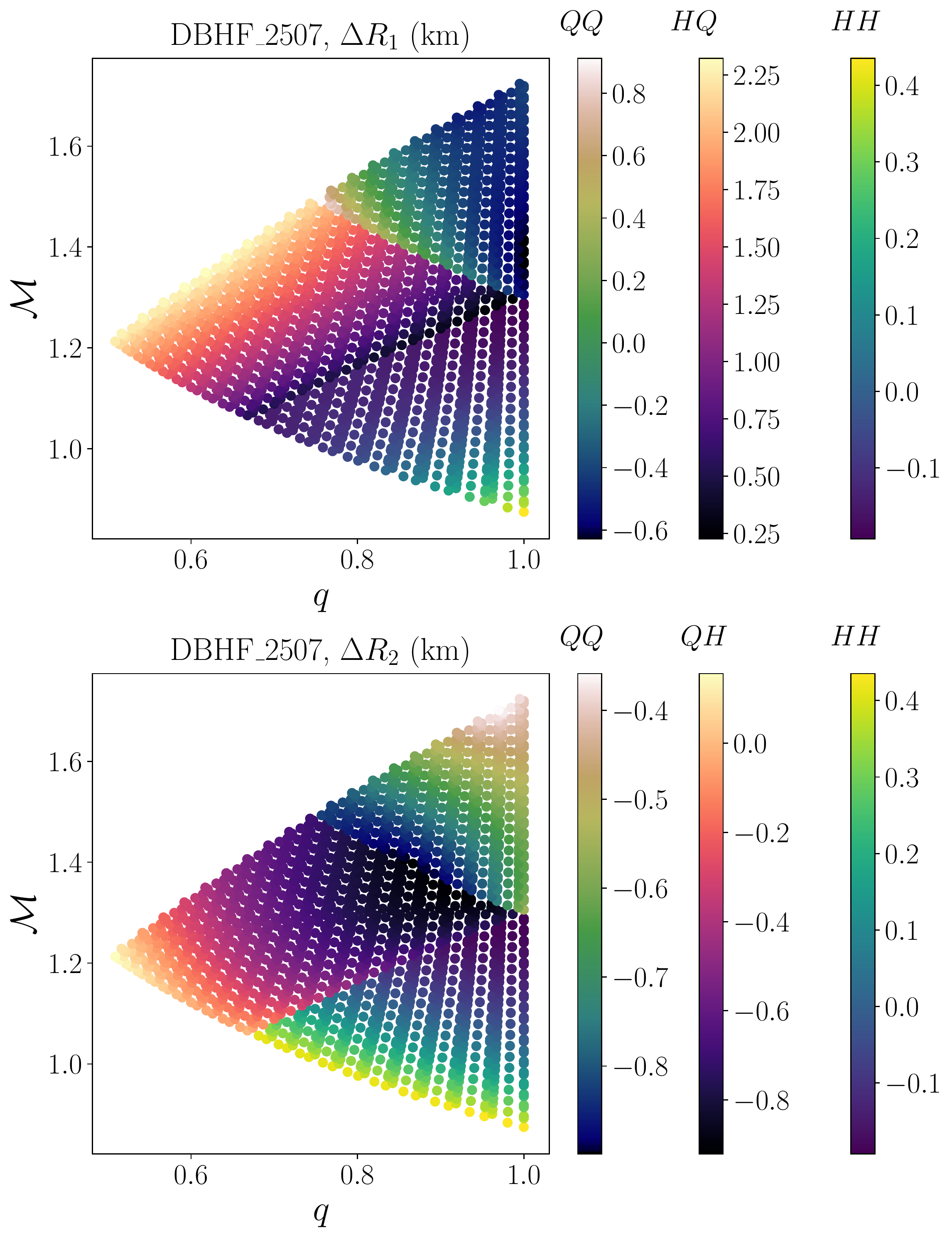}
\includegraphics[width=.67\columnwidth,clip=true]{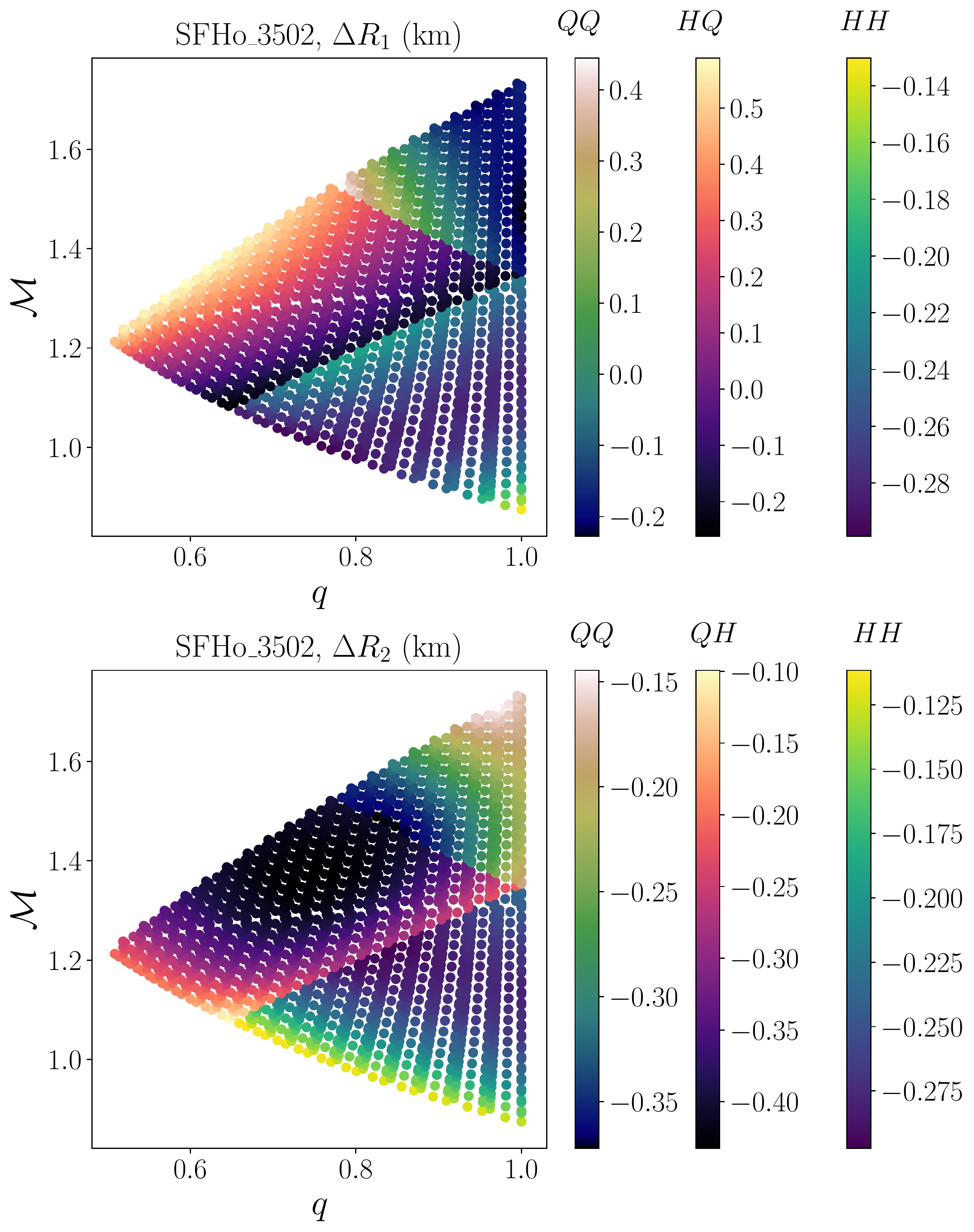}
\caption{Absolute value of the difference between the true radius and the inferred radius for different EoSs with and without phase transitions
when analyzed with EoS-insensitive relations that assume hadronic EoSs. Different color maps correspond to
different binary combinations of hadronic and hybrid stars. In all cases here and in Fig.~\ref{fig:DeltaR2}, the radius error is less than $\sim 2.5$~km.}
\label{fig:DeltaR}
\end{figure*}

The results are in Figs.~\ref{fig:DeltaR} and~\ref{fig:DeltaR2} for different EoSs with phase transitions. Each plot presents the difference $\De R$ between the true and the inferred radius for each binary component for a particular EoS (see plot title). The different color maps correspond to different types of binaries with each EoS: ``HH'' -- two hadronic stars, ``HQ'' -- one hadronic and one hybrid star, ``QQ'' -- two hybrid stars. In all cases, purely hadronic binaries (``HH'') have a radius bias of less than 500~m, which is consistent with the quoted systematic error in the EoS-insensitive relations~\cite{Yagi:2015pkc,Yagi:2016bkt}. 
This is most evident on the left panel for the purely hadronic DBHF EoS: all BNSs have a radius error of less than $\sim0.4$~km. 
We obtain similar radii errors in the ``HH'' case when instead using the relations proposed in~\cite{Zhao:2018nyf}.

When the binary contains at least one hybrid star (``HQ'' and ``QQ'' cases) the systematic error in the inferred radius can be much larger depending on the EoS and the masses of the system. EoSs that lead to large deviations with respect to their hadronic baseline in Fig.~\ref{fig:MR} are also expected to lead to large values of $\De R_i$.
For example, DBHF\_2010 (purple line in the left panel of Fig.~\ref{fig:MR}) contains a hybrid branch that leads to a radius reduction of $2-3$~km compared to the hadronic baseline DBHF. Moreover, the hybrid branch is qualitatively different from the hadronic one as it has a large slope of $dR/dM$. As a result, DBHF\_2010 leads to some of the largest systematic errors in the radius measurement (middle panel of Fig.~\ref{fig:DeltaR}), exceeding $2$~km for certain systems. 
On the other hand, SFHo\_3502 (orange line in the right panel of Fig.~\ref{fig:MR}) leads to a radius reduction of less than $500$~m, and only for stars heavier than $\sim 1.6\,\Msolar$. Unsurprisingly, the systematic error from using hadronic-based EoS-insensitive relations with this hybrid EoS is also small (right panel of Fig.~\ref{fig:DeltaR}). Additional plots for the other EoSs is included in Appendix~\ref{AppA}. 
Interestingly, in all cases we find that the error in the radius is smaller than the inferred statistical uncertainty for GW170817~\cite{TheLIGOScientific:2017qsa} of $3.7$~km using this method, which was derived under the assumption of normal hadronic EoS throughout all densities. 
Nevertheless, we emphasize that such estimates are not directly comparable to predictions from EoS models that have incorporated phase transitions.

The above results give an indication of what type of phase transitions and EoSs can be detectable with future GW measurements.
Cases where the systematic error is small are either too weak to measure, or the hybrid branch can be mistaken for a hadronic branch possibly with a small systematic error in the radius. In the next section, we investigate the detectability of phase transitions with populations of BNSs and how well the parameters of the hybrid EoSs can be constrained.

\section{A population of BNS signals}
\label{sec:popModel}

A single loud event, such as GW170817, can offer a reliable measurement of $\tilde{\Lambda}$. However, in order to identify the effect of a phase transition in the EoS and estimate its parameters, we need to consider multiple such detections. In this section, we describe how we study phase transitions with a population of BNS signals, as well as how we simulate such possible future detections.

\subsection{Chirp radius}
\label{sec:chirpR}

\begin{figure*}[]
\parbox{0.48\hsize}{
\includegraphics[width=\hsize]{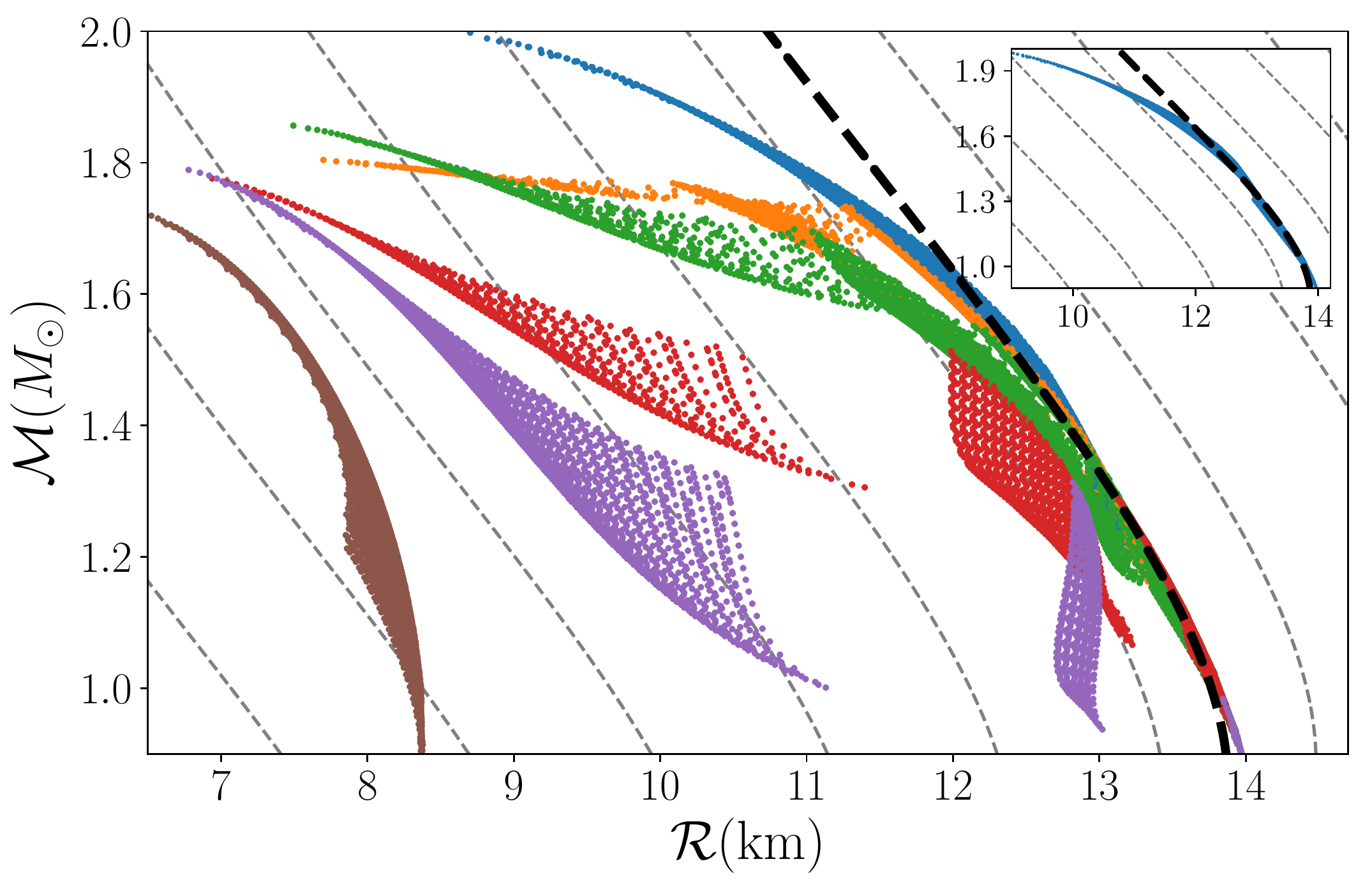}\\[-2ex]
}\parbox{0.48\hsize}{
\includegraphics[width=\hsize]{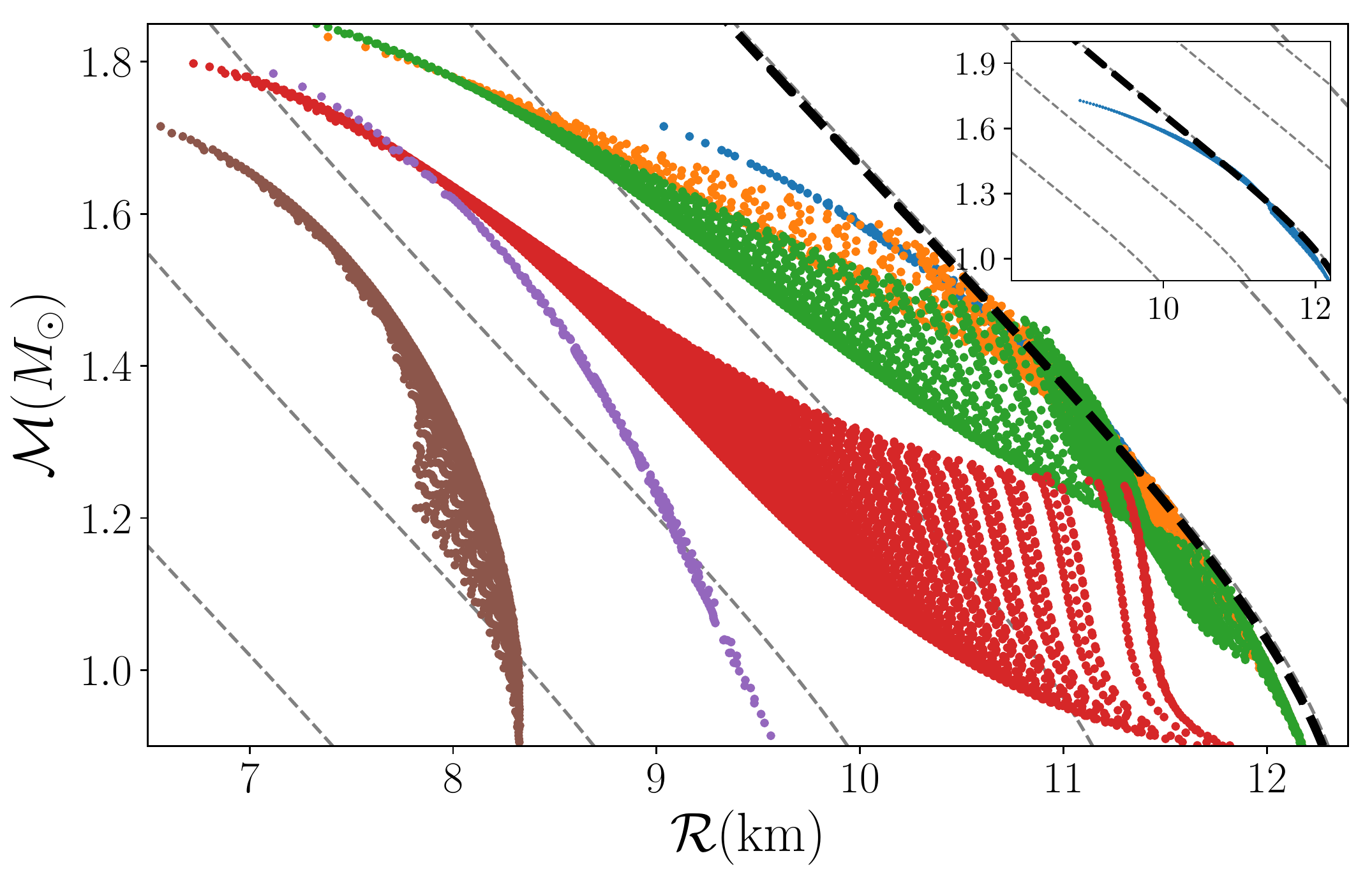}\\[-2ex]
}
\caption{Chirp mass and chirp radius for binary systems with the EoSs from Fig.~\ref{fig:MR} with the DBHF (left) and the SFHo (right) hadronic EoS baseline. Each dot represents a BNS system with masses randomly selected in $[1\,\Msolar,\Mmax]$; the color scheme follows Fig.~\ref{fig:MR}. The grey dashed lines are the ${\cal{M}}({\cal{R}};R)$ fit for purely hadronic EoSs for increments of $R$ of $1$~km. The thick black dashed line is the fit for $R=\Rtyp$, the radius of a $1.4\,\Msolar$ NS with each hadronic baseline model. The inset in each plot shows the hadronic EoS (blue dots) and the fits. The ${\cal{M}}({\cal{R}};R)$ fit is based on EoS-insensitive relations and it can model the purely hadronic systems up to a chirp mass of $\sim1.6\,\Msolar$. 
}
\label{fig:ChirpRFit}
\end{figure*}

In order to study the detectability of phase transitions from a population of BNSs, we consider the ``chirp radius" parameter
introduced in Wade \textit{et al.}~\cite{Wade:2014vqa} and defined as
\begin{equation}
{\cal{R}} \equiv 2 {\cal{M}} \tilde{\Lambda}^{1/5},
\label{def:Rchirp}
\end{equation}
where the chirp mass is $\Mchirp=(m_{1}m_{2})^{3/5}/(m_1+m_2)^{1/5}$. Figure~\ref{fig:ChirpRFit} shows the chirp mass and the chirp radius for various EoSs with phase transitions (
colors follow the scheme of Fig.~\ref{fig:MR}) and a purely hadronic EoS (blue). For this plot we select BNSs with random masses uniformly distributed in $[1\,\Msolar,\Mmax]$, compute the chirp mass ${\cal{M}}$ and chirp radius ${\cal{R}}$ for a given EoS, and plot them with a dot. For the hadronic EoS (blue dots, see plot inset) the possible BNS systems fall on a curve with a small width caused by the subdominant dependence of $\tilde{\Lambda}$, and by extension ${\cal{R}}$, on the binary mass ratio. This approximate independence of $\tilde{\Lambda}$ on the mass ratio $q$ was also observed in~\cite{Wade:2014vqa} and later used to translate the GW170817 bound on $\tilde{\Lambda}$ into a bound on the NS radius~\cite{Raithel:2018ncd}.

The EoS-insensitive relations discussed in the previous section can be used to construct an approximate relation ${\cal{R}}({\cal{M}},R)$ that holds for hadronic EoSs, where $R$ is the (approximately constant) radius of the hadronic stars. Specifically, for a given $R$ and two NSs of mass $m_1=m_2$ (we take the two masses to be equal due to the approximate independence of ${\cal{R}}$ on the mass ratio $q=m_2/m_1$) we use the $C(\Lambda)$ relation to compute
their respective tidal deformabilities $\Lambda_1=\Lambda_2$. From those we compute $\tilde{\Lambda}$ through \Eqn{def:lambdaT} and ${\cal{R}}$ through \Eqn{def:Rchirp}.
We plot the relation ${\cal{R}}({\cal{M}},R)$ in Fig.~\ref{fig:ChirpRFit} for different values of $R$ incremented by $1$~km with grey dashed lines. The black dashed line is the fit for $R=\Rtyp$, the radius of a $1.4\,\Msolar$ star for DBHF (left) and SFHo (right). We find that the fit accurately reproduces the behavior of the hadronic EoSs (blue dots) up to a chirp mass of about $\Mchirp=1.6\,\Msolar$.

The presence of a strong phase transition in the EoS has a large effect on the tidal parameter $\tilde{\La}$ and the chirp radius $\Rchirp$. 
All EoSs with phase transitions shown in Fig.~\ref{fig:ChirpRFit} deviate from the hadronic fit by multiple kms, with some systems even forming completely disjointed islands. This observation suggests a way to identify the presence of phase transitions from GW observations of a population of BNSs. Each system will result in an estimate of the chirp mass and the chirp radius. 
If the BNS is composed of two hadronic stars, then we expect its ${\cal{M}}$ and ${\cal{R}}$ to fall on, or close to, the theoretically expected curve within measurement errors. If, on the other hand, one or two binary components are hybrid stars, then the measured ${\cal{M}}$ and ${\cal{R}}$ will deviate from the expected curve to some degree that depends on both the masses of the stars and the details of the EoS.
For a population of BNSs with different chirp masses, we expect the hadronic binaries to follow a unique curve parametrized by their almost constant hadronic radius $R$, while the hybrid systems will deviate from that curve.

\subsection{The hierarchical model}
\label{hiermodel}

Most EoS models with phase transitions predict normal hadronic stars up to some transition mass $\Mtrans$ and hybrid stars containing quark cores above that mass\footnote{We neglect the effects from ``rising twins''~\cite{Schertler:2000xq,Christian:2018jyd} where the heaviest hadronic stars are slightly more massive than the lightest hybrid stars, as the relevant mass range is typically very small and hardly discernible by tidal deformability measurements~\cite{Montana:2018bkb,Han:2018mtj}.}. 
In a population of BNS systems, we expect that systems with $m_2<m_1<\Mtrans$ approximately follow the expected ${\cal{R}}({\cal{M}},R)$ curve, while systems with $m_2<\Mtrans<m_1$ or $\Mtrans<m_2<m_1$ deviate from it to some degree that depends on the masses and the EoS. In order to estimate $\Mtrans$ and $R$ from the BNSs data, we employ a hierarchical analysis along the lines laid out in~\cite{Stevenson:2017dlk}.

We consider $N$ detections of GW signals from BNSs, and assume that the GW data for each event lead to a posterior measurement for the component masses and the tidal parameter $\tilde{\Lambda}$ similar to that of GW170817~\cite{TheLIGOScientific:2017qsa}; the latter can be trivially transformed into a posterior for the chirp radius ${\cal{R}}$. The GW data for each detection $j$ are denoted $d_j$. We define $\theta_j\equiv({\cal{M}}_j,q_j,{\cal{R}}_j)$ to be the source parameters of interest, and $x\equiv(\Mtrans,R,\sigma_1,{\cal{R}}_2,\sigma_2)$ are the population parameters, some of which will be defined shortly.

The BNS events are independent, so the likelihood for the data $d=\{d_j\}$ is the product of the individual event likelihoods ${\cal{L}}(d|x)=  \prod_j^N {\cal{L}}(d_j|x)$. The latter are given by~\cite{Mandel:2009pc}
\begin{equation}
{\cal{L}}(d_j|x)=\int d \theta_j p(d_j|\theta_j)p(\theta_j|x),\label{L}
\end{equation}
where $p(d_j|\theta_j)$ is the likelihood of the source parameters computed by the GW data, and $p(\theta_j|x)$ corresponds to the population model. For the latter we use
\begin{align}
p(\theta_j|x)&=p({\cal{M}}_j,q_j,{\cal{R}}_j|R,\Mtrans,\sigma_1,{\cal{R}}_2,\sigma_2)\nn \\
&\sim
 \begin{cases}
      {\cal{N}}[{\cal{R}}({\cal{M}}_j,R),\sigma_1]({\cal{R}}_j), & \text{if}\ m_{1j}< \Mtrans\\
      {\cal{N}}[{\cal{R}}_2,\sigma_2]({\cal{R}}_j), & \text{otherwise}
    \end{cases}\label{popL}
\end{align}
where ${\cal{N}}[\mu,\sigma](y)$ is a normal distribution for the parameter $y$ with a mean $\mu$ and variance $\sigma^2$.

Our population model in \Eqn{popL} consists of two components which are designed to follow the hadronic and hybrid branches of an EoS. If the heavier component is lighter than $\Mtrans$, then the binary is hadronic and its chirp radius ${\cal{R}}_i$ follows a Gaussian distribution centered at ${\cal{R}}({\cal{M}}_i,R)$ and with a standard deviation $\sigma_1$.
Otherwise, the binary is considered to be hybrid and the chirp radius is distributed according to ${\cal{N}}[{\cal{R}}_2,\sigma_2]$. The parameters of the population model are: (i) $\Mtrans$, the mass above which NSs have a quark core, (ii) $R$, the typical radius of the hadronic branch of the EoS, (iii) $\sigma_1$, the standard deviation of the inferred chirp radii around the hadronic fit ${\cal{R}}({\cal{M}},R)$, (iv) ${\cal{R}}_2$, the mean, and (v) $\sigma_2$, the standard deviation of the distribution of the chirp radii for the hybrid BNSs. Additionally, the true chirp radii (as opposed to those measured with the GW data) are also taken to be free parameters of the hierarchical model, as described in~\cite{Hogg_2010}.

We compute the posterior distribution for these parameters employing the following priors. The prior for the transition mass $\Mtrans$ is taken to be uniform in $[1\,\Msolar,\Mmax]$. Although a number of EoS models we test have a lower transition mass, we assume that the detected NSs have masses above $1\,\Msolar$ and therefore restrict $\Mtrans$ accordingly. The hadronic radius $R$ is taken to be flat in $[10,14]$~km. The upper limit is consistent with existing constraints and does not affect our results. The lower limit, on the other hand, is critical as we explain below. We choose a lower limit of $10$~km which is the smallest hadronic EoS radius that can support a $2\,\Msolar$ NS~\cite{Antoniadis:2013pzd} as shown in Fig. 1 of~\cite{Annala:2017llu}\footnote{The study of~\cite{Annala:2017llu} used two parametrizations for the hadronic EoS, one with three and the other with four polytropic segments.
The lower radius they arrive at is $\sim 11$~km and $\sim 10$~km, respectively. We choose $10$~km to be conservative.}, which is also consistent with the results of~\cite{Abbott:2018exr}. The parameter $\sigma_1$ captures the scatter of hadronic binaries around the expected ${\cal{R}}({\cal{M}},R)$ fit, so we determine a suitable prior by fitting hadronic-only populations with a hierarchical model that includes the first component of \Eqn{popL} only. We find that a log-normal distribution with parameters $\mu=\log(0.15/\sqrt{N})$ and $\sigma^2=0.4$  is a reasonable fit. For $\sigma_2$ we choose a wider log-normal distribution, since this parameter captures the much larger scatter of the hybrid chirp radii. We empirically choose parameters $\mu=\log(3)$ and $\sigma^2=1$. For the mean of the chirp radii of the hybrid stars we choose a uniform distribution  $[7,{\cal{R}}(1.2,R)]\,\Msolar$. The upper limit is chosen as to impose that the hybrid binaries have smaller chirp radii than the hadronic binaries, as expected from Fig.~\ref{fig:ChirpRFit}. We also choose to evaluate the upper limit of ${\cal{R}}(1.2,R)$ at a chirp mass of $\Mchirp=1.2\,\Msolar$, typical of the values expected for galactic NS binaries~\cite{Tauris:2017omb}.

\subsection{Simulated population}

We apply the above model to simulate populations of BNS GW signals. The parameters of interest to be extracted from the GW data are $\theta\equiv({\cal{M}},q,{\cal{R}})$. Posterior distributions for all source parameters can be computed by sampling the high-dimensional parameter spaces~\cite{Veitch:2014wba}, but this is computationally intractable for the thousands of simulated signals we investigate here. For this reason we describe below how we approximate the measured uncertainty in the source parameters. 

For a given EoS model from Sec.~\ref{sec:CSS}, we select BNS masses according to two mass distributions. In the first case, both masses are drawn uniformly in $[1\,\Msolar,\Mmax]$. In the second case, the component masses follow the observed galactic BNS distribution, which is a Gaussian with a mean of $1.32\,\Msolar$ and the standard deviation of $0.1\,\Msolar$~\cite{Kiziltan_2013}. In both cases we impose $m_2<m_1$. The tidal parameter $\tilde{\La}$ of the binary and the chirp radius are computed from the masses and the EoS model. The signal-to-noise ratio $\rho$ of the system is drawn from a $\sim 1/\rho^{-4}$ distribution~\cite{Chen:2014yla}, the expected distribution for local events for which cosmological effects are subdominant.

The uncertainty in each inferred source parameter, i.e. the term $p(d_i|\theta_i)$ in \Eqn{L}, is inversely proportional to the signal-to-noise of the event. We assume a perfect measurement of the chirp mass $\Mchirp$, since the typical expected relative measurement errors are ${\cal{O}}(10^{-3})$~\cite{Farr:2015lna}. For the mass ratio $q$ we simulate a relative measurement error of $0.15$ at the 1-$\sigma$ level for a binary with $\rho=10$~\cite{Farr:2015lna}. 
To obtain an estimate of the uncertainty in $\tilde{\Lambda}$ we consider the parameter estimation studies presented in~\cite{Wade:2014vqa,Dudi:2018jzn,Samajdar:2018dcx}. Table I of~\cite{Wade:2014vqa} quotes absolute measurement uncertainties for $\tilde{\Lambda}$ for simulated BNS signals of different NS mass combinations. They report a measurement uncertainty between 100 and 700 for an SNR of $30$ at the $90\%$ level (which we obtain by converting their quoted 2-$\sigma$ error bars). 
For our simulated population, we use an uncertainty of $400$ at the $90\%$ level for SNR $30$ for $\tilde{\Lambda}$ which is slightly larger than the median uncertainty in~\cite{Wade:2014vqa}. This uncertainty should in principle depend on the true masses, but in the absence of a model to simulate that, we keep it fixed for all masses and use the approximate median value of~\cite{Wade:2014vqa}. 
The uncertainty we use is also consistent with the results of~\cite{Dudi:2018jzn,Samajdar:2018dcx}, while the fact that the absolute uncertainty on $\tilde{\Lambda}$ is not a strong function of the true value of $\tilde{\Lambda}$ is also shown in~\cite{Dudi:2018jzn}. The uncertainty in the chirp radius $\Rchirp$ can be obtained from the $\tilde{\Lambda}$ uncertainty through error propagation.

We approximate all likelihood distributions for the relevant source parameters as Gaussians centered at the true value of each parameter plus a random shift due to noise realization, and a width determined by the uncertainties described above. The shift due to noise realization in the GW detectors is random and distributed according to a zero-mean Gaussian with a standard deviation equal to the parameter measurement uncertainty.

\section{Results and Discussion}
\label{sec:results}

We create random populations of BNSs with the EoSs described in Sec.~\ref{sec:CSS}, and compute posterior distributions for the population parameters $\Mtrans$ and $R$ using the hierarchical model described in Sec.~\ref{sec:popModel}. We discuss what constraints we can place on the parameter space of the phase transition and how many detections are needed.

\subsection{Characteristic cases}
\label{cases}

We expect that our ability to identify or constrain phase transitions depends sensitively on the properties of the EoS, specifically the onset density and transition strength, as well as the masses of the observed BNSs. For this reason we begin by examining single populations of 100 simulated BNSs with different EoSs, and discuss below the different characteristic cases possible depending on the values of $\Mtrans$, $\De \ep/\etrans$, and the observed BNS masses. Unless otherwise noted, examples presented in this subsection are obtained with a uniform NS mass distribution. In Table~\ref{tab:pt-MR} we summarize which case each EoS belongs in. Posterior densities for the transition mass $\Mtrans$ (left) and the hadronic radius $R$ (right) for all cases are presented in Figs.~\ref{fig:Pop-CaseA}--\ref{fig:Pop-CaseD}.

\subsubsection{Case A: Purely hadronic EoS}

\begin{figure}[]
\includegraphics[width=.46\columnwidth,clip=true]{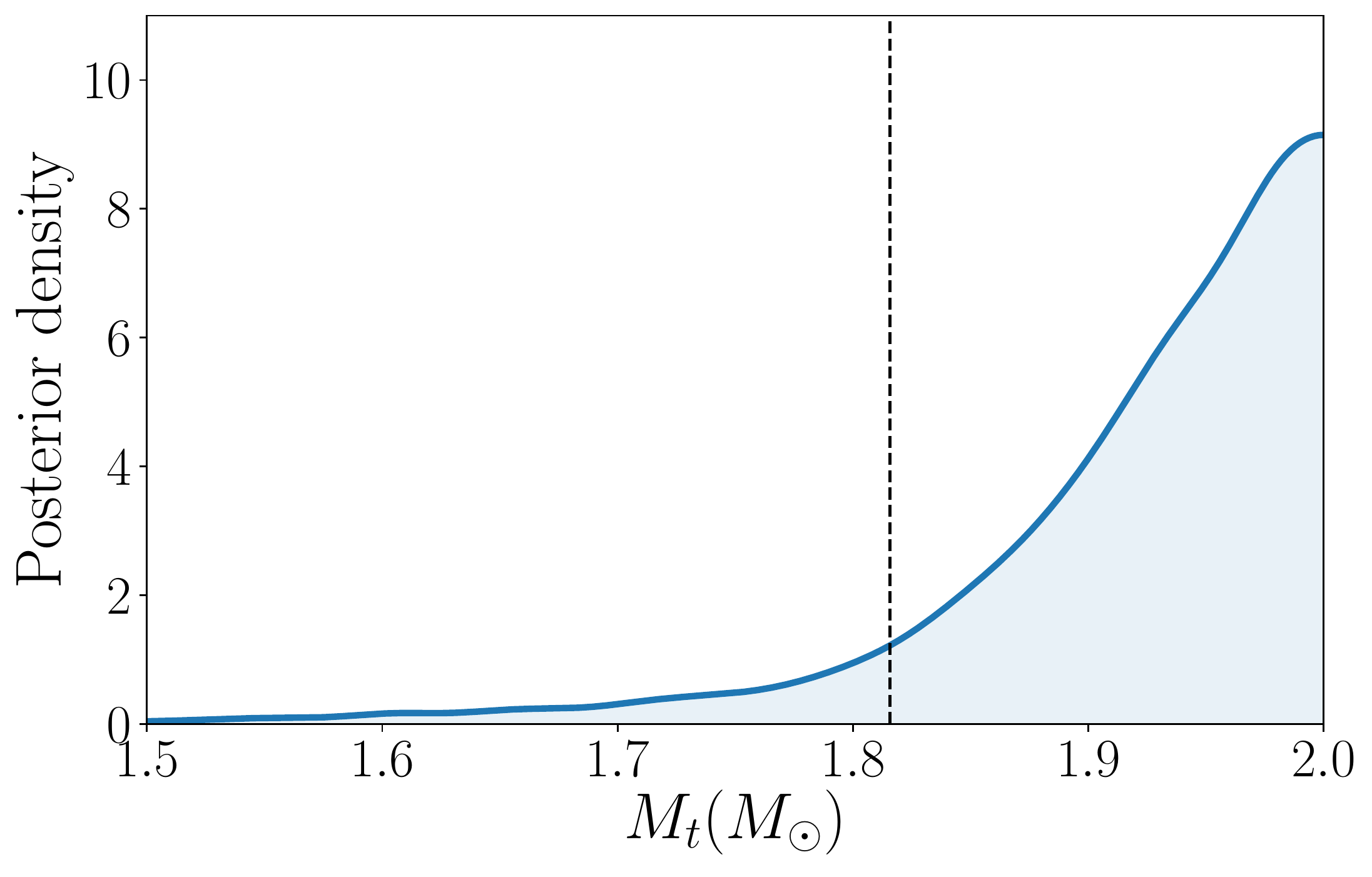}
\includegraphics[width=.45\columnwidth,clip=true]{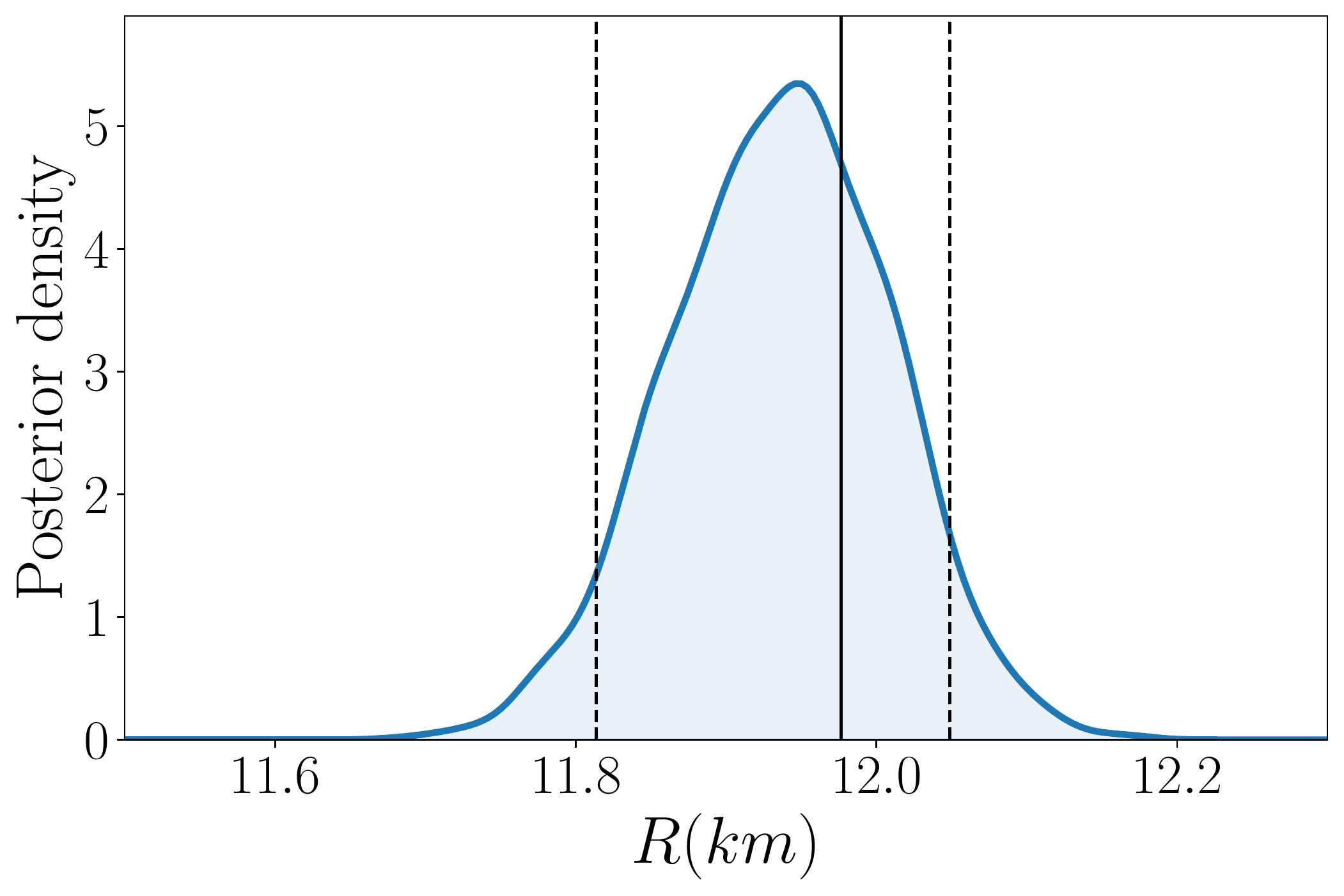}\\
\caption{Case A: Marginalized 1-dimensional posterior density for the transition mass $\Mtrans$ (left) and the hadronic radius $R$ (right) for a random population of 100 BNSs with the purely hadronic
EoS SFHo. Solid vertical lines denote the true parameters, where applicable. Dashed vertical lines denote the $90\%$ credible lower limit (left) and symmetric interval (right).
}
\label{fig:Pop-CaseA}
\end{figure}

The most simple case is that of a purely hadronic EoS for all densities. In our set of EoSs we have two such models, namely DBHF and SFHo, which also serve as the hadronic baselines for the hybrid EoSs within the CSS parametrization. 
We simulate 100 BNS detections from a uniform mass distribution, and plot  in Fig.~\ref{fig:Pop-CaseA} the posterior density for the transition mass $\Mtrans$ and the hadronic radius $R$ for the case of the SFHo EoS. 
We obtain qualitatively similar results with DBHF, though note that this EoS is mildly inconsistent with GW170817. 

We find that the hadronic radius $R$ is reliably extracted with an accuracy of $\sim 200$~m at the 90\% level, suggesting that the systematic error from assuming a universal ${\cal{R}}({\cal{M}},R)$ fit is subdominant. In this case of the absence of a phase transition, the posterior for $\Mtrans$ can be used to place a lower limit on the onset mass/density; we find that $\Mtrans>1.81\,\Msolar$ at the $90\%$ level. 
This limit on $\Mtrans$ is lower than the heaviest NS in our population, suggesting that some heavy systems cannot be confidently classified as hadronic or hybrid. Indeed, the tidal deformability of a NS decreases rapidly with increasing mass, making the measurement of the tidal properties of heavy systems challenging. 
Moreover, as we discuss below, a similar constraint on $\Mtrans$ would be obtained for an EoS with a phase transition at high enough mass (case D).

\subsubsection{Case B: EoS with a phase transition below the minimum NS mass}
\label{sec:caseB}

\begin{figure}[]
\includegraphics[width=.45\columnwidth,clip=true]{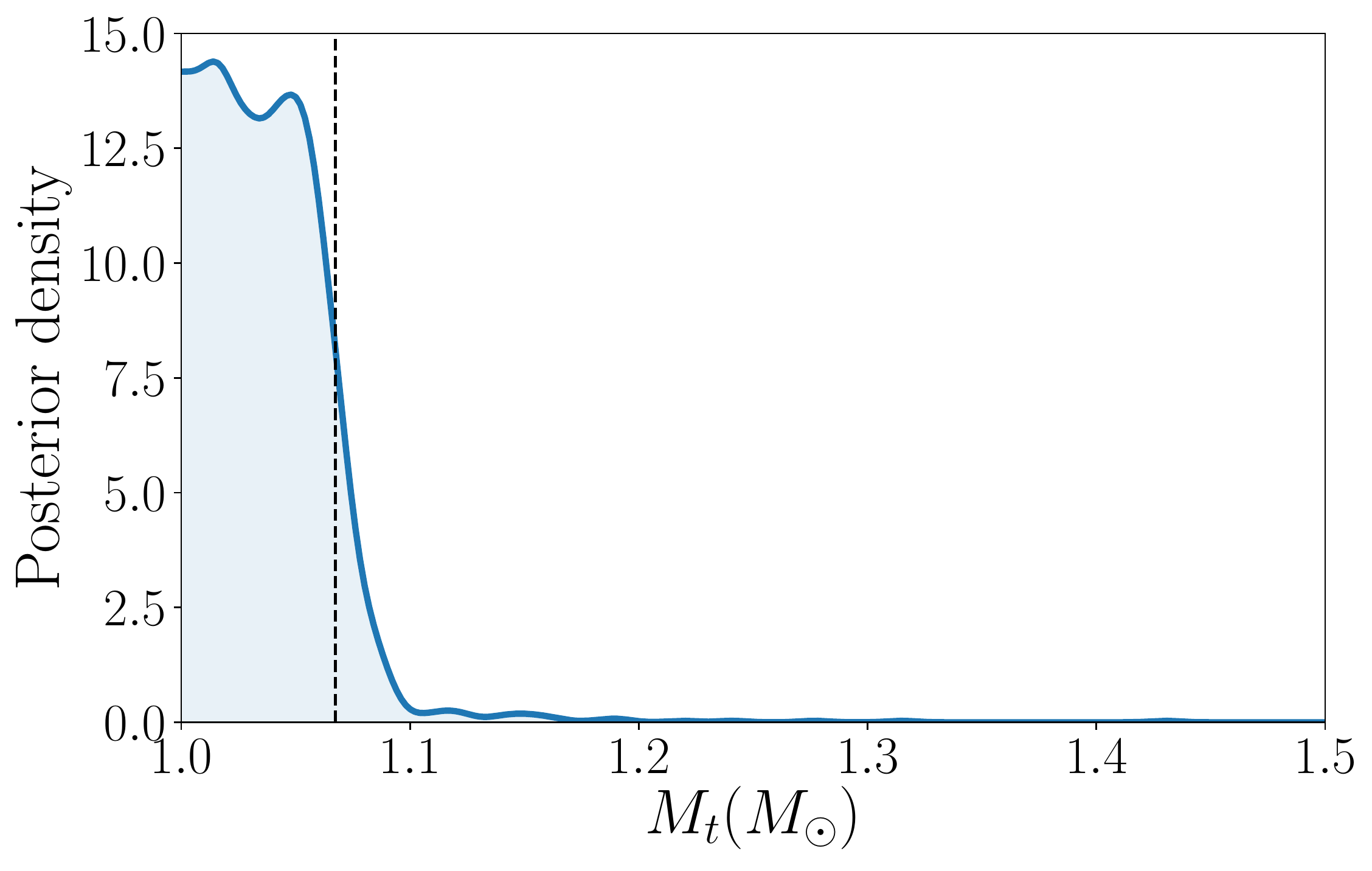}
\includegraphics[width=.45\columnwidth,clip=true]{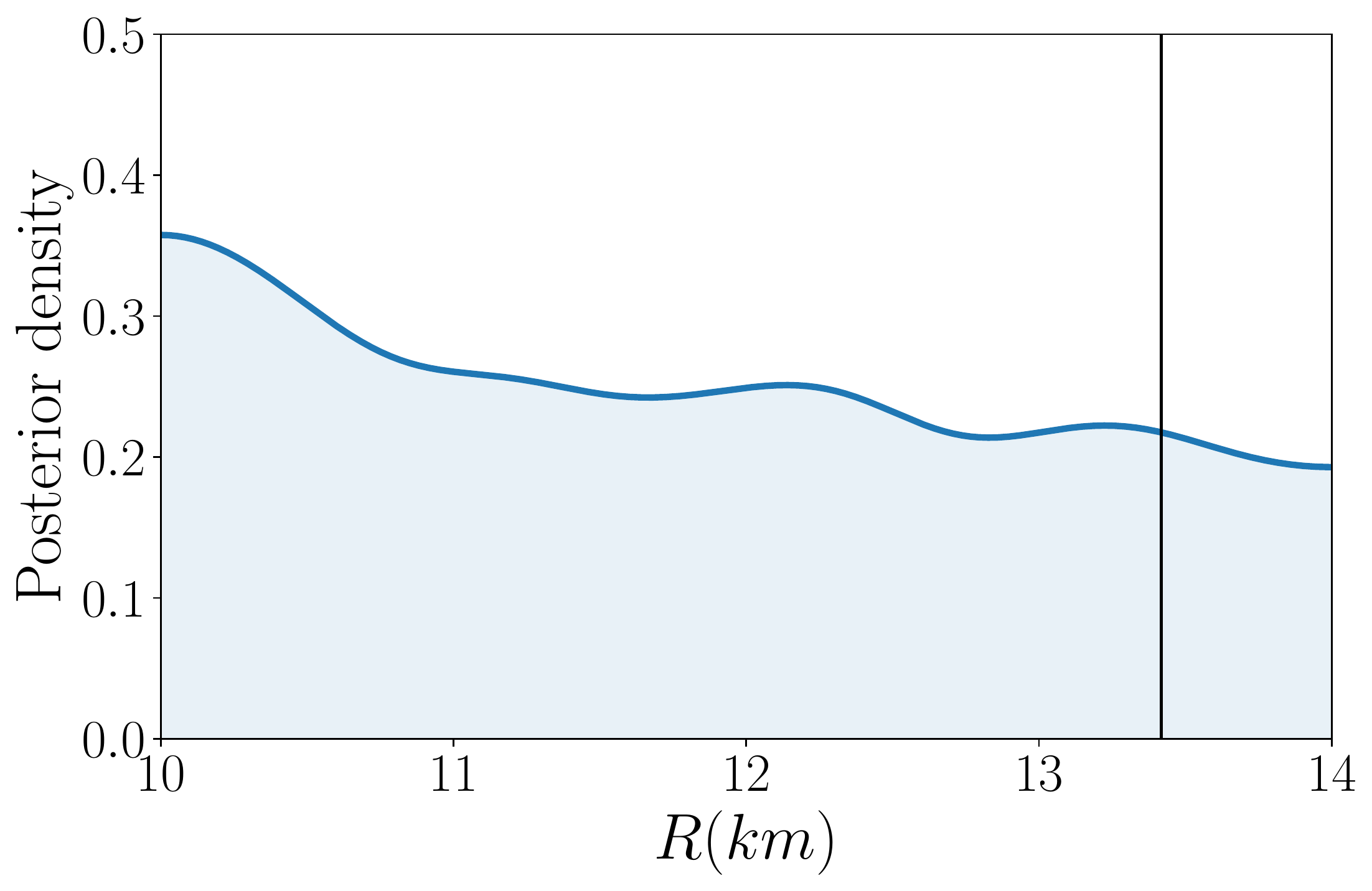}\\
\includegraphics[width=.45\columnwidth,clip=true]{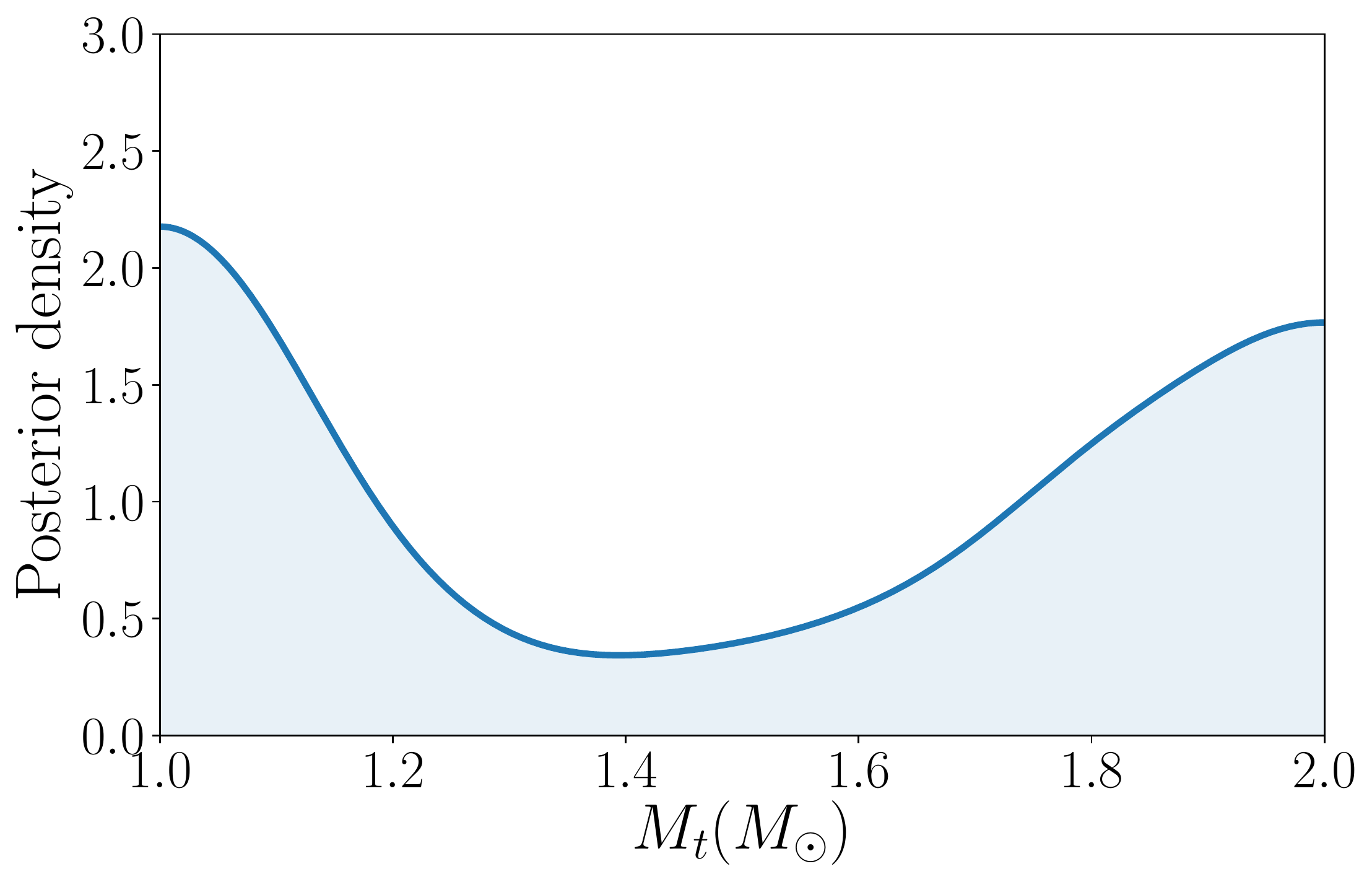}
\includegraphics[width=.45\columnwidth,clip=true]{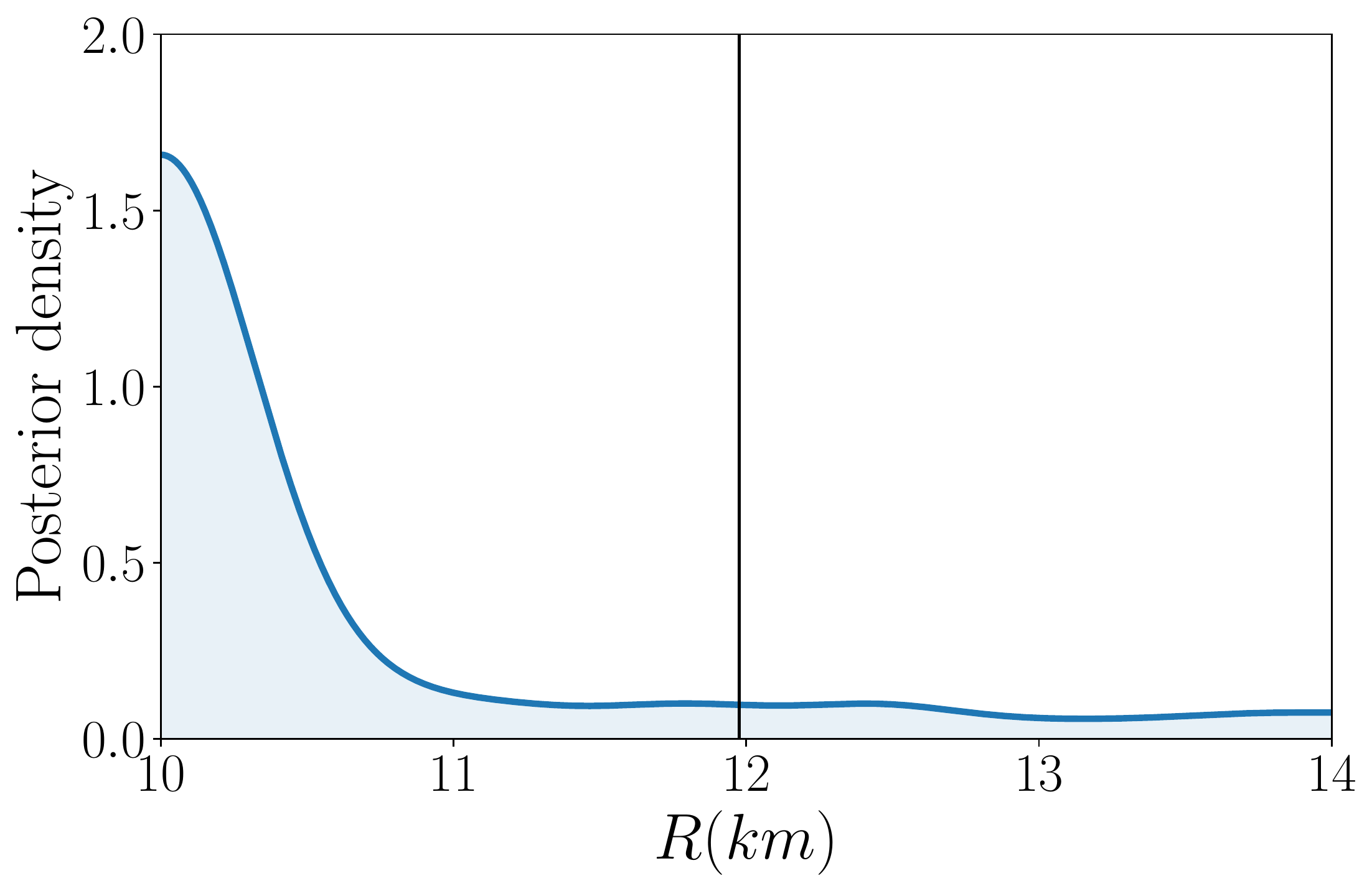}\\
\includegraphics[width=.45\columnwidth,clip=true]{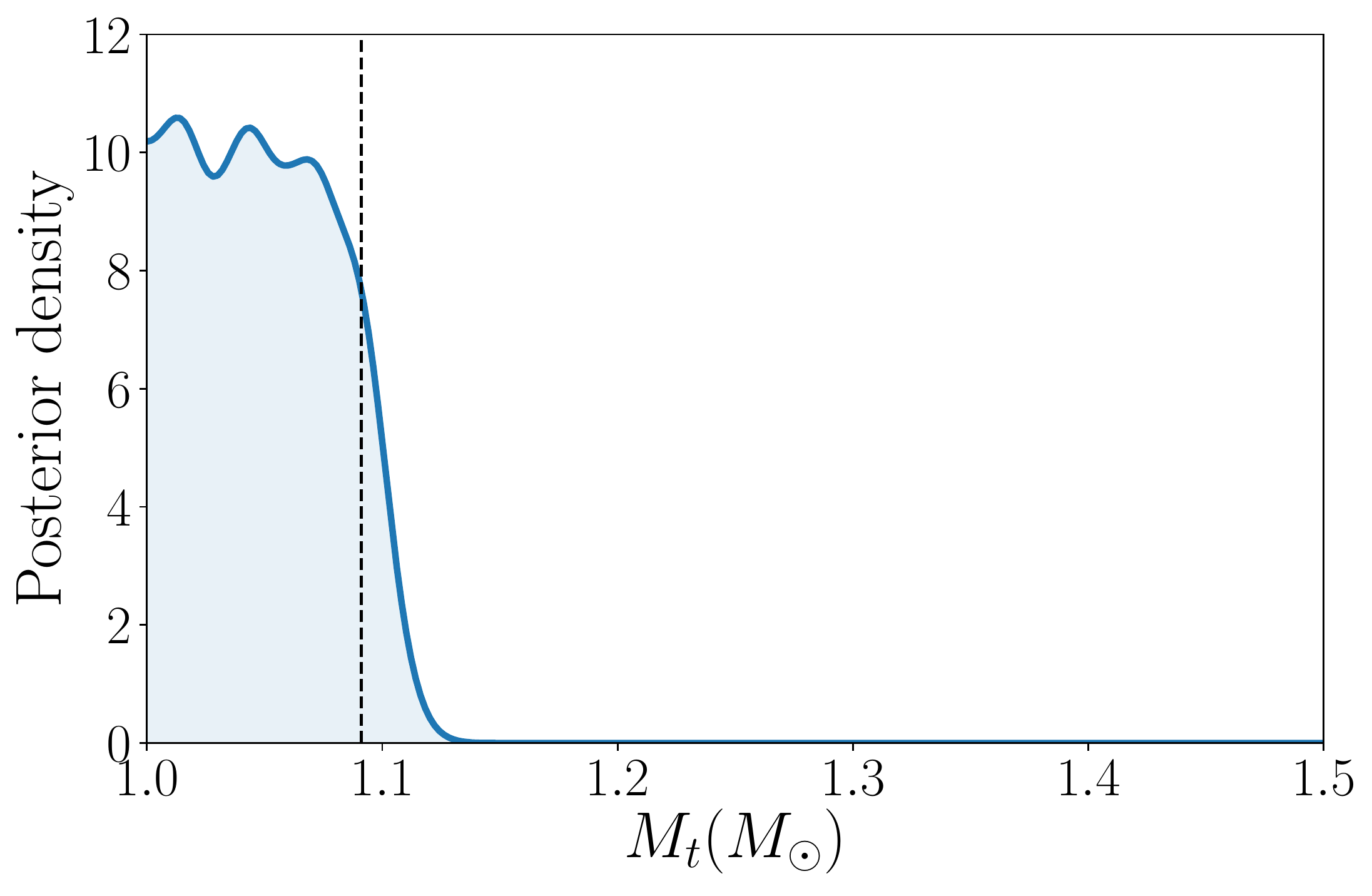}
\includegraphics[width=.45\columnwidth,clip=true]{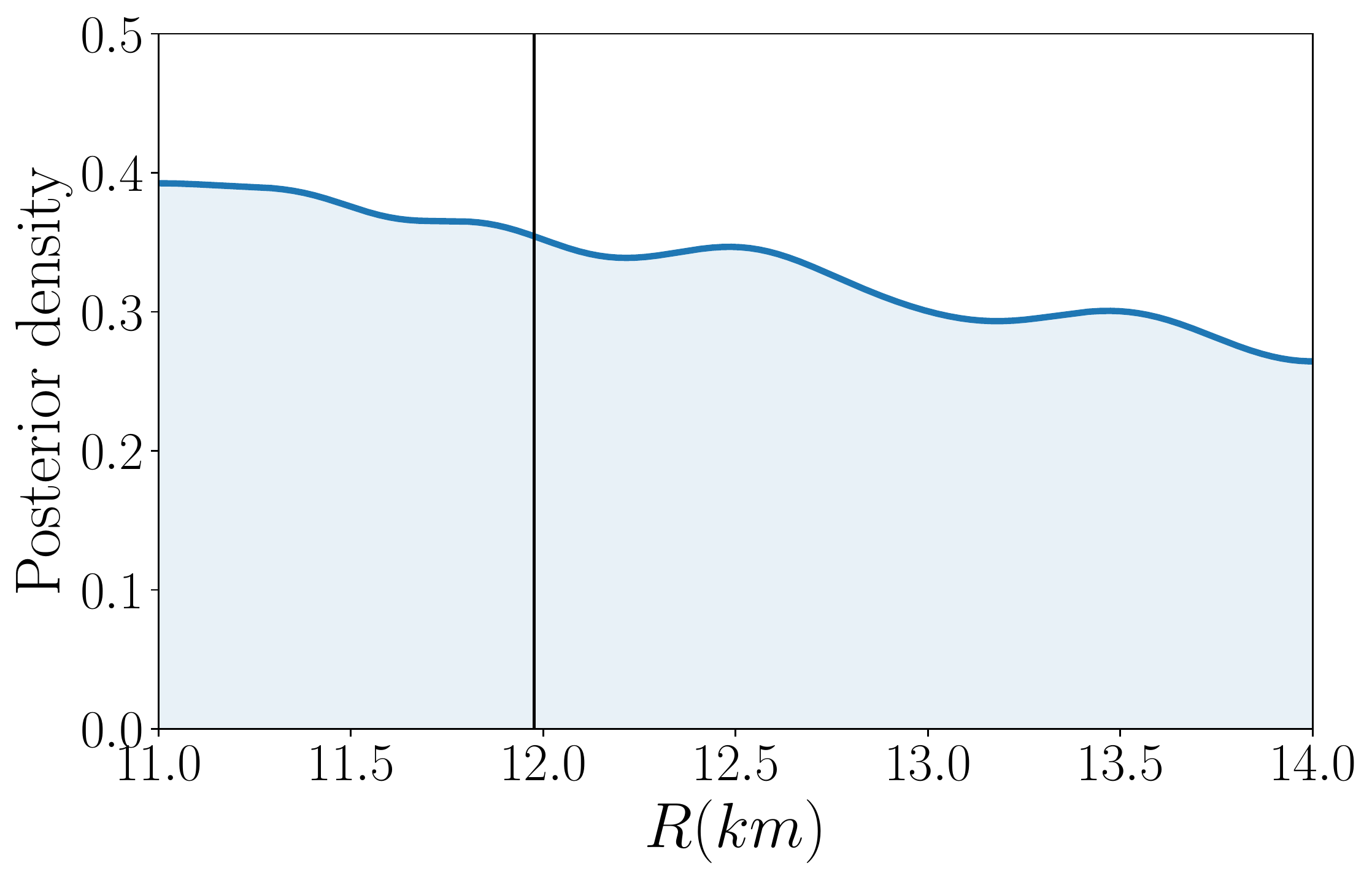}
\includegraphics[width=.45\columnwidth,clip=true]{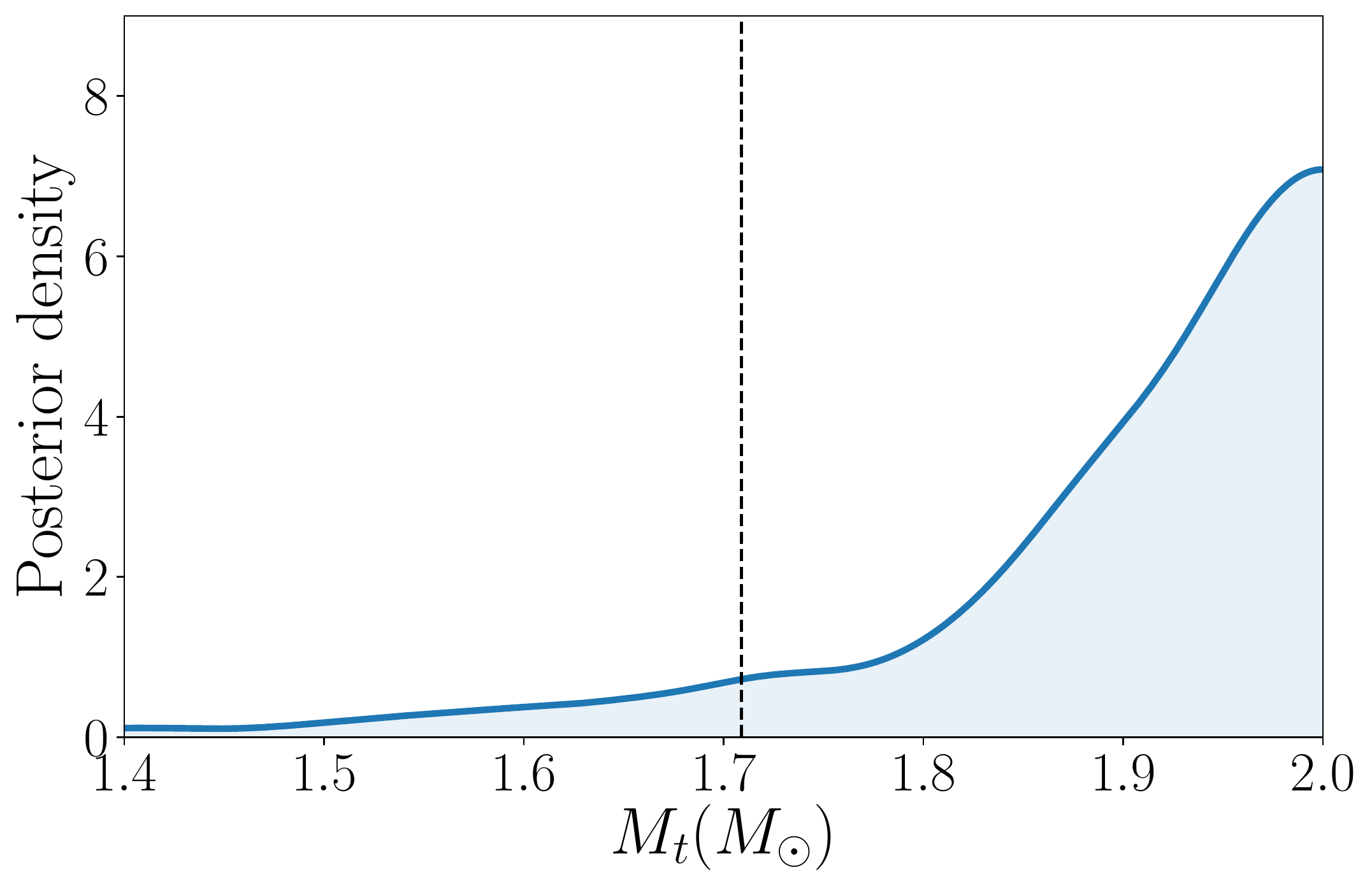}
\includegraphics[width=.45\columnwidth,clip=true]{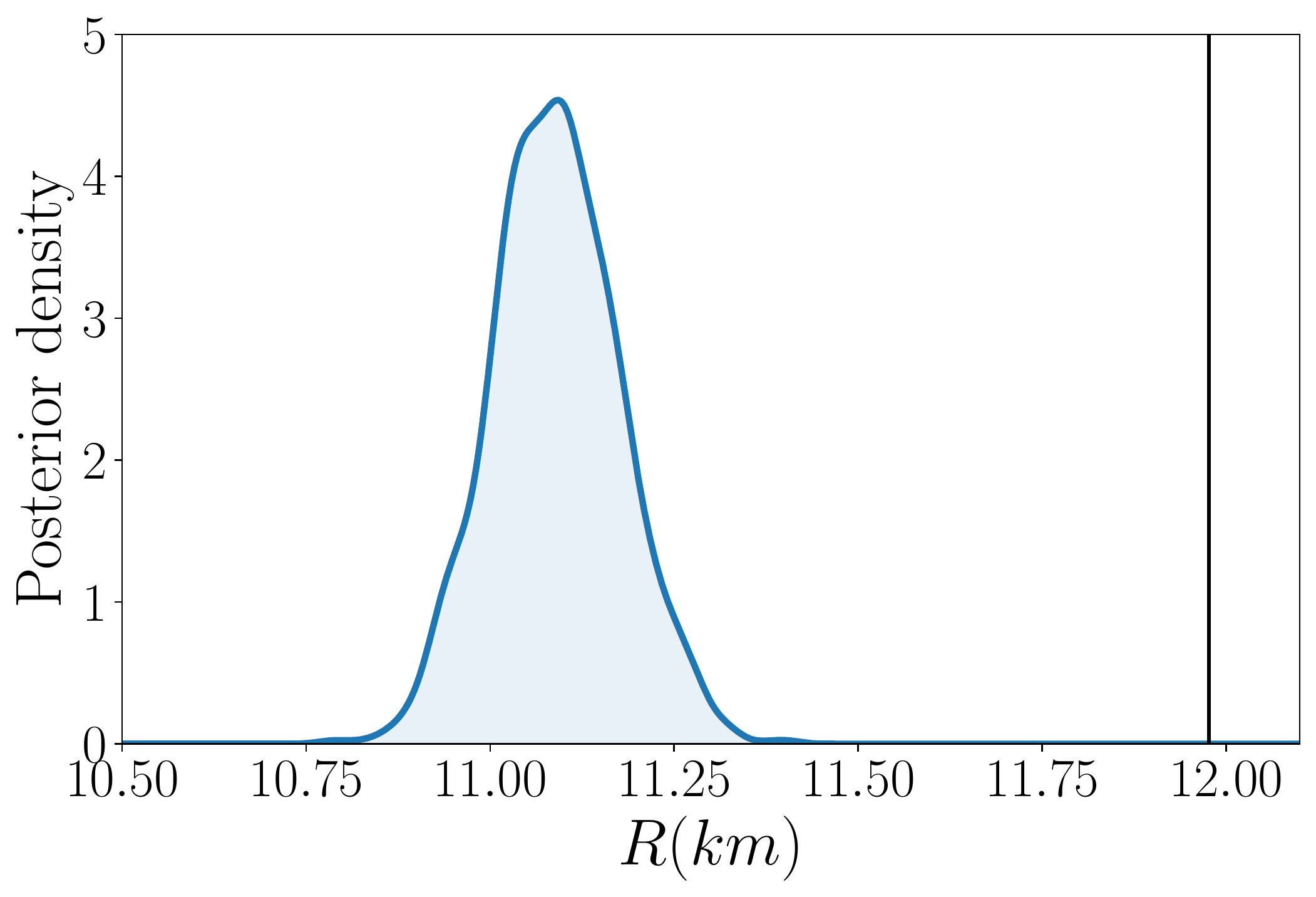}
\caption{Case B: Marginalized 1-dimensional posterior density for the transition mass $\Mtrans$ (left) and the hadronic radius $R$ (right) for a random population of 100 BNSs with EoSs whose transition
mass is lower than the detected population masses, $\Mtrans<1\,\Msolar$. Solid vertical lines denote the true parameters, where applicable. The dashed vertical line denotes the $90\%$ credible upper or lower limit as appropriate (left).
We show results with DBHF\_1032 (top row), SFHo\_2009 with a hadronic radius prior lower limit of $10$~km (second row), SFHo\_2009 with a hadronic radius prior lower limit of $11$~km (third row),
and SFHo\_2506 (bottom row).
}
\label{fig:Pop-CaseB}
\end{figure}

The standard astrophysical formation scenario of NSs suggests their masses above $\sim 1\,\Msolar$~\cite{Lattimer:2012nd,Suwa:2018uni}, while the smallest NS mass measurement corresponds to the PSR J0453+1559 companion with $1.174\pm 0.004\, \Msolar$ \cite{Martinez:2015mya}. 
It is therefore possible that the EoS exhibits a phase transition at densities low enough that all NSs in the Universe contain both hadronic and quark matter. As Fig.~\ref{fig:MR} shows, such EoSs (specifically DBHF\_1032, SFHo\_1031, and SFHo\_2009) deviate from their respective hadronic baselines at low NS masses. More importantly, it is possible for these hybrid EoSs to qualitatively resemble some much softer hadronic EoSs on the mass-radius plane, i.e. hadronic EoS with a much smaller radius.

This apparent degeneracy between hybrid EoSs with low-density phase transitions and very soft hadronic EoSs can be broken by imposing a lower limit on the hadronic radius inferred from massive pulsar observations~\cite{Demorest:2010bx,Antoniadis:2013pzd,Fonseca:2016tux,Cromartie:2019kug}. As already discussed in Sec.~\ref{sec:popModel}, most of our results assume $R>10$~km, which was obtained in~\cite{Annala:2017llu} after imposing the requirement that their hadronic EoSs can support $2\,\Msolar$ stars~\cite{Antoniadis:2013pzd}.
Interestingly, a probably even heavier pulsar has recently been reported, though with a larger error bar of $2.14^{+0.20}_{-0.18}\,\Msolar$ at the 2-$\sigma$ level~\cite{Cromartie:2019kug}. The existence of such a heavy NS implies an even more stringent lower bound for the hadronic NS radius $R$. As expected we find that detection of heavy pulsar is crucial for identifying low-density phase transitions; this highlights once again the interdisciplinary nature of the NS EoS measurement project.

Figure~\ref{fig:Pop-CaseB} shows posterior distributions for the transition mass $\Mtrans$ and the hadronic radius $R$. The top row corresponds to the DBHF\_1032 EoS (similar results are obtained with SFHo\_1031 so we omit them), which has a typical radius of $\sim 9$~km in the mass range probed by our simulated BNS population. This radius value is below the prior bound for the hadronic radius of 10~km, as purely hadronic EoSs with a radius of $9$~km would not be able to support $2\,\Msolar$ pulsars. Therefore in this case our hierarchical analysis correctly identifies that the entire BNS population consists of hybrid stars and results in an upper limit on the transition mass $\Mtrans$ (left panel). Indeed the transition mass posterior rails heavily against its $1\,\Msolar$ prior bound, indicating a small transition mass; for this simulated population we find $\Mtrans<1.07\,\Msolar$ at the 90\% credible level. Additionally, since we infer that there are no hadronic stars in the detected population, the posterior for the hadronic radius is uninformative and similar to its flat prior (right panel).

The second row of Fig.~\ref{fig:Pop-CaseB} corresponds to SFHo\_2009, and shows a case where the applied prior lower limit on the hadronic radius cannot fully break the degeneracy between a low-density phase transition and the hadronic EoS softness. Indeed from Fig.~\ref{fig:MR} we see that SFHo\_2009 has a typical radius of $\sim10$~km in the mass range of interest, which is marginally consistent with our prior. The resulting posterior for both the transition mass and the hadronic radius exhibits a ``dual" behavior. Part of the posterior is characteristic of a purely hybrid population and resembles the top row of the figure, suggesting an upper limit on $\Mtrans$ coupled with an uninformative $R$ estimate. The other part of the posterior is more similar to the case of a purely hadronic population and Fig.~\ref{fig:Pop-CaseA}, resulting in a lower limit on the $\Mtrans$ and a measurement of $R$ which is around $10$~km, marginally consistent with the existence of a $2\,\Msolar$ pulsar.

For the third row of Fig.~\ref{fig:Pop-CaseB} we consider the same EoS and BNS population, but increase the prior lower bound on the hadronic radius from $10$~km to $11$~km. As already alluded to, we can now correctly conclude that all the detected BNSs involve hybrid stars, as we recover an upper limit on $\Mtrans$ and an uninformative $R$. 
Both posteriors are qualitatively similar to those on the first row of Fig.~\ref{fig:Pop-CaseB}. As already known, the discovery of heavier pulsars is invaluable for the EoS inference and we here encounter one more example of this.

EoS SFHo\_2506 represents a borderline case as it predicts $\Mtrans=1.01\,\Msolar$, which is marginally above our lowest possible BNS mass. Given the number of BNS detections expected in the coming years, it is fairly unlikely that a hadronic binary would be observed under this EoS; we therefore discuss it here. 
Figure~\ref{fig:MR} shows that SFHo\_2506 (red line on the right panel) results in masses and radii that are qualitatively similar to that of a softer, purely hadronic EoS with $R$ around $11$~km. The bottom panel of Fig.~\ref{fig:Pop-CaseB} confirms this expectation, as our inference finds that the detected population is consistent with a hadronic EoS (lower limit on $\Mtrans$ on the left panel) with a biased radius (right panel). This is perhaps the  worst case scenario for GW-only observations and would result in a misidentification of the EoS properties. As discussed earlier, this degeneracy between a low-density phase transition and a softer hadronic EoS can be broken through the detection of very massive pulsars, such as the one in~\cite{Cromartie:2019kug}.

\subsubsection{Case C: EoS with a phase transition in the detected population}
\label{sec:caseC}

\begin{figure}[]
\includegraphics[width=.45\columnwidth,clip=true]{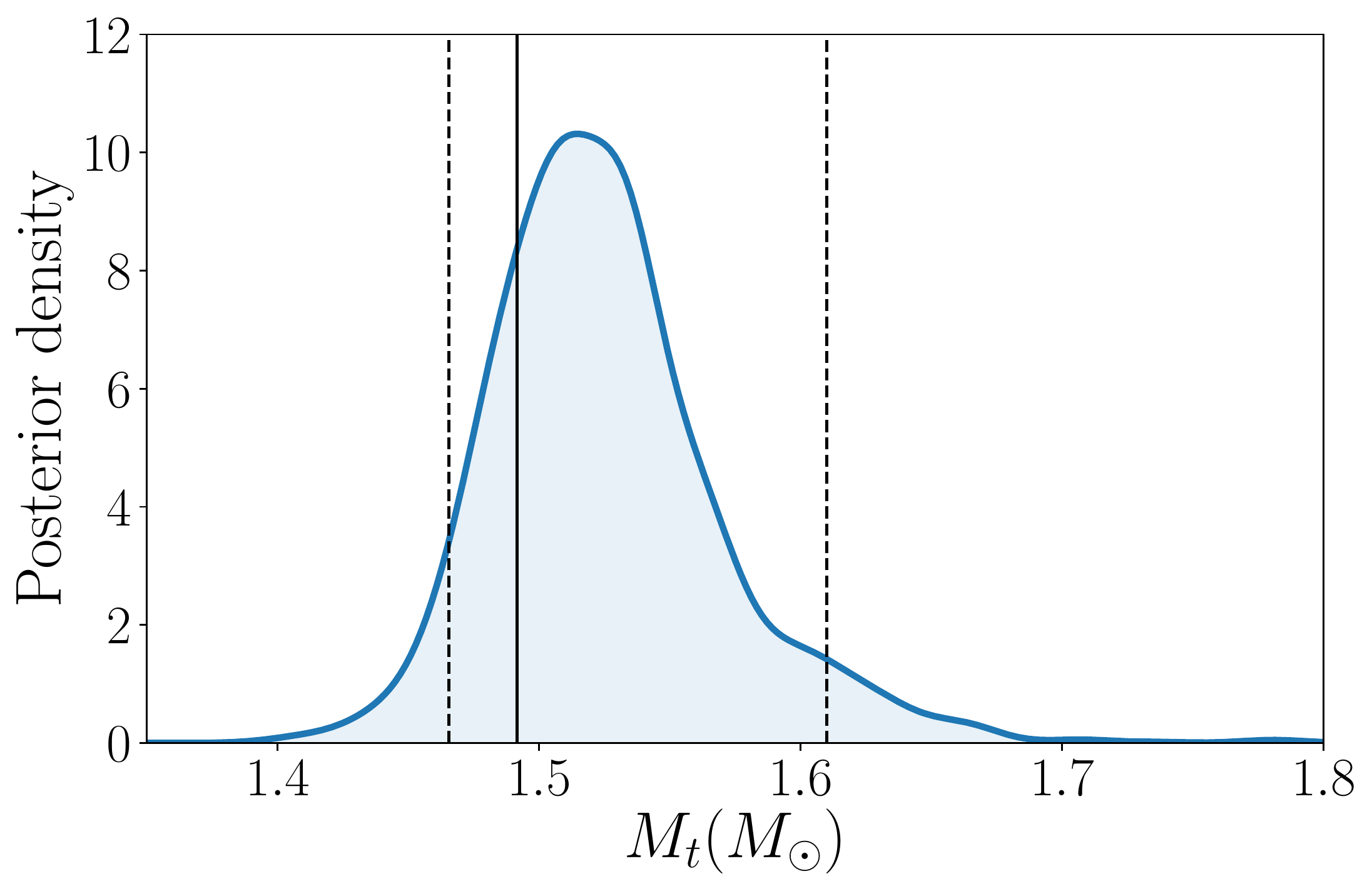}
\includegraphics[width=.45\columnwidth,clip=true]{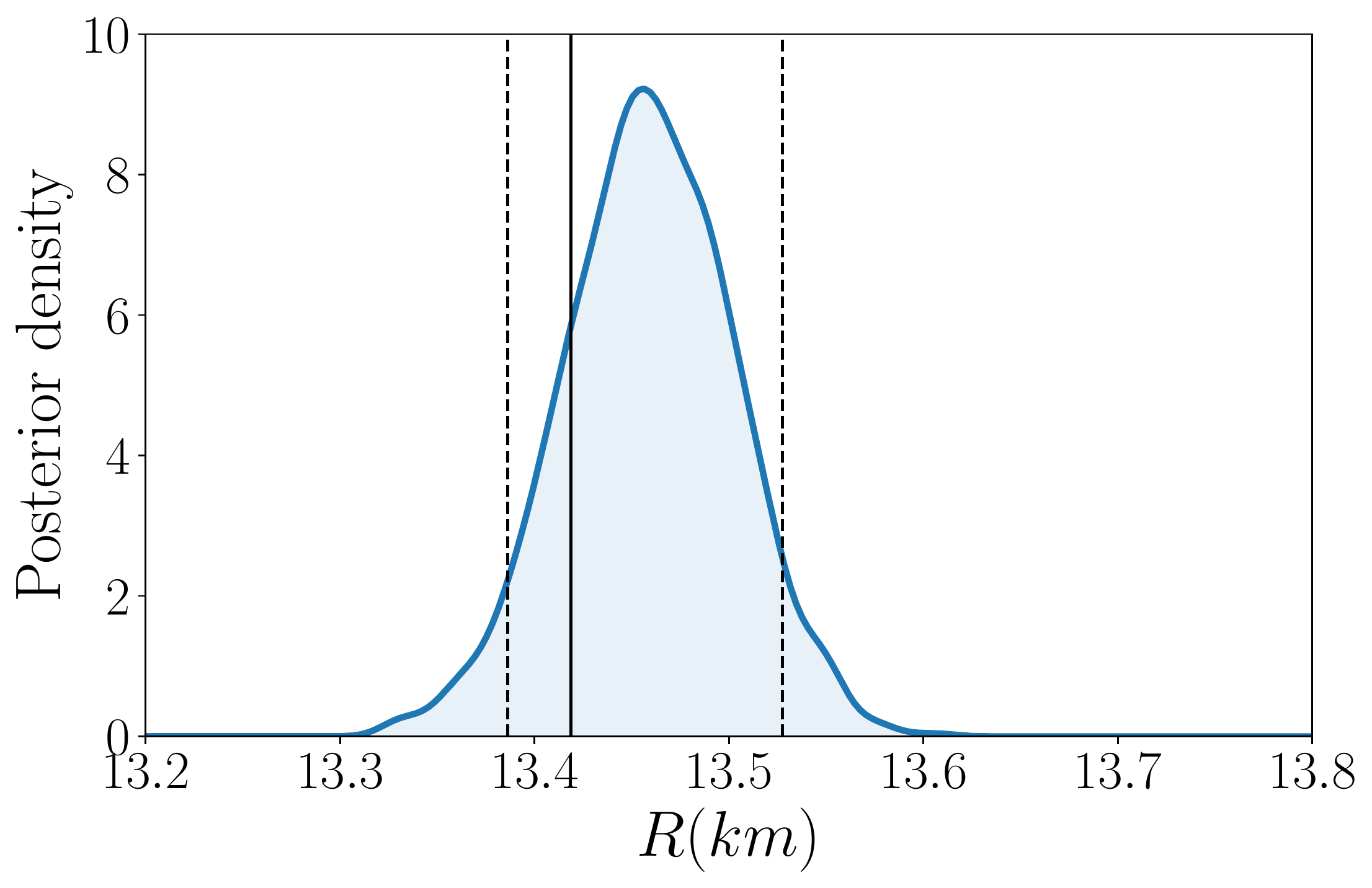}\\
\includegraphics[width=.45\columnwidth,clip=true]{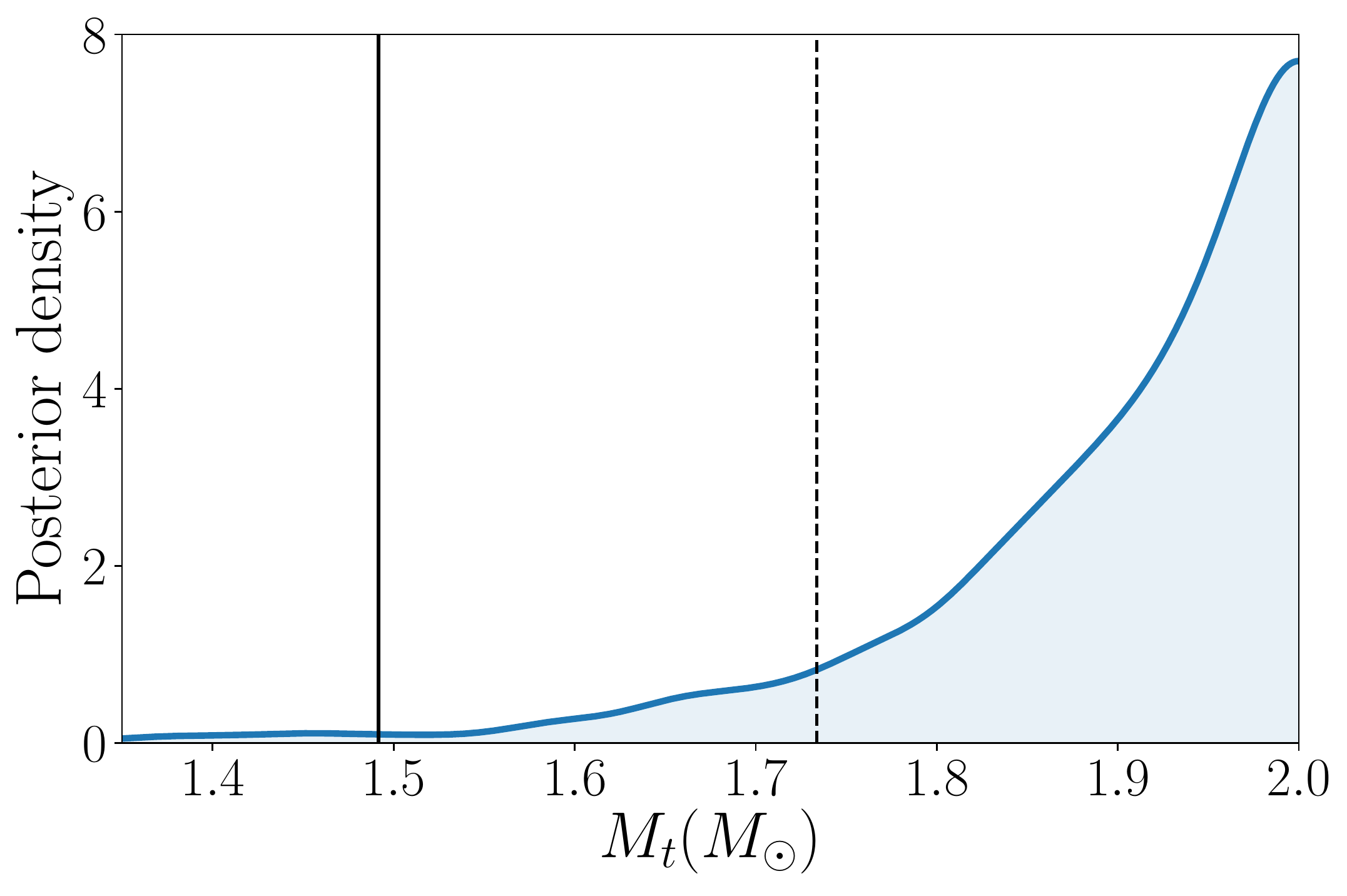}
\includegraphics[width=.45\columnwidth,clip=true]{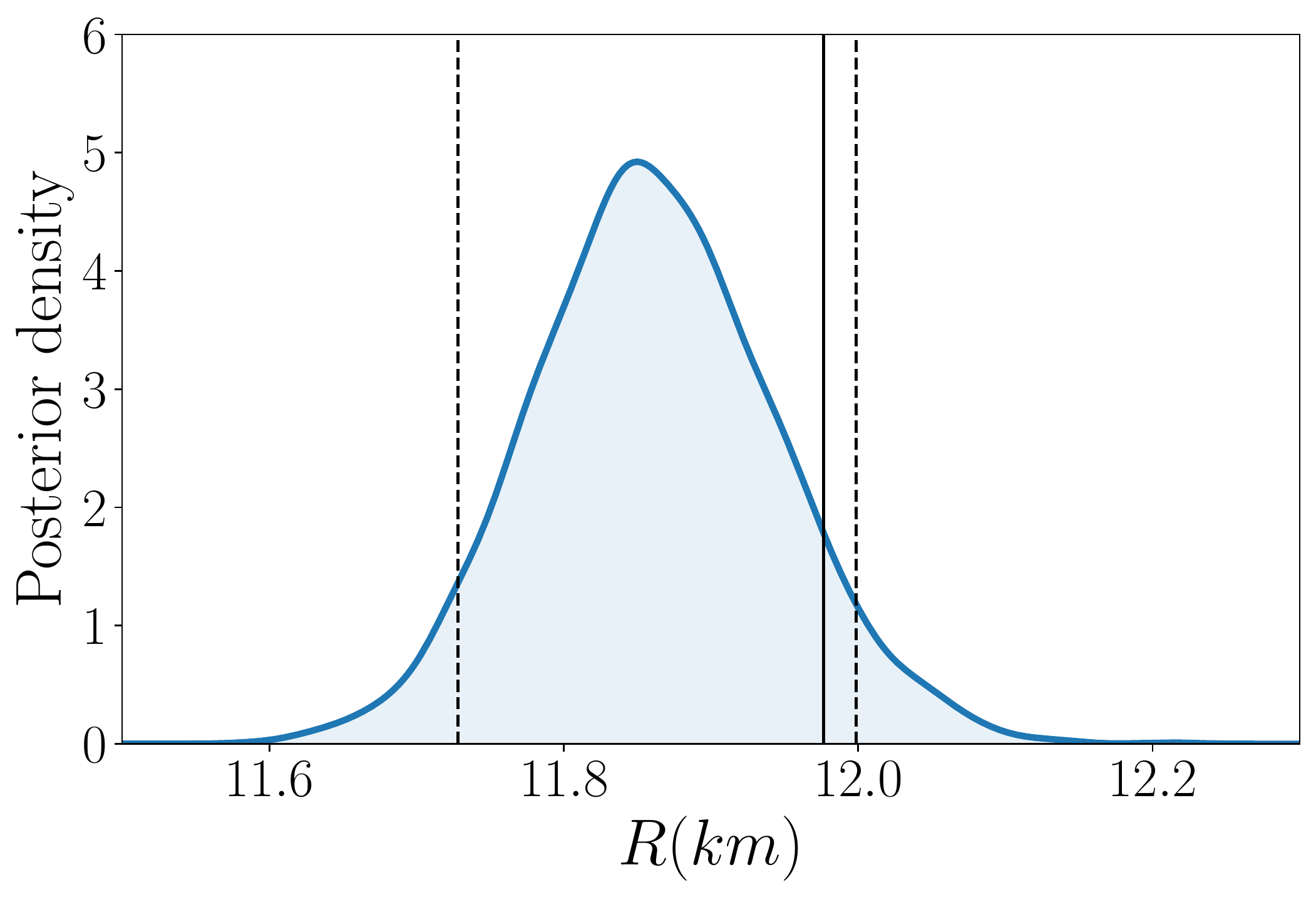}\\
\caption{Case C: Marginalized 1-dimensional posterior density for the transition mass $\Mtrans$ (left) and the hadronic radius $R$ (right) for a random population of 100 BNSs with EoSs whose transition mass
is comparable to the typical NS masses in the detected population. Solid vertical lines denote the true parameters. 
The dashed vertical lines denote the $90\%$ credible upper limit or symmetric interval as appropriate.
We show results with DBHF\_2507 (top row), and SFHo\_3003 (bottom row).
}
\label{fig:Pop-CaseC}
\end{figure}

Perhaps the most interesting case is the one where the detected BNS population contains both hadronic and hybrid stars. In this scenario, the detectability of the phase transition is determined by the exact value of $\Mtrans$ as well as the strength of the transition $\De \ep/\etrans$. The dimensionless tidal deformability $\La$ of a star decreases steeply with its mass~\cite{Hinderer:2009ca}, therefore heavier stars (that are more likely to be hadron-quark hybrids) undergo weaker tidal interactions. This means that if $\Mtrans$ is too high, binaries involving hybrid stars do not deviate from the expected ${\cal{R}}({\cal{M}},R)$ hadronic curve considerably and could be mistaken for purely hadronic. At the same time, the value of $\De \ep/\etrans$ affects how much the radius and tidal parameters of hybrid stars deviate from their hadronic counterparts~\cite{Alford:2013aca,Alford:2015gna,Han:2018mtj}. One pertinent example is SFHo\_3502 which has a relatively low $\Mtrans\sim 1.6\,\Msolar$, but the hybrid star radius is merely $\sim 500$~m smaller than the hadronic one at best (see Fig.~\ref{fig:MR}) due to its small value of $\De \ep/\etrans$. 

Figure~\ref{fig:Pop-CaseC} shows the possible posteriors for the transition mass $\Mtrans$ (left) and the hadronic radius $R$ (right) for EoSs belonging in this category. The top row corresponds to DBHF\_2507, an EoS with $\Mtrans\sim 1.5\,\Msolar$. Such value of $\Mtrans$ suggests that a considerable portion of the detected BNS population is comprised of hybrid stars. At the same time, the transition strength $\De\ep/\etrans$ is high enough that the chirp radius of the hybrid binaries deviates from the hadronic ones by $\gtrsim1$~km, see Fig.~\ref{fig:ChirpRFit}. As expected, therefore, we are able to infer that the analyzed BNS population is not purely hadronic and measure $\Mtrans$ to within $0.15\,\Msolar$ at the 90\% level and $R$ to within $130$~m at the same credible level.
We obtain qualitatively similar results with DBHF\_2010.

The second row of Fig.~\ref{fig:Pop-CaseC} corresponds to SFHo\_3003, an example of EoSs with a phase transition in the masses observed in the population but whose strength is too small to be detectable. In this scenario all BNSs have a similar chirp radius, and we obtain an estimate of $R$ (right panel) consistent with the EoS's typical hadronic radius $\Rtyp$ at the $90\%$ level, though visibly shifted towards lower values. 
However, the $\Mtrans$ posterior (left panel) places all the support at high values and looks qualitatively similar to Fig.~\ref{fig:Pop-CaseA} which refers to purely hadronic EoSs. This means all detected BNSs have been found to agree with a single hadronic ${\cal{R}}({\cal{M}},R)$ curve and thus interpreted as not containing a quark core. We also find qualitatively similar results with EoSs SFHo\_3502 and DBHF\_3004.
Figure~\ref{fig:ChirpRFit} suggests that we would need a fairly loud and heavy system to identify a phase transition in these EoS with weak transitions. For example, in the case of SFHo\_3003 we find that a ``lucky'' system with $m_1=m_2=1.7\,\Msolar$ and $\rho\sim 600$ could result in the detection of the phase transition. Though such loud systems are unlikely for second generation detectors, they might be within reach of third generation ground-based detectors~\cite{2010CQGra..27h4007P,2011CQGra..28i4013H}.

A key finding of the analyses of cases B and C is that to achieve a convincing identification of the phase transition from observations, both a low enough $\Mtrans$ and a high enough $\De\ep/\etrans$ are necessary. 
The former is essential as it guarantees the existence of quark cores in the detected component NSs, while the latter ensures that the tidal properties of the hybrid branch are sufficiently distinguishable from that of the purely hadronic branch.
This is expected since the ``disconnected'' third-family stars are the ones that exhibit maximal deviation from their hadronic counterparts, and these stars are realized through sufficiently strong transitions $\De\ep/\etrans\gtrsim0.6$ at relatively low densities $\ntrans\lesssim2.5-3\,n_0$~\cite{Han:2018mtj,Alford:2015gna}. 
Our results confirmed this prediction from generic mass-radius topology for stable hybrid stars, and show quantitatively the difficulties in revealing phase transition from observations when varying the EoS parameters.

\subsubsection{Case D: EoS with a phase transition at high NS masses}
\label{sec:caseD}

\begin{figure}[]
\includegraphics[width=.45\columnwidth,clip=true]{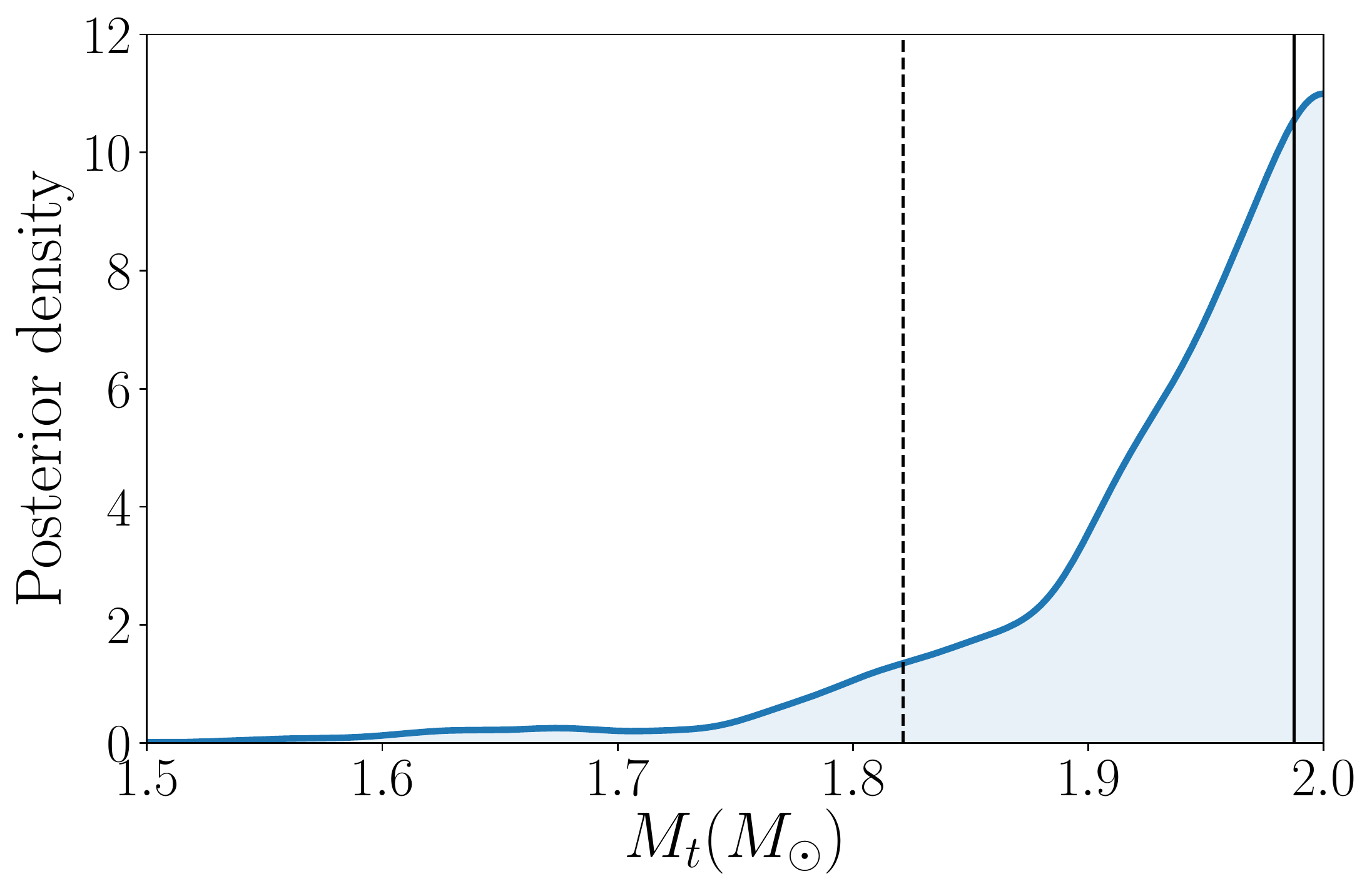}
\includegraphics[width=.45\columnwidth,clip=true]{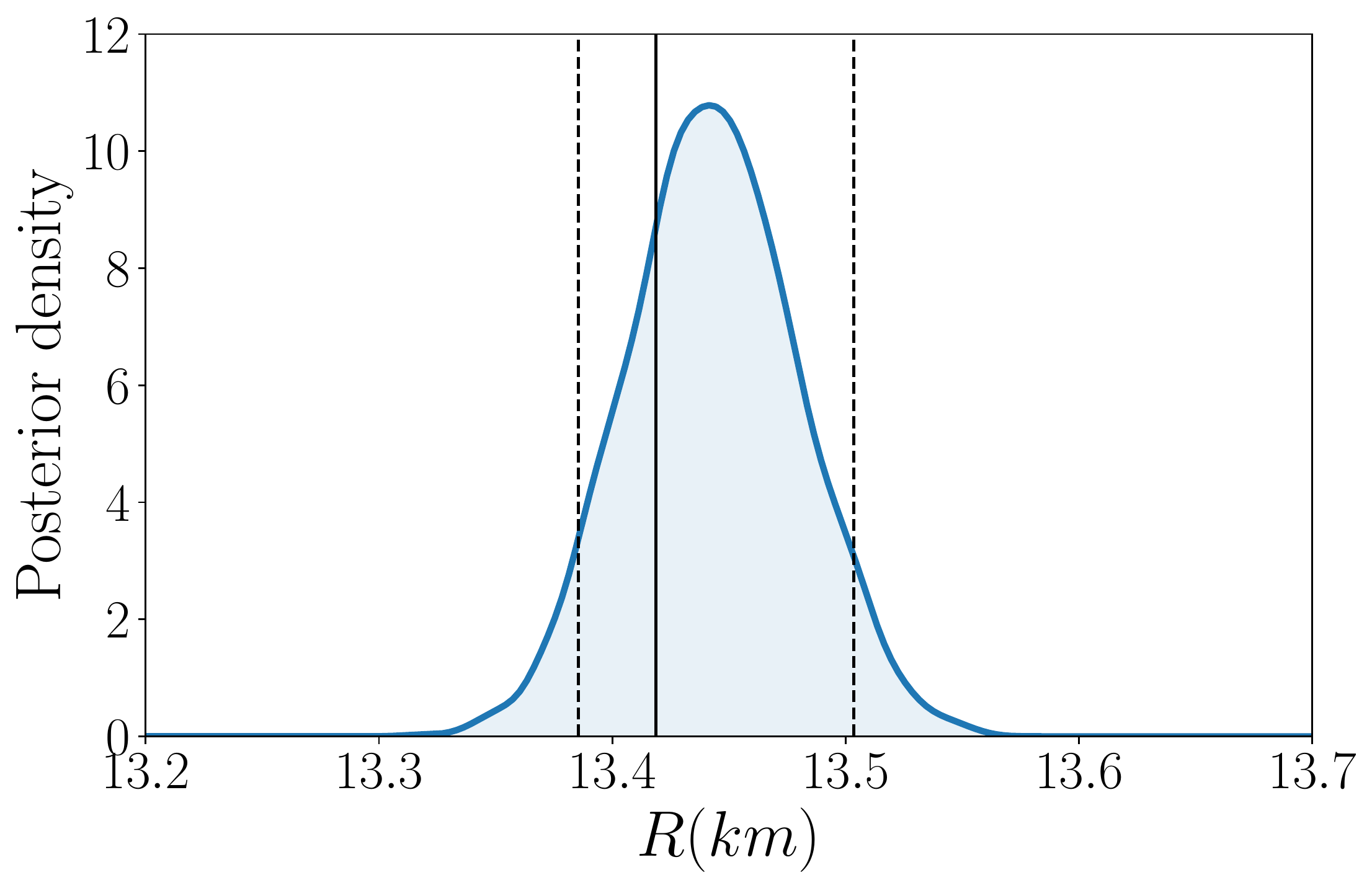}\\
\includegraphics[width=.45\columnwidth,clip=true]{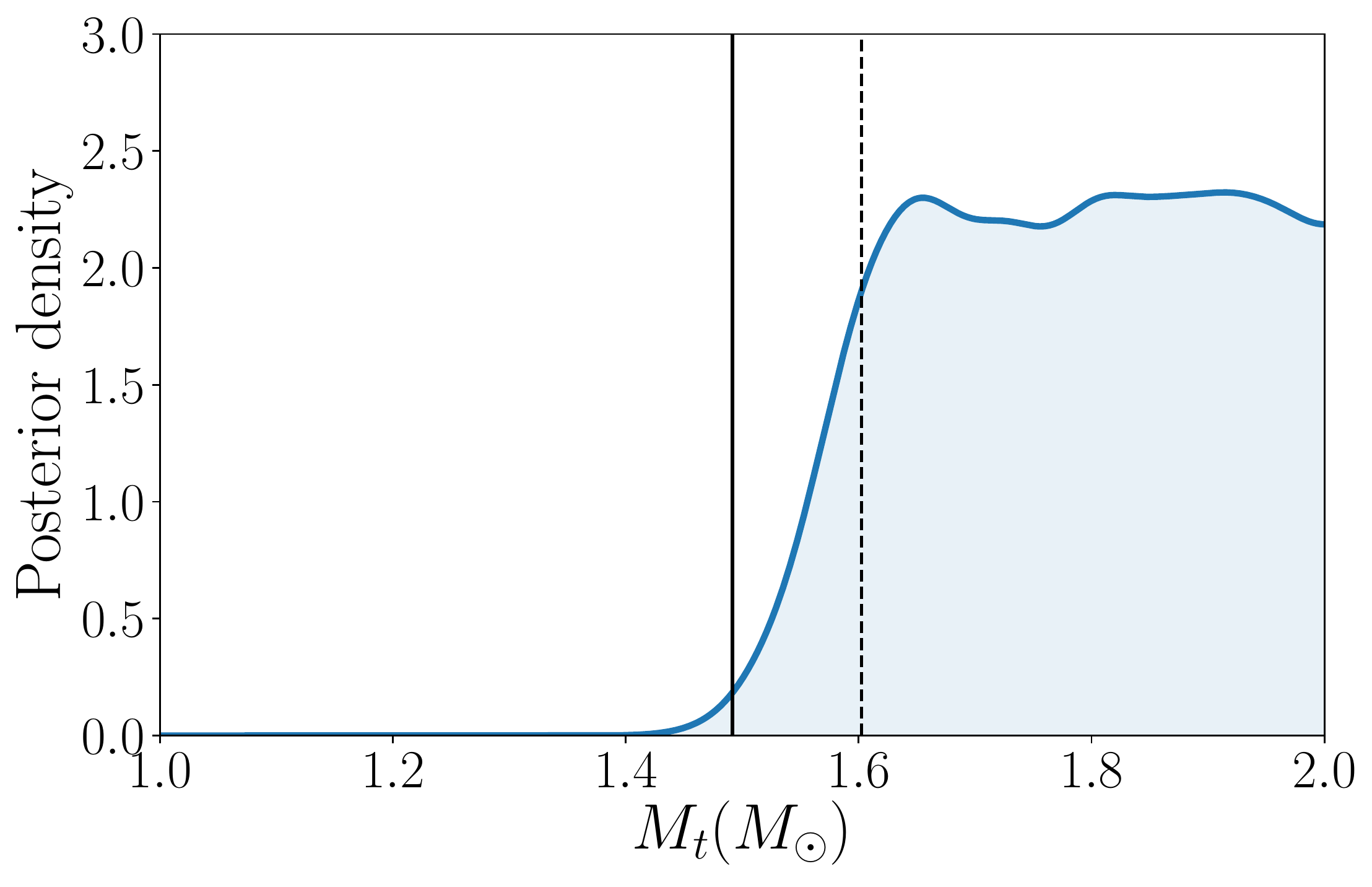}
\includegraphics[width=.45\columnwidth,clip=true]{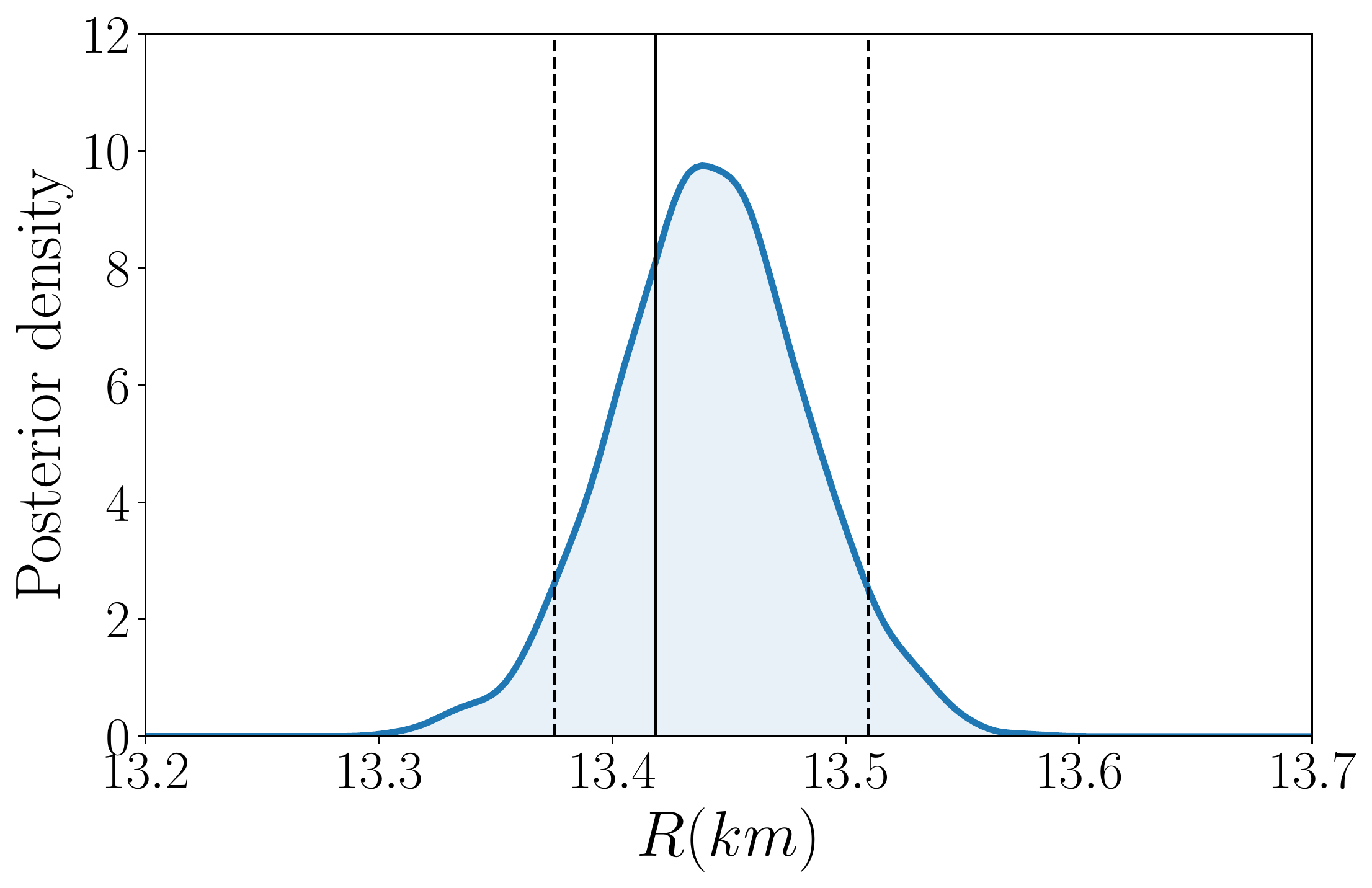}\\
\caption{Case D: Marginalized 1-dimensional posterior density for the transition mass $\Mtrans$ (left) and the hadronic radius $R$ (right) for a random population of 100 BNSs with EoSs whose transition mass is at 
relatively high masses and the phase transition is not detectable. Solid vertical lines denote the true parameters. 
The dashed vertical lines denote the $90\%$ credible upper limit (left) or symmetric interval (right).
We show results with DBHF\_3504 (top row), and DBHF\_2507 with a BNS mass distribution that is tightly centered around $\sim 1.32\,\Msolar$ (second row).
}
\label{fig:Pop-CaseD}
\end{figure}

The final case is an extension of case C where the transition mass is so high that it is not measurable, while the hadronic radius can still be reliably extracted. Figure~\ref{fig:Pop-CaseD} shows two such examples. The top row corresponds to EoS DBHF\_3504 which has $\Mtrans \sim2\,\Msolar$. 
At such a high transition mass, almost the entire population is hadronic so the $\Mtrans$ and $R$ posteriors are similar to those of Fig.~\ref{fig:Pop-CaseA} for the purely hadronic EoS. We again find that it is difficult to tell apart a hadronic EoS and a hybrid EoS with a phase transition at high densities or of a weak strength.
In the next subsection we elaborate on this and show how we can derive an upper limit on the transition strength.

The second row of Fig.~\ref{fig:Pop-CaseD} corresponds to the scenario where the transition mass might not be too high, but it is higher than the observed masses. Assuming the DBHF\_2507 EoS with $\Mtrans\sim 1.5\,\Msolar$, we draw NS masses from a Gaussian distribution peaked at $1.32\,\Msolar$ and with a standard deviation of $0.1\,\Msolar$, and obtain a BNS population with no hybrid stars. 
Since the population contains only hadronic stars, we obtain a lower limit on $\Mtrans$ that is consistent with the highest mass observed in the binary systems. Unsurprisingly the $\Mtrans$ posterior is uninformative for masses above that, since we do no have any observed data in the high mass range. The hadronic radius is extracted accurately. Therefore if the BNS population we observe with GWs does not contain heavy NSs, our observations can only rule out phase transitions up to the masses observed.

\subsection{Constraining the phase transition strength}
\label{sec:pt_para}

While directly identifying phase transitions and measuring their properties hinges on a low enough $\Mtrans$ and a large enough $\De\ep/\etrans$ (i.e. on the existence of a ``disconnected'' category of hybrid stars), it might still be possible to constrain the parameter space of EoSs involving such transitions with GW observations. 
In this subsection we discuss what can be inferred about the strength of the transition $\De \ep/\etrans$ for each of the three possible outcomes for the transition mass $\Mtrans$: a lower limit, an upper limit, or a measurement of $\Mtrans$. 

\subsubsection{Lower limit on $\Mtrans$}
\label{sec:lower}

Beginning with the scenario of a lower limit on $\Mtrans$, an EoS with a phase transition at large masses cannot have arbitrarily large transition strength if it is required to also support NS masses of some high value; see Fig. 5 of~\cite{Alford:2015gna} and Fig. 3 of~\cite{Han:2018mtj}.
Therefore if our analysis of the detected BNSs has resulted in a lower limit on $\Mtrans$ and an accurate determination of the hadronic radius $R$, the requirement of the existence of heavy pulsars can result in an upper limit constraint on $\De \ep/\etrans$, or even the elimination of a phase transition.

\begin{figure}[]
\parbox{\hsize}{
\includegraphics[width=0.9\hsize]{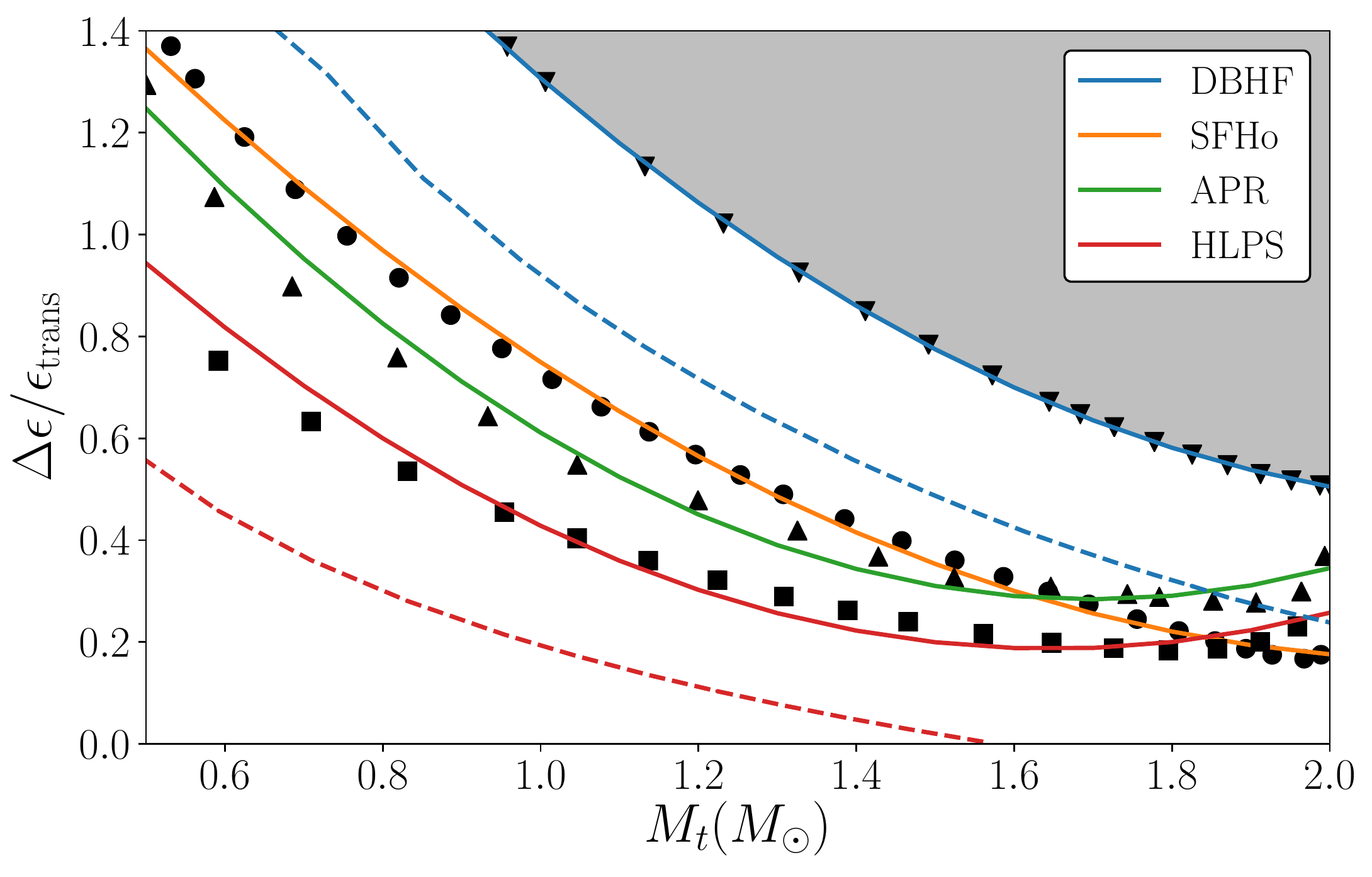}\\[-1ex]
}
\caption{Relation between the strength of the phase transition $\De\ep/\etrans$ and the onset mass for quarks $\Mtrans$ at fixed maximum mass of hybrid stars $\Mmax=2.0\,\Msolar$ (solid) and $\Mmax=2.2\,\Msolar$ (dashed for selected EoSs). Points and dashed curves correspond to numerical solutions to the TOV equations, while solid curves are fits with the coefficients from Table~\ref{tab:fit-function}. The gray-shaded regions mark the area excluded by the corresponding heavy pulsar measurements $\Mmax\geq 2.0\,\Msolar$ for DBHF. The detection of even heavier pulsars will further decrease the allowed parameter space.
}
\label{fig:Mt_De_fit}
\end{figure}

Figure~\ref{fig:Mt_De_fit} displays contours of the NS maximum mass, $\Mmax=2.0\,\Msolar$ (solid) and $\Mmax=2.2\,\Msolar$ (dashed), of stable hybrid stars on the ($\Mtrans$, $\De\ep/\etrans$) plane for various baseline hadronic EoSs. 
For $\Mmax=2.0\,\Msolar$ the symbols are obtained by numerically solving the Tolman-Oppenheimer-Volkov (TOV) equations~\cite{Tolman:1939jz,Oppenheimer:1939ne} while solid curves are analytical approximate fits (see Table~\ref{tab:fit-function}). The dashed curves track the numerical TOV solutions for $\Mmax=2.2\,\Msolar$, which we do not individually show with symbols for clarity.
Besides the familiar DBHF and SFHo baseline we also use the APR~\cite{Schneider:2019vdm} and HLPS~\cite{Hebeler:2010jx} models, which are representative of even softer hadronic EoSs, with smaller radii ($\Rtyp^{\rm HLPS}=10.88$~km and $\Rtyp^{\rm APR}=11.31$~km) that are compatible with the $2\,\Msolar$ constraint ($\Mmax^{\rm HLPS}=2.15\,\Msolar$ and $\Mmax^{\rm APR}=2.18\,\Msolar$). The region to the upper right of each curve (gray-shaded zone for DBHF) refers to hybrid EoSs for which the maximum-mass stars are below $2\,\Msolar$ and hence are considered ruled out. 
If the quark matter was softer than assumed here ($\cQMsq<1$) the allowed space would be even smaller~\cite{Alford:2013aca,Alford:2015gna}, therefore the limits shown are conservative. 

\begin{table}[]
\begin{center}
\begin{tabular}{c|@{\quad}c@{\quad}|@{\quad}c@{\quad}|@{\quad}c@{\quad}}
\hline 
&&&\\[-2ex]
 & $c_0$  & $c_1$  & $c_2$   \\[0.5ex]
\hline  
&&&\\[-2ex]
  DBHF & 3.15835 & --2.37617 & 0.524739  \\[0.5ex]
 \hline 
&&&\\[-2ex]
 SFHo & 2.19855 & --1.88657 & 0.437564 \\[0.5ex]
\hline
&&&\\[-2ex]
 APR & 2.22027 & --2.28079 & 0.671553  \\[0.5ex]
\hline
&&&\\[-2ex]
 HLPS & 1.746 & --1.89115 & 0.573446  \\[0.5ex]
\hline
\end{tabular}
\end{center}
\caption{Coefficients of the fits for $\De\ep/\etrans=f (\Mtrans/\Msolar)$ at $\Mmax=2.0\,\Msolar$ (see Fig.~\ref{fig:Mt_De_fit}), where $f(x)=c_0+c_{1} x+c_{2} x^{2}$. The parabolic form works reasonably well within the component mass range $1.1-1.7\,\Msolar$ unless a rather soft hadronic part is subject to the more stringent $\Mmax\geq2.2\,\Msolar$ constraint.
}
\label{tab:fit-function}
\end{table} 

The contours in Fig.~\ref{fig:Mt_De_fit} suggest that, the combination of a maximum-mass constraint with a lower bound on $\Mtrans$, as well as an inference on the (radius of) hadronic baseline EoS from premerger GW detections (as shown e.g. in Figs.~\ref{fig:Pop-CaseA} and~\ref{fig:Pop-CaseD}), can lead to a robust upper bound on $\De\ep/\etrans$. 
Additionally, the higher the $\Mmax$ imposed by pulsar mass measurements, the more severe constraints on the phase transition parameters will be. For instance, if future BNS detections indicate that $\Mtrans>1.5\,\Msolar$ together with $\Mmax>2.2\,\Msolar$ inferred from heavy pulsars, then the hadron-to-quark first-order phase transition scenario is almost completely incompatible with mature stable NSs, if the hadronic baseline EoS has also been determined to be as soft as HLPS with $\Rtyp^{\rm HLPS}=10.88$~km. In this case the only feasible option left would be that either deconfined quarks emerge at densities beyond cold massive NS cores\footnote{Such a phase could be attainable in hot dense remnants of supernovae or mergers.}, or the transition is a continuous hadron-quark crossover instead of a first-order one. 

Among all hadronic models tested, the stiffest one, DBHF, leads to the largest allowed parameter space for a first-order transition; however, the hadronic branch itself is at the edge of exclusion by GW170817's upper bound on the effective tidal deformability $\tilde\La$ for $\Mchirp\sim 1.2\,\Msolar$~\cite{Abbott:2018exr}. This suggests that if future GW data were to further lower that upper bound, the transition mass cannot be too high for such a stiff hadronic baseline EoS, presumably $\Mtrans<m_1\in [1.36, 1.60]\, \Msolar$. This implies a somewhat low transition density ($\ntrans\lesssim 2.7\,n_0$), connecting to constraints from heavy-ion collisions \cite{Danielewicz:2002pu,Stone:2017jpa} and nuclear matter calculations \cite{Hebeler:2010jx,Gandolfi:2011xu} at relevant densities.

\subsubsection{Upper limit on $\Mtrans$}
\label{sec:upper}

Next, we turn to the scenario of an upper limit on $\Mtrans$, which could be obtained if the transition density is low and all detected NSs are hadron-quark hybrids (case B).
As discussed previously, the combination of a lower bound on the hadronic radius by virtue of the heavy pulsar measurements and an even lower $R$ inferred from the GW data lead to the conclusion that no hadronic NSs exist in the detected population. The resulting upper limit on $\Mtrans$ is driven by the lightest systems observed.
Such a low-density phase transition ($\Mtrans<1\,\Msolar$) typically starts at a rather large radius $\Rtrans$ (see Fig.~\ref{fig:MR} and Table~\ref{tab:pt-MR}). This suggests that for the hybrid branch to reach such surprisingly small radii (and thus small tidal deformabilities) that the EoS is not misinterpreted as a soft hadronic EoS, a sufficiently large transition strength $\De\ep/\etrans$ is imperative. 
Said transition strength, however, cannot be too large as it would violate the upper bound limited by $\Mmax$ (Fig.~\ref{fig:Mt_De_fit}).

\begin{figure}[]
\parbox{\hsize}{
\includegraphics[width=0.9\hsize]{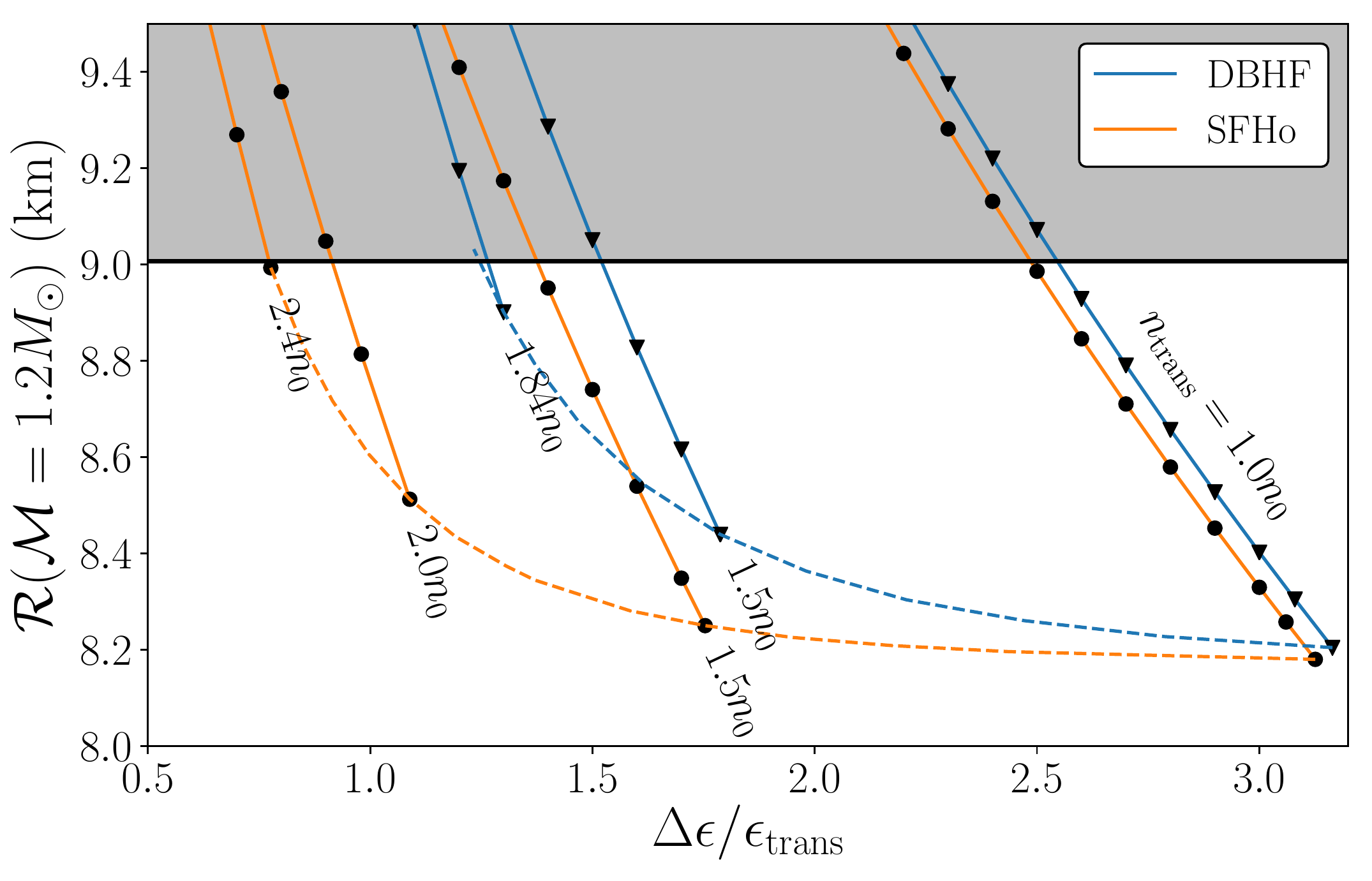}\\[-1ex]
}
\caption{Dependence of the chirp radius $\Rchirp^{\rm QQ} (\Mchirp=1.2\,\Msolar)$ on the strength of the phase transition $\De\ep/\etrans$ for a small transition mass $\Mtrans\lesssim 1\,\Msolar$ (case B); equal-mass binaries are assumed. We again consider two hadronic baseline EoSs and vary the onset density of the transition $\ntrans$ (indicated next to each line). At the lowest transition density $\ntrans=1.0 \,n_0$, the hadronic baseline EoS has little effect though its influence increases with increasing $\ntrans$. 
The horizontal line denotes $\Rchirp^{\rm HH}(\Mchirp=1.2\,\Msolar,R=10\,\rm{km})\approx9$~km. 
For fixed $\ntrans$, the maximum value of $\De\ep/\etrans$ is determined by $\Mmax=2.0\,\Msolar$ (dashed curves), 
which also corresponds to the leftmost part of the boundary contours in Fig.~\ref{fig:Mt_De_fit} when $\Mtrans\lesssim1\,\Msolar$. Note that for a given $\ntrans$, different hadronic base EoSs reflect different values of $\Mtrans$.
}
\label{fig:Rc_De}
\end{figure}

Figure~\ref{fig:Rc_De} shows in the case of a low-density phase transition with $\Mtrans<1\,\Msolar$, the chirp radius $\Rchirp^{\rm QQ}$ for a hybrid binary as a function of the transition strength $\De\ep/\etrans$ for different onset densities $\ntrans$; the chirp mass is held fixed at $\Mchirp=1.2\,\Msolar$. 
Since all NSs considered here are hybrids, we again expect that the chirp radius does not sensitively depend on the mass ratio (see for example the brown and purple dots on the right panel of Fig.~\ref{fig:ChirpRFit}), and we therefore restrict to equal-mass systems. The largest possible values of $\De\ep/\etrans$ on each curve are limited by $\Mmax\geq2.0\,\Msolar$, tracked by the dashed lines as $\ntrans$ is varied.
If the transition happens at saturation density $\,n_0$ (right-most solid lines), then the hadronic baseline EoS has little effect on the tidal properties of the hybrid star as it is mostly composed of quark matter. 
As the transition density increases, the hadronic part of the EoS also has an increasing influence. For a transition at twice the saturation density, the stiff DBHF EoS already results in $\Mtrans>1\,\Msolar$.

The solid horizontal line corresponds to $\Rchirp^{\rm HH} (\Mchirp=1.2\,\Msolar,R=10\,\rm{km})\approx9$~km, the minimum chirp radius for this chirp mass if the binary were hadronic and the EoS must support a $2\,\Msolar$ NS. As discussed in Sec.~\ref{sec:caseB}, it is the identification of binaries with a chirp radius less than this value that establishes the fact that the entire observed population consists of hybrid stars, and that a phase transition has occurred at sufficiently low densities. Figure~\ref{fig:Rc_De} suggests that if we measure the chirp radius of a binary and it is below $\sim 9$~km, then we would be able to estimate $\De \ep/\etrans$ if we knew the transition density and stiffness of the hadronic EoS. 

As already discussed in Sec.~\ref{sec:caseB} though, neither the transition density nor the properties of the hadronic EoS can be constrained with GWs in this case, as that would require detecting unexpectedly light NSs, with a masses around $0.5\,\Msolar$. Nevertheless, Fig.~\ref{fig:Rc_De} suggests then, that we can still place a conservative lower limit on $\De\ep/\etrans$ if the presence of a low-density phase transition (together with an upper limit on $\Mtrans$) has been established, by considering the highest transition density $\ntrans$ such that $\Mtrans\lesssim1\,\Msolar$ for a soft hadronic EoS. 
This conservative lower limit is specified by the intersection of the $\Rchirp=9$~km horizontal line with either the $\ntrans (\Mtrans=1\,\Msolar)$ curve or the $\Mmax=2.0\,\Msolar$ borderline, whichever gives higher $\De\ep/\etrans$ for the softer hadronic EoS. For Fig.~\ref{fig:Rc_De}, these two criteria lead to $\De\ep/\etrans\gtrsim 0.78$. 
If $\De\ep$ were smaller, the chirp radii would be $\Rchirp>9$~km, which would lead to the case where the phase transition cannot be identified, and we cannot obtain an upper limit on $\Mtrans$\footnote{It is worth mentioning that for any softer quark phase $\cQMsq<1$, the required strength of $\De\ep$ to achieve small (chirp) radii is also higher, so the limit discussed here is indeed conservative.}. 
This conservative bound might further improve if we can combine non-GW information about hadronic matter at those lower densities, for example from heavy-ion collisions or nuclear matter calculations.

\subsubsection{Measurement of $\Mtrans$}
\label{sec:measurement}

The final scenario corresponds to EoSs like DBHF\_2507 for which the phase transition has taken place in the detected population,  and we can measure $\Mtrans$ with enough BNS observations (case C).
Besides the transition mass, for these systems we also have access to the chirp radii of individual BNSs, as well as the mean chirp radius of the hybrid binaries $\Delta {\cal{R}}_2$; see Sec.~\ref{hiermodel}. The relation between the chirp radii of the binaries containing hybrid stars and hadronic stars depends on the strength of the transition $\De \ep/\etrans$, with stronger transitions in general leading to a larger reduction in the chirp radii of hybrid binaries.

\begin{figure}[]
\parbox{\hsize}{
\includegraphics[width=0.9\hsize]{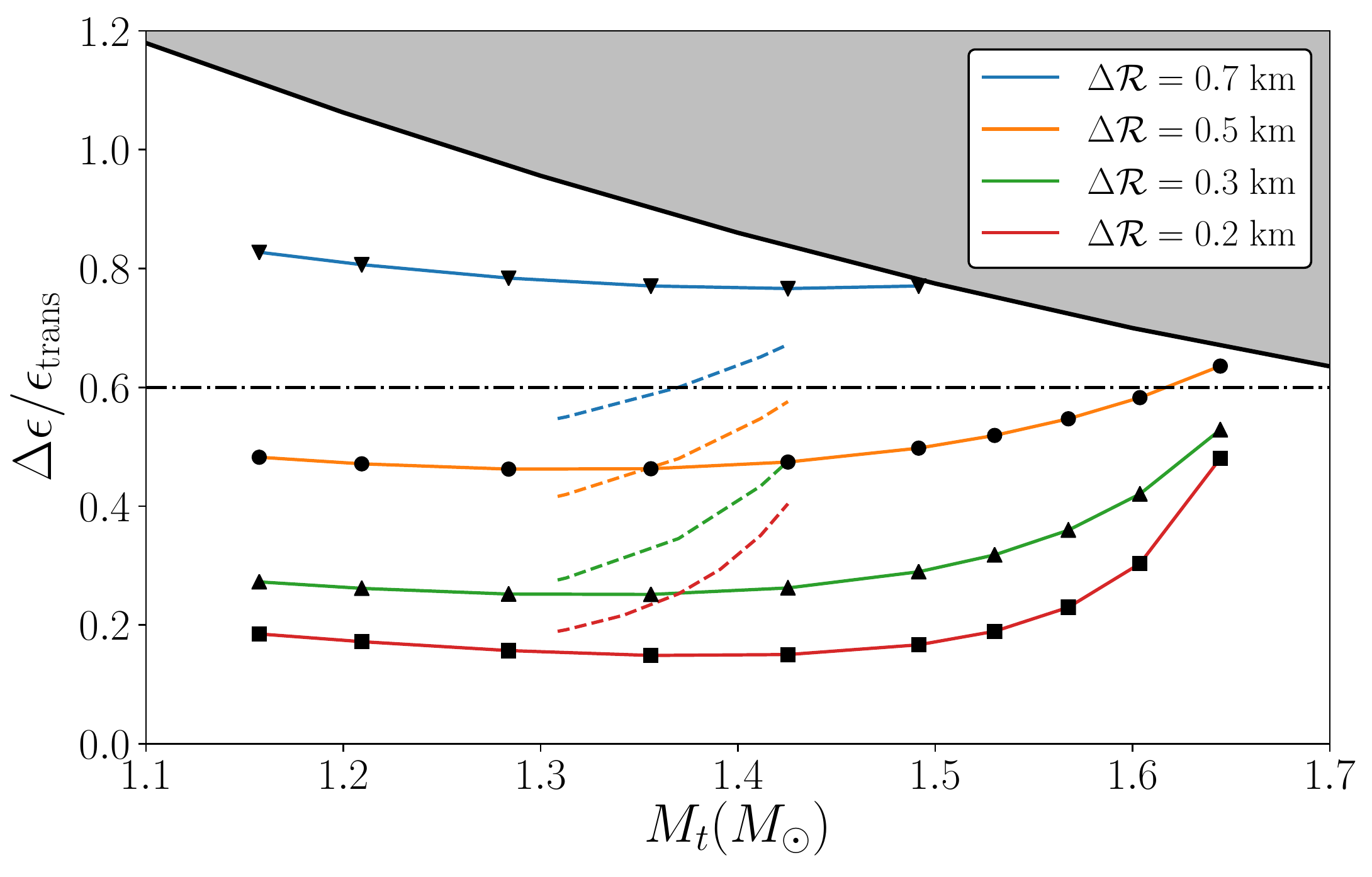}\\[-1ex]
}
\caption{Contours of constant reduction in chirp radius $\Delta{\cal{R}}\equiv\Rchirp^{\rm HH}(\Mchirp=1.2\,\Msolar)-\Rchirp^{\rm HQ}(\Mchirp=1.2\,\Msolar)$ for ``HQ'' binaries with mass ratio $q=0.7$ (solid) or $q=0.9$ (dashed) on the $(\Mtrans, \De\ep/\etrans)$ plane. We use the DBHF baseline for which the chirp radius is $\Rchirp^{\rm HH}(\Mchirp=1.2\,\Msolar)\approx 13.5$~km for purely hadronic binaries. 
The dash-dotted horizontal line denotes $\De \ep/\etrans=0.6$, the approximate threshold for a separate hybrid branch to form, while the shaded region is excluded by $\Mmax\geq 2.0\,\Msolar$.
}
\label{fig:Rc_Mt_De}
\end{figure}

In Fig.~\ref{fig:Rc_Mt_De} we plot on the $(\Mtrans, \De\ep/\etrans)$ plane the contours of $\De\Rchirp\equiv\Rchirp^{\rm HH}(\Mchirp=1.2\,\Msolar)-\Rchirp^{\rm HQ}(\Mchirp=1.2\,\Msolar)$, where $\Rchirp^{\rm HH}(\Mchirp=1.2\,\Msolar)$ and $\Rchirp^{\rm HQ}(\Mchirp=1.2\,\Msolar)$ refer to the chirp radii of ``hadronic-hadronic'' and ``hadronic-hybrid'' binaries, respectively, with the same chirp mass. We show results with two mass ratio values, $q=0.7$ and $q=0.9$, as there is no equal-mass ``HQ'' binary because of the assumption that masses on the hybrid branch are always higher than on the hadronic one.
For the more symmetric binary systems ($q=0.9$, dashed curves), $\De\Rchirp$ is more sensitive to $\Mtrans$, although there is a much narrower range of $\Mtrans$ for which $m_2<\Mtrans<m_1$. The shaded region is excluded by the requirement that $\Mmax\geq2.0\,\Msolar$.
 
Figure~\ref{fig:Rc_Mt_De} suggests that with a measurement of $\Mtrans \approx1.5\,\Msolar$ and the hadronic radius (see e.g. the first row of Fig.~\ref{fig:Pop-CaseC}, case C), an estimate of $\Rchirp^{\rm HQ} (\Mchirp=1.2\,\Msolar)$ from the detected population can potentially further constrain the preferred range of $\De\ep/\etrans$. For instance, if $\Rchirp^{\rm HQ} (\Mchirp=1.2\,\Msolar)$ is measured to be $\sim0.7$~km smaller than the corresponding hadronic chirp radius, this favors $\De\ep/\etrans\gtrsim 0.8$. 
Such constraint is also consistent with the generic expectation that a disconnected hybrid branch is detectable when the transition strength is above $\sim 0.6$ (see discussion in Sec.~\ref{sec:caseC}), therefore the identification of a phase transition in the detected population can be used to conservatively infer $\De \ep/\etrans \gtrsim0.6$.

\subsection{Measurement accuracy with multiple detections}
\label{sec:detect}

In Sec.~\ref{cases} we discussed the possible measurement outcomes in detail, and showed how they depend on properties of the EoS and the mass distribution of the observed BNSs by studying a single realization of a population of $100$ detected binaries. The precise numerical bounds we obtained on $\Mtrans$ and $R$ depend to some extent on the specific population realization, i.e. the exact masses and the SNRs of the systems detected. In this section, we simulate multiple population realizations of $N$ detections, and compute the expected accuracy with which we can measure the transition mass $\Mtrans$ and hadronic radius $R$ averaged over the different populations. 
We also discuss how these measurements can lead to estimates on the transition strength $\De \ep/\etrans$, along the lines of the discussion of Sec.~\ref{sec:pt_para}. 
In all cases, we find that the averaged posterior agrees qualitatively with the single-population posteriors discussed in Sec.~\ref{cases}.

\begin{figure*}[]
\subfloat[SFHo]{
  \includegraphics[width=0.5\columnwidth,clip=true]{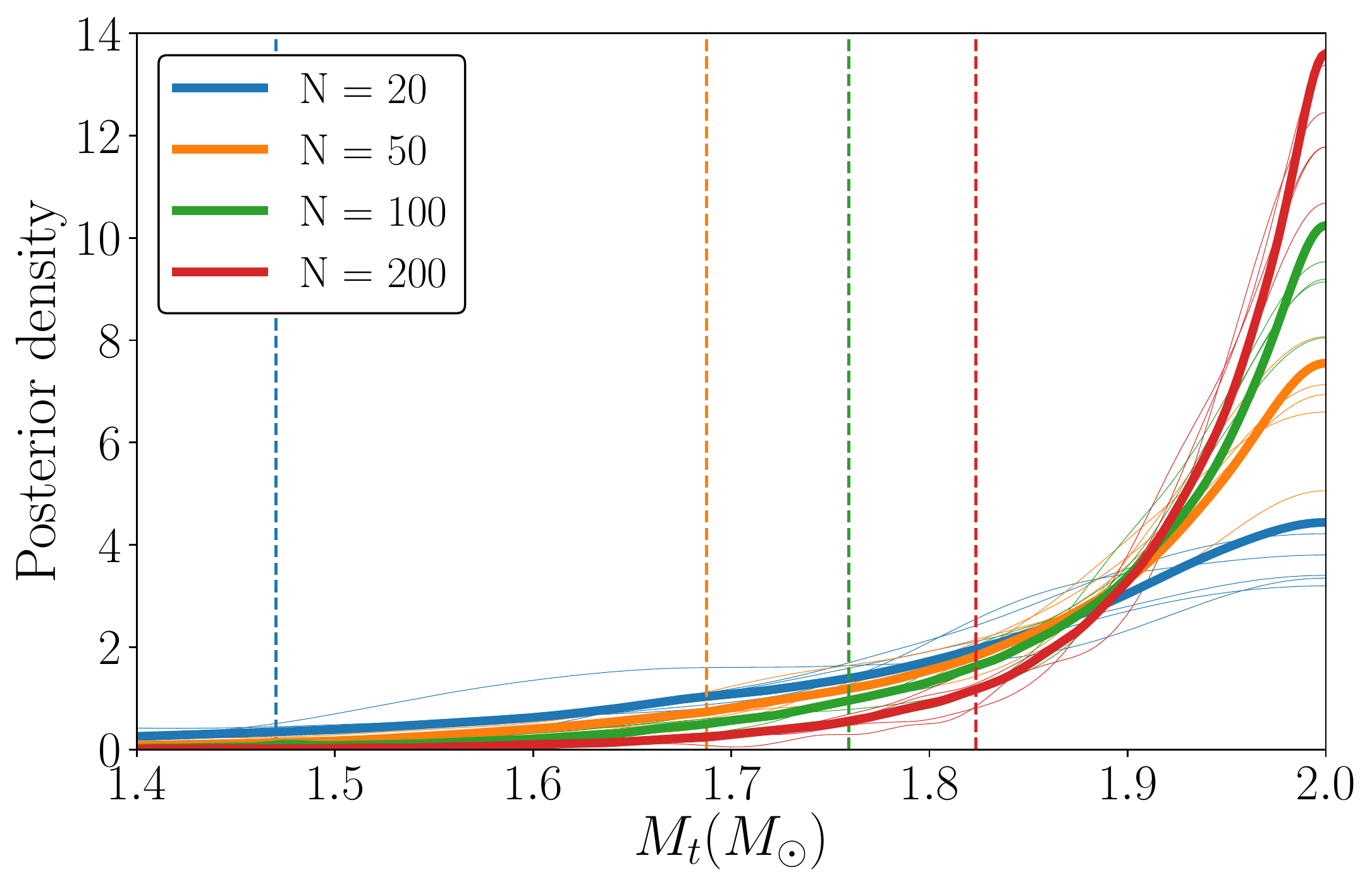}
 \includegraphics[width=0.52\columnwidth,clip=true]{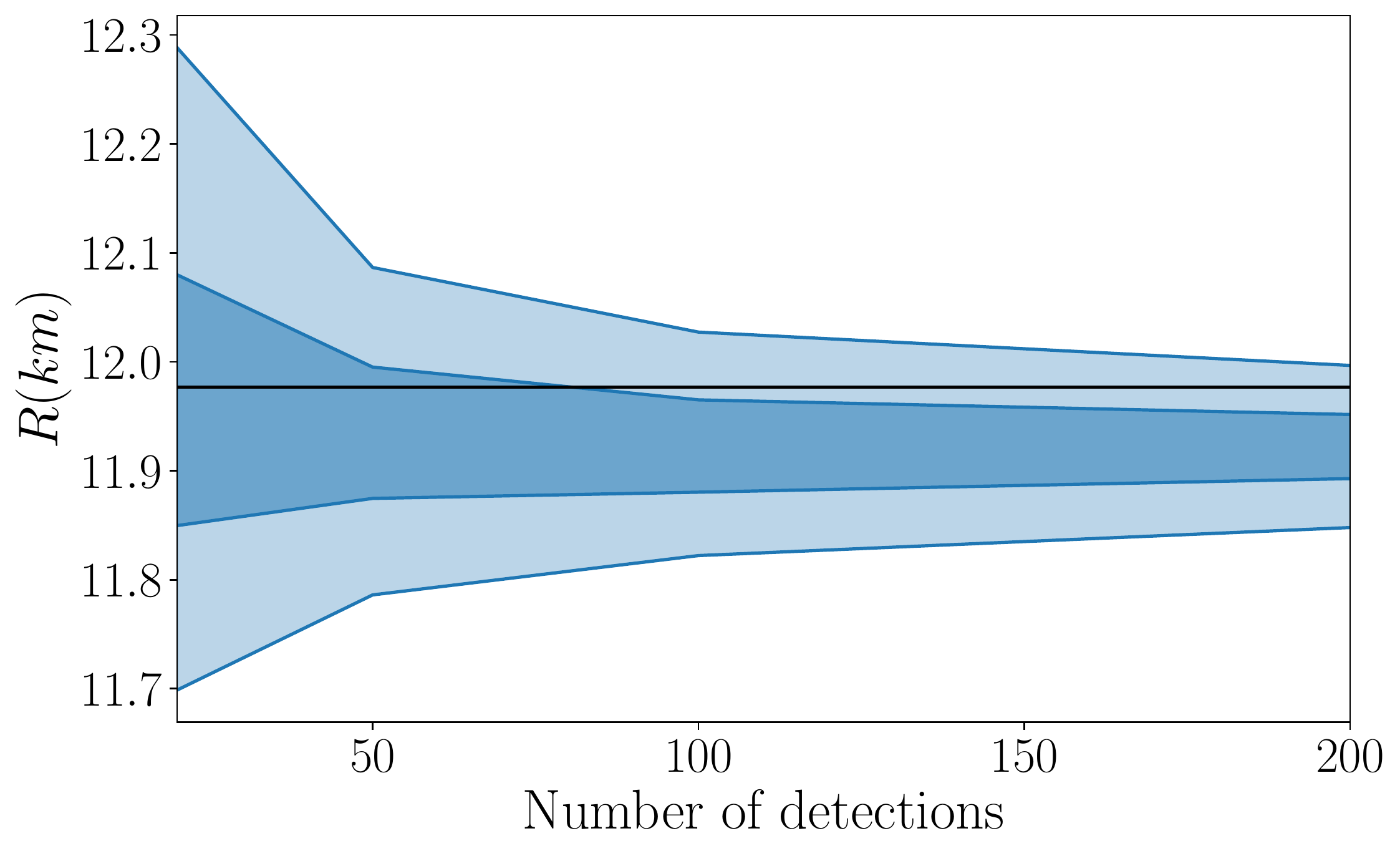}
}
\subfloat[DBHF\_1032]{
  \includegraphics[width=0.5\columnwidth,clip=true]{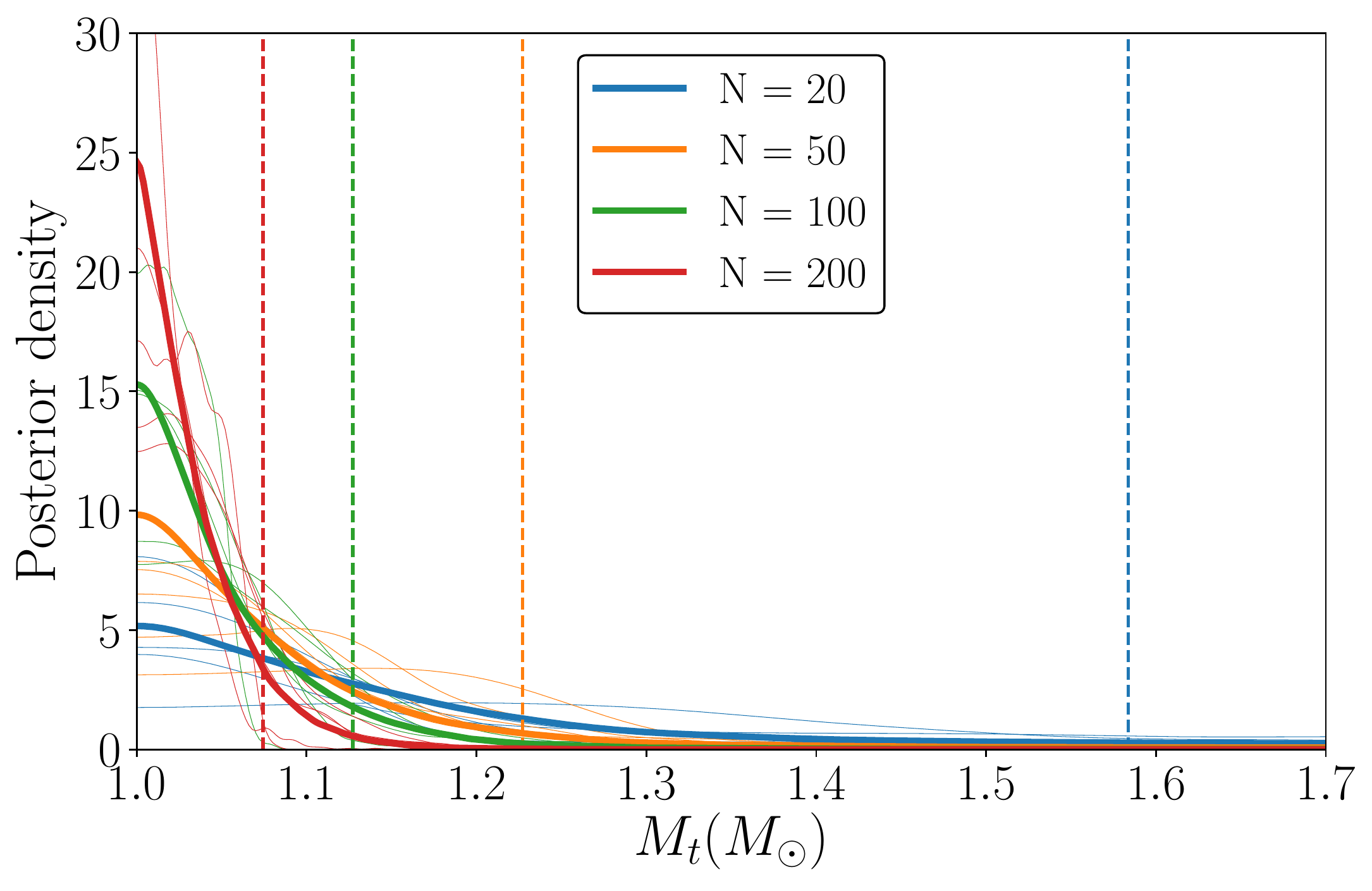}
  \includegraphics[width=0.51\columnwidth,clip=true]{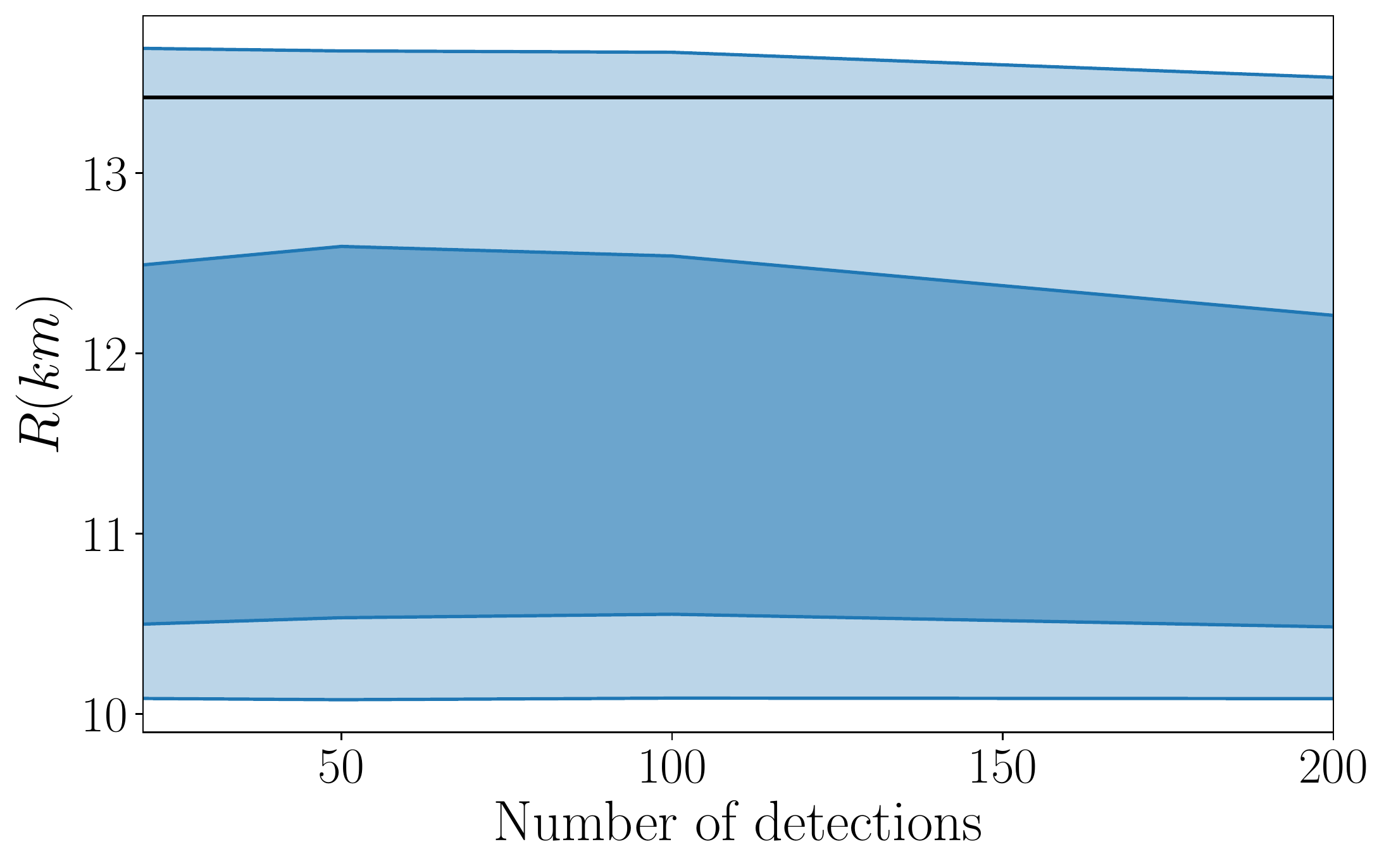}
}
\\
\subfloat[SFHo\_2506]{
  \includegraphics[width=0.5\columnwidth,clip=true]{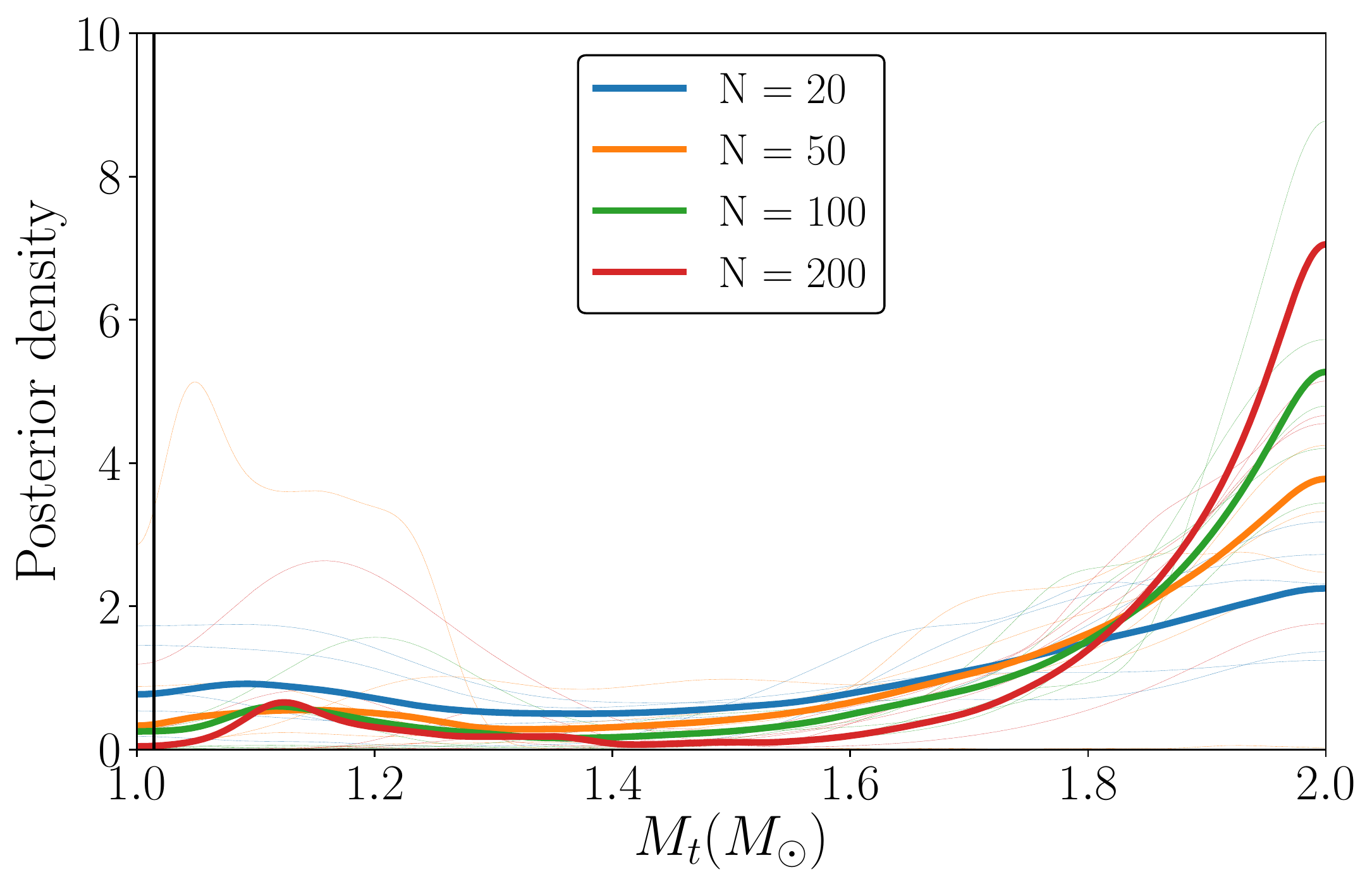}
  \includegraphics[width=0.52\columnwidth,clip=true]{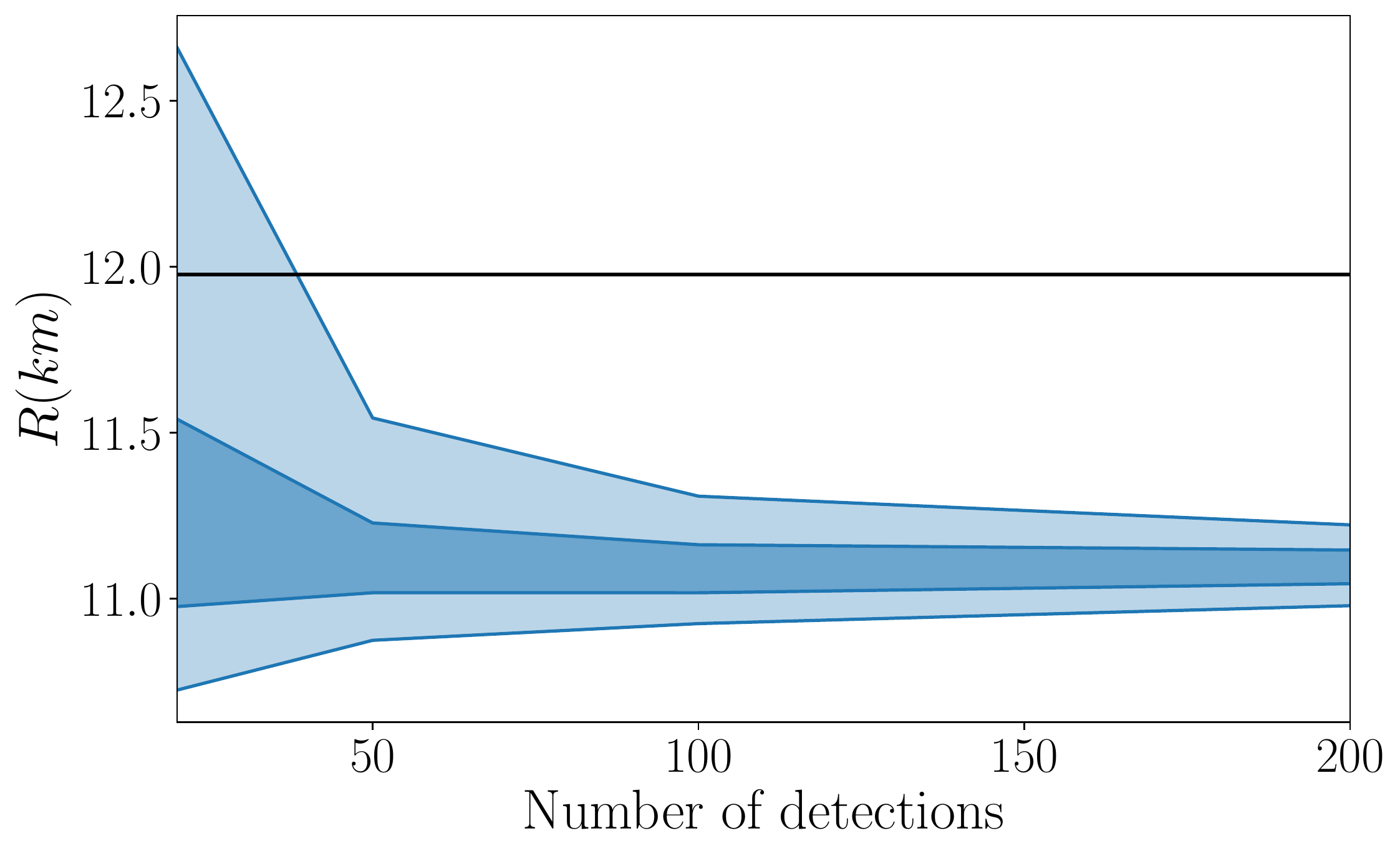}
}
\subfloat[DBHF\_2507]{
  \includegraphics[width=0.5\columnwidth,clip=true]{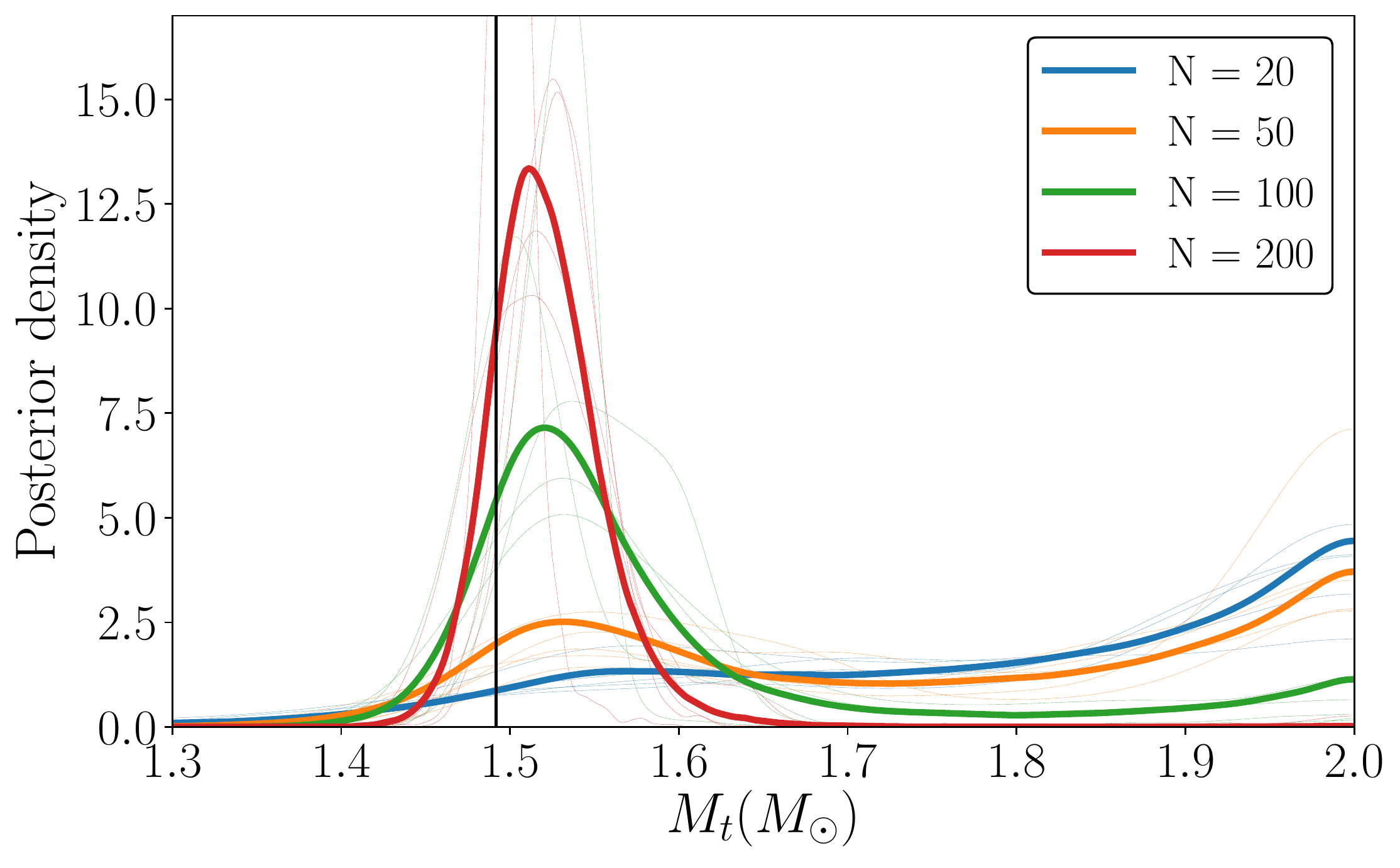}
  \includegraphics[width=0.51\columnwidth,clip=true]{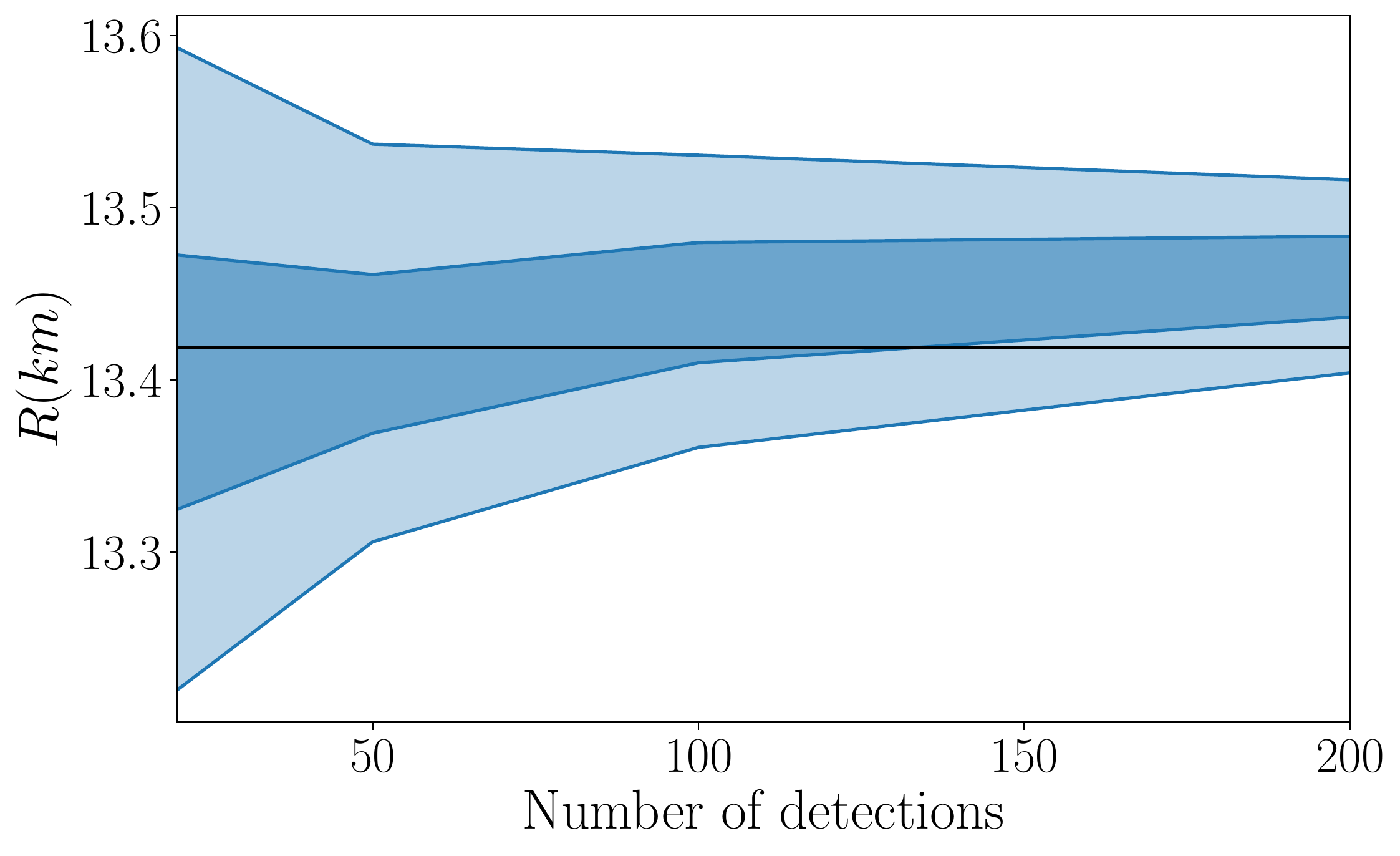}
}
\\
\subfloat[SFHo\_3003]{
  \includegraphics[width=0.5\columnwidth,clip=true]{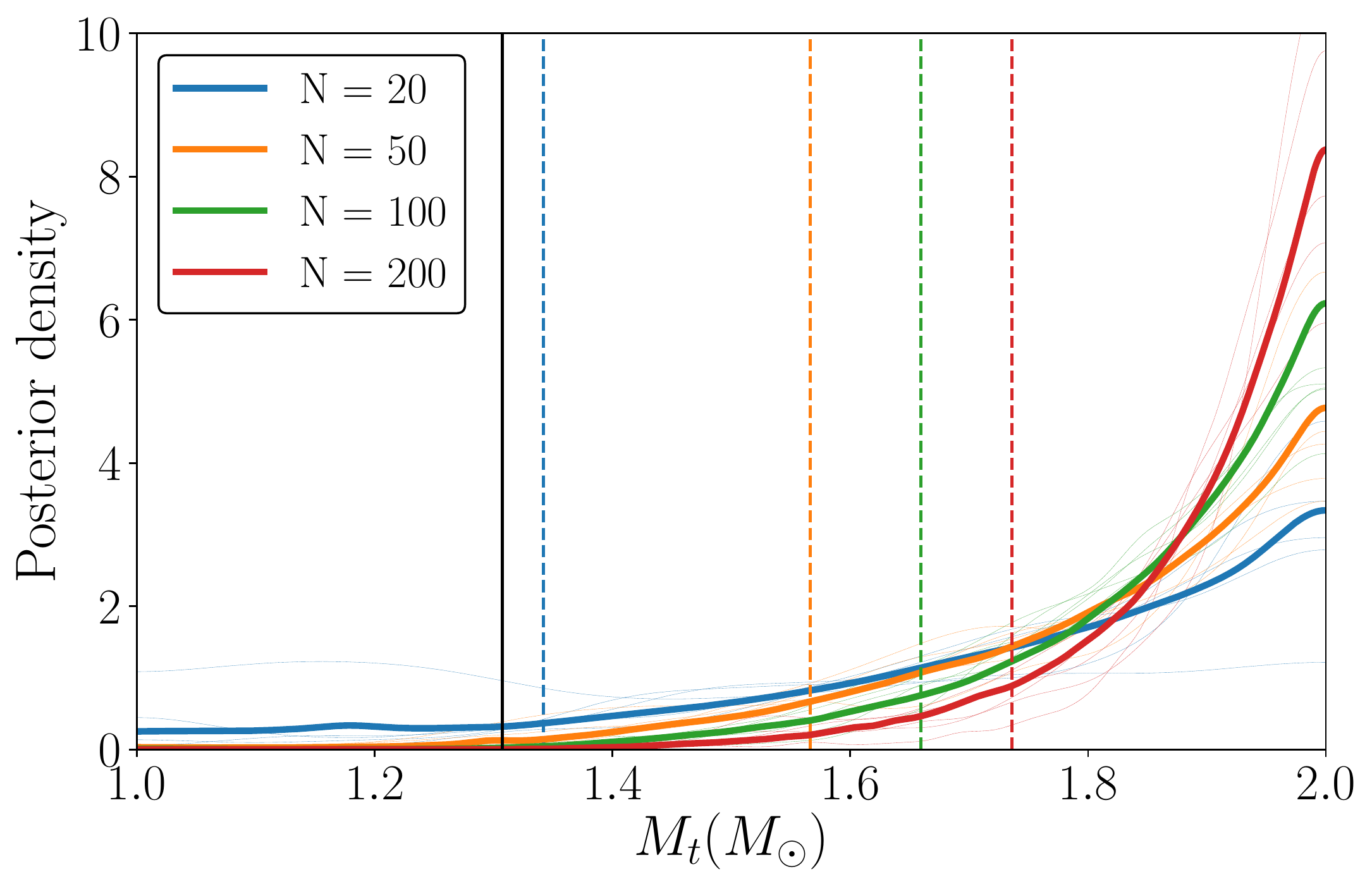}
  \includegraphics[width=0.52\columnwidth,clip=true]{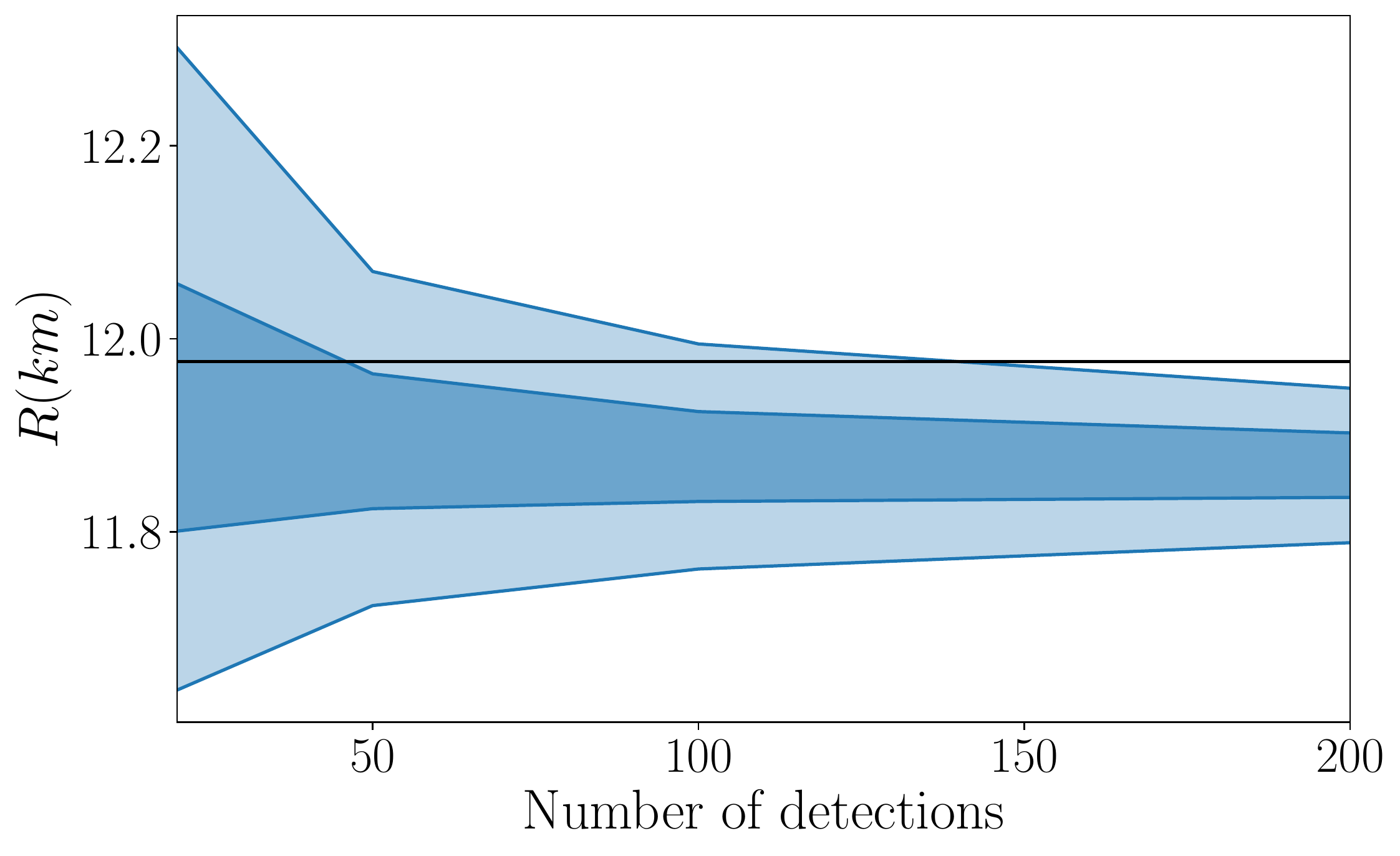}
}
\subfloat[DBHF\_3504]{
  \includegraphics[width=0.5\columnwidth,clip=true]{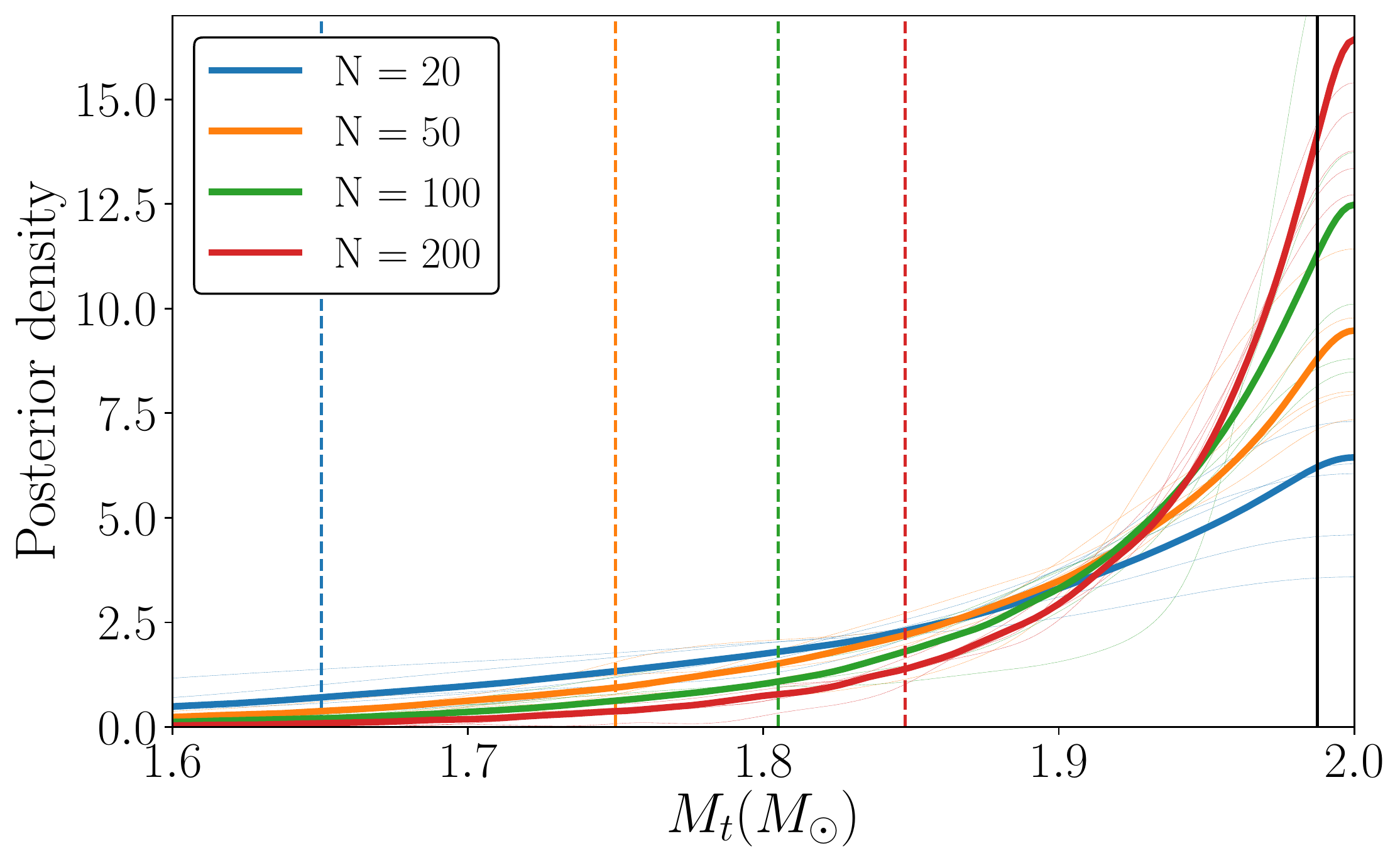}
  \includegraphics[width=0.51\columnwidth,clip=true]{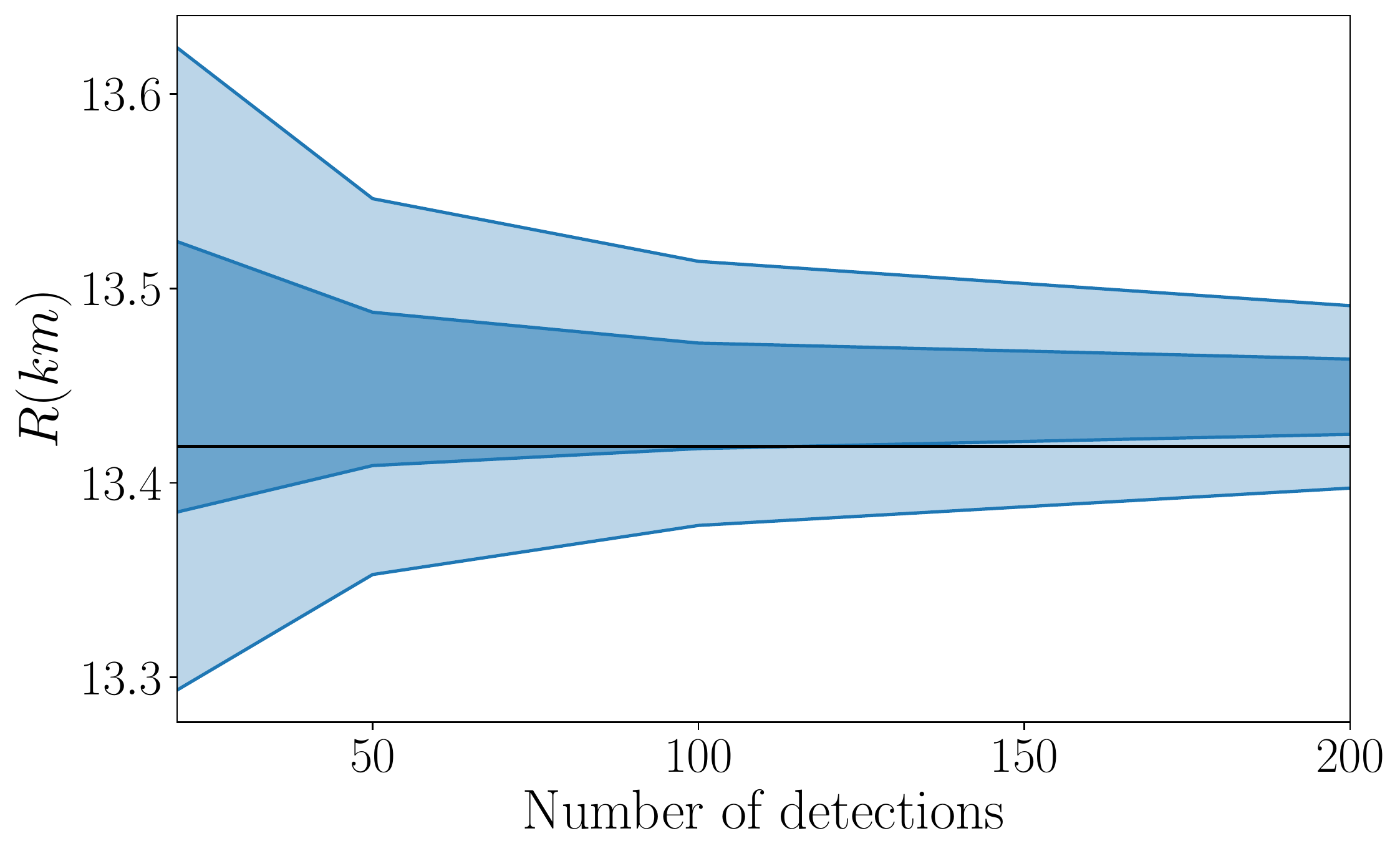}
}
\caption{Marginalized posterior distributions for $\Mtrans$ (left subplot) and median $90$\% (light shading) and $50$\%
(dark shading) credible interval  for $R$ (right subplot) for different numbers 
of detected BNSs $N$ for various EoSs (see subplot caption). In the left subplots solid curves show the posterior
averaged over 100 random population realizations, while thin dashed lines show the posterior for $5$ random population realizations for each $N$ (same color as the averaged posterior).
Solid vertical or horizontal black lines denote the injected $\Mtrans$ and $\Rtyp$, where applicable. Dashed vertical lines in the $\Mtrans$ posteriors denote upper or lower $90\%$ credible intervals, 
where appropriate (same color as the averaged posterior).
}
\label{fig:Population}
\end{figure*}

For each number of BNS detections $N$ we simulate $100$ random population realizations and compute the posterior distribution for $\Mtrans$ and $R$. Figure~\ref{fig:Population} shows our result for selected characteristic EoSs, while in Table~\ref{tab:measurement} we present numerical estimates for the measurement accuracy for $\Mtrans$, $R$, and $\De \ep/\etrans$ for all EoSs. 
The left panel in each subplot of Fig.~\ref{fig:Population} shows the posterior for $\Mtrans$ averaged over population realizations\footnote{We obtain the average posterior over population realizations by combining the posterior samples from all realizations.} (thick solid lines) as well as for $5$ random population realizations (thin dashed lines) for different values of $N$. The right panel in each subplot corresponds to the hadronic radius $R$ and shows the median of the $90\%$ (light shaded region) and $50\%$ (dark shaded region) credible upper and lower limits over the different population realizations.

The expected number of BNS detections in the next years are estimated in~\cite{Aasi:2013wya}, which concludes that two years of detector operation in the expected fourth observing run network sensitivity could yield a few dozens of BNS detections. 
Additional planned improvements, network expansion, and multiyear observing runs are expected beyond that, so we choose to simulate up to $N=200$ BNS detections. 
Estimates from further detections can be easily obtained based on the fact that measurement accuracy roughly scales as $1/\sqrt{N}$ in the regime where the data are highly informative. Accordingly, in Table~\ref{tab:measurement} we quote expected measurement accuracies for  $N=100$.

In the case of a purely hadronic EoS [top left, subplot (a), SFHo], we obtain a tight measurement of the hadronic radius as well as a lower limit on the transition mass that improve with increasing number of detections. 
The radius measurement contains the injected $\Rtyp$ value at the $90\%$ credible level, showing that systematic errors in the ${\cal{R}}({\cal{M}},R)$ fit are subdominant for this EoS even for $N=200$ binary systems. We reach the same conclusion in the case of the stiffer DBHF EoS. 
Additionally, we obtain qualitatively similar results in the case of an EoS with a phase transition at very high masses [bottom right, subplot (f), DBHF\_3504]. As discussed earlier, this is due to the fact that either the entire detected population is hadronic, or the few hybrid systems are heavy and undergo only weak tidal interactions.
The tight measurement of $R$ and the lower limit on $\Mtrans$ can be used to place an upper limit on the transition strength $\De \ep/\etrans$ using Fig.~\ref{fig:Mt_De_fit}. For example for SFHo [subplot (a)] we can constrain $\De \ep/\etrans <0.3$; see Table~\ref{tab:measurement}.
  
The top right and the middle left subplots correspond to the case of a low transition mass compared to the detected NS masses. In the case where the low-density phase transition softens the EoS to radii below $10$~km [subplot (b), DBHF\_1032], we obtain an upper limit on $\Mtrans$ that improves with $N$, while the hadronic radius remains unmeasurable regardless of how many BNSs are detected; indeed the posterior $90\%$ credible interval for the radius spans almost 90\% of the prior range. 
Detecting such a low-density phase transition with $\Mtrans<1\,\Msolar$ can place a conservative lower bound on 
$\De \ep/\etrans\gtrsim0.7$; see Fig.~\ref{fig:Rc_De}.

On the other hand, if the low-density transition results in an EoS that is similar to an allowed softer hadronic EoS, then radius inference will be biased [subplot (c), SFHo\_2506]. In that case we obtain an increasingly precise estimate of the radius $R$, which, however, is inconsistent with the true hadronic radius for $N\sim 50$ and above. 
The transition mass $\Mtrans$ posterior generally favors large values, consistent with the absence of a phase transition, though the average posterior does not scale with $N$ as strongly as the one in subplot (a).  Indeed we find certain selected population realizations where the phase transition is identified due to lucky detections of some low-mass hadronic BNS.

Results with EoSs that give rise to BNS populations that include both hadronic and hybrid stars are shown in the middle right and bottom left plots. If the transition is strong enough [subplot (d), DBHF\_2507], we can identify it and measure its properties with enough detections. 
The left panel shows that with $\sim 20$ BNSs we can rule out a transition at low masses. With $\sim 50$ detections the posterior starts showing increased support for the correct value of $\Mtrans$ resulting in a slight bimodal structure. The measurement of $\Mtrans$ becomes increasingly precise with more detections as the posterior mode that contains the injected value gets more favored. 
At the same time, the hadronic radius $R$ is measured to $\sim 200$~m thanks to the hadronic binaries in the detected population. The phase transition strength can be conservatively inferred as $\De \ep/\etrans\gtrsim0.6$, a requirement for the emergence of a sufficiently disconnected hybrid branch.
Additionally, measurement of a small chirp radius $\Rchirp^{\rm HQ}$ from ``hadronic-hybrid'' binaries can also potentially constrain the transition strength; see Fig.~\ref{fig:Rc_Mt_De}.

If the phase transition is moderately weak, then the radius difference between hybrid and hadronic stars might be too small to detect. This case is shown in subplot (e) for SFHo\_3003, which shows that the resulting posterior distribution for the transition mass $\Mtrans$ increasingly favors large values, similar to the case where the entire population is hadronic. 
The hadronic radius $R$ is again measured to high precision, though the bias is slightly larger than if the population was indeed hadronic [subplot (a), SFHo]. 
For the EoS studied here, we find that the inferred radius agrees with $\Rtyp$ up to $N\sim 100$ at the 90\% level. The combination of a lower limit on $\Mtrans$ and a measurement of the hadronic radius $R$ leads to an upper limit on the transition strength of $\De\ep/\etrans\lesssim 0.3$ for this EoS; see Table~\ref{tab:measurement}.

The cause of this small radius bias is the fact that hybrid NSs with a smaller radius than their hadronic counterparts are interpreted as hadronic, overall bringing the radius estimate to lower values. 
The radius bias depends on the strength of the phase transition and if the transition is very weak, such as the case studied here, the radius difference remains small and the inferred hadronic radius is not strongly biased. 
If the transition strength increases beyond a certain point, typically $\De\ep/\etrans\gtrsim 0.6$ when $\Mtrans\gtrsim 1\,\Msolar$ (with modest dependence on the underlying hadronic EoS and the transition density) that ensures a stable disconnected branch, then the effect becomes directly detectable and the hadronic radius can be estimated accurately; see subplot (d) for DBHF\_2507.

\begin{table}[]
\begin{center}
\begin{tabular}{c|@{\quad}c|@{\quad}c|@{\quad}c}
\hline 
&&&\\[-2ex]
 EoS& $\Mtrans$  & $R$ &$\De \ep/\etrans$  \\[0.5ex]
  \hline 
&&&\\[-2ex]
 DBHF&$>1.75\,\Msolar$&$140$~m &  $\lesssim 0.6$\\[0.5ex]
\hline  
&&&\\[-2ex]
  DBHF\_1032 & $<1.13\,\Msolar$ & N/A  & $\gtrsim0.7$ (hyb)\\[0.5ex]
 \hline 
&&&\\[-2ex]
 DBHF\_2010 & $0.14\,\Msolar$ & $530$~m & $\gtrsim0.6$ (det)\\[0.5ex]
 \hline
 &&&\\[-2ex]
 DBHF\_2507 & $0.25\,\Msolar$& $170$~m & $\gtrsim0.6$ (det)\\[0.5ex]
 \hline 
&&&\\[-2ex]
 DBHF\_3004 & $>1.78\,\Msolar$& $140$~m & $\lesssim0.6$ \\[0.5ex]
  \hline 
  &&&\\[-2ex]
 DBHF\_3504 & $>1.81\,\Msolar$ & $140$~m  & $\lesssim0.5$\\[0.5ex]
  \hline 
&&&\\[-2ex]
 SFHo& $>1.76\,\Msolar$& $210$~m  & $\lesssim0.2$\\[0.5ex]
 \hline 
&&&\\[-2ex]
 SFHo\_1031 & $<1.13\,\Msolar$ & N/A & $\gtrsim0.7$ (hyb)  \\[0.5ex]
\hline
&&&\\[-2ex]
 SFHo\_2009 & dual & dual & $\gtrsim0.6$ (det)\\[0.5ex]
\hline
&&&\\[-2ex]
 SFHo\_2506 & $>1.23\,\Msolar$ &$380$~m (biased) & $\lesssim0.6$ (deg) \\[0.5ex]
\hline
&&&\\[-2ex]
 SFHo\_3003 & $>1.66\,\Msolar$ & $230$~m & $\lesssim0.3$\\[0.5ex]
\hline
&&&\\[-2ex]
 SFHo\_3502 & $>1.7\,\Msolar$& $220$~m & $\lesssim0.3$\\[0.5ex]
\hline
\end{tabular}
\end{center}
\caption{Expected measurement accuracy for the transition mass $\Mtrans$ and the hadronic radius $R$ at the 90\% credible level for $N=100$ BNS detections, as well as corresponding conclusion about the transition strength $\De\ep/\etrans$. 
We present results for all EoSs studied here; ``det'' refers to a phase transition confirmed in the detected population, ``hyb'' refers to a phase transition below the minimum NS mass for which all NSs are identified as hybrid stars, and ``deg'' refers to a low-density transition present completely degenerate with a softer hadronic EoS.
}
\label{tab:measurement}
\end{table} 

\section{Conclusions}
\label{sec:con}

The detection of GW170817 along with the measurement of its tidal parameters has offered a first example of how GW observations can constrain the NS EoS, while further observations are expected to tighten those constraints. Besides improvements on the measurement accuracy of EoS parameters, in this paper we consider future detections of BNS inspirals with GWs and study their potential for constraining features of the EoS such as possible strong first-order phase transitions. 
We show that within the generic CSS framework, combining information from premerger GW signals with limits on the NS maximum mass provided by heavy pulsars can significantly narrow down the parameter space of first-order phase transitions. We utilize the fact that binaries with hadron-quark hybrid stars exhibit qualitatively different tidal effects than their hadronic counterparts and model the entire detected BNS population in a hierarchical way.

Our ability to constrain parameters of the hadronic baseline EoS, in particular the nearly constant hadronic radius $R$, and parameters of the phase transition such as the onset mass $\Mtrans$ and the transition strength $\De \ep/\etrans$, depends sensitively on both the type of EoS and on the intrinsic mass distribution of NSs in binaries in the Universe. 
Regarding the hadronic baseline EoS, it first relates the transition density $\ntrans$ to the transition mass $\Mtrans$, which should be low enough so that the detected BNS population contains a significant fraction of hybrid stars; note that at given $\Mtrans$, $\ntrans$ is smaller if the hadronic EoS is stiffer, a caution to check the compatibility with theoretical/experimental constraints that are available at relevant densities. Additionally, the hadronic baseline EoS affects the parameter space for a first-order phase transition to be consistent with the heavy pulsar observations, for which stiffer hadronic EoSs are in general favored. Regarding the phase transition parameters, a phase transition is more easily detectable when hybrid stars deviate considerably from their purely hadronic counterparts in the mass-radius plane, ideally forming disconnected branches. The latter is typically realized in hybrid EoSs with both relatively low transition mass $\Mtrans$ and large transition strength $\De\ep$.

Depending on the properties of the phase transition and the NS mass distribution we identify four possible scenarios:
\begin{enumerate}

\item If the EoS is purely hadronic, we can place a lower limit on $\Mtrans$ and measure the hadronic radius $R$. 
These estimates from the BNS observations can be combined with the heavy pulsar measurements and be translated to an upper limit on the transition strength $\De \ep/\etrans$.

\item If the onset mass is lower than the detected BNS masses, $\Mtrans<1\,\Msolar$, then we expect to observe only hybrid stars, a situation that might be degenerate with BNSs that obey a softer purely hadronic EoS. In this case the observation of heavy pulsars is crucial, as they can lead to lower limits on the softness of the hadronic EoS and thus break the degeneracy. With such observations we can place an upper limit on $\Mtrans$, and a conservative lower limit on the transition strength $\De \ep/\etrans$ given detections of sufficiently small chirp radii, while the hadronic radius is naturally unmeasurable. This is based on the requirement that hybrid stars with canonical masses satisfying the constraint of small tidal deformability from~\cite{Abbott:2018wiz} also favors larger values of $\De \ep/\etrans$ (see e.g. Fig. 7 of~\cite{Han:2018mtj}), but not too large to violate the $\Mmax\geq2.0\,\Msolar$ limit.
 
\item If $\Mtrans$ is comparable to the NS masses observed, then a strong enough transition can be inferred after $50-100$ observations, while the hadronic BNSs that were detected can lead to an accurate estimate of the hadronic radius. Additional information on $\De\ep/\etrans$ can possibly be extracted by measuring $\De\Rchirp$ between ``HH'' and ``HQ'' binaries from multiple detections, or conservatively from the generic condition $\De \ep/\etrans\gtrsim0.6$ for the presence of a disconnected hybrid branch. If, on the other hand, the transition turns out to be weak, i.e. $\De \ep/\etrans$ is small, then the tidal properties of the hybrid stars are identical to those of hadronic stars to within observational error, and the presence of the phase transition would be challenging to establish. This scenario can only be realized if the quark phase is reasonably stiff (in this work we consider $\cQMsq=1$). A stiff quark phase is necessary not only for the hybrid EoSs to be able to support $2\,\Msolar$ stars, but also for hybrid and hadronic stars to have sufficiently different radii such that they are distinguishable with GW observations. This point is further elaborated on in Appendix~\ref{AppB}.

\item Finally, if $\Mtrans$ is high, then the phase transition occurs only for the heaviest mass systems that exhibit weak tidal interactions to begin with. In that case we can place a lower limit on $\Mtrans$ and an upper limit on $\De \ep/\etrans$, but measuring either will be challenging with second generation GW detectors. Though such phase transitions are inaccessible to the BNS inspiral signals considered here, they might have a more observable imprint in the dynamic evolutions of the merger product if $\Mtrans$ is close to or even above $\Mmax$. 
 
\end{enumerate}

The above results assume that the tidal parameter $\tilde{\Lambda}$ can be measured without further systematic biases from future, potentially loud, observations. This might not generically be the case with currently available waveform models and stronger or multiple signals~\cite{Dudi:2018jzn,Samajdar:2018dcx}. However, efforts are underway to further reduce waveform systematics, see e.g.~\cite{Dietrich:2018uni,Dietrich:2019kaq,Hinderer:2016eia,Messina:2019uby,Akcay:2018yyh}; 
such improvements will be essential for realizing the potential of BNS inspirals as probes of distinct features on the QCD phase diagram, such as the study presented here.

Throughout the paper we have emphasized the importance of incorporating EoS constraints derived from mass measurements of heavy pulsars, but other constraints can also be folded in. Further information from terrestrial experiments and nuclear calculations can also be included in our analysis to better understand the behavior of the EoS at densities around saturation. 
For example, the future PREX-II~\cite{Abrahamyan:2012gp,Fattoyev:2017jql} experiments aim to accurately measure the neutron skin thickness of heavy nuclei that correlates with the pressure/stiffness of neutron-rich matter near or below saturation density; modern computations based on chiral effective field theory~\cite{Hebeler:2010jx} and quantum Monte Carlo methods~\cite{Gandolfi:2011xu} hold the promise of improving the uncertainty analysis in the nuclear EoS at 1--2 times the saturation density.
Moreover, forthcoming results from NICER would provide complementary constraints on the NS radii~\cite{Miller:2014mca,Watts:2016uzu} which could help serve as prior information for the analysis of GW data~\cite{Forbes:2019xaz}.

Our work has established the conditions under which GW inspiral signals from BNSs can be used to place constraints on the parameter space of phase transitions. 
We further highlight the importance of incorporating constraints from different observations, and demonstrate the crucial role of measuring the masses of heavy pulsars. 
The determination of the properties of dense matter is a firmly interdisciplinary project that will benefit from future BNS observations, not only by measuring the EoS parameters with better precision, but also by constraining scenarios of the QCD phase diagram such as strong first-order transitions from hadronic to quark matter. \\

\section*{Acknowledgements}

We thank Will Farr for discussions on the hierarchical model and we thank Madappa Prakash for comments and suggestions on an earlier version of the manuscript.
The Flatiron Institute is supported by the Simons Foundation. S.H. is supported by National Science Foundation, Grant PHY-1630782, and the Heising-Simons Foundation, Grant 2017-228. S.H. would like to thank the Yukawa Institute for Theoretical Physics at Kyoto University, where part of this work was performed during the workshop YITP-T-19-04 ``Multi-Messenger Astrophysics in the Gravitational Wave Era''.
Plots in this manuscript have been made with {\tt matplotlib}~\cite{Hunter:2007} and we have used {\tt stan}~\cite{JSSv076i01} to sample the hierarchical model.

\appendix

\section{$\De R$ plots for all EoSs}
\label{AppA}
 
\begin{figure*}[]
\includegraphics[width=.65\columnwidth,clip=true]{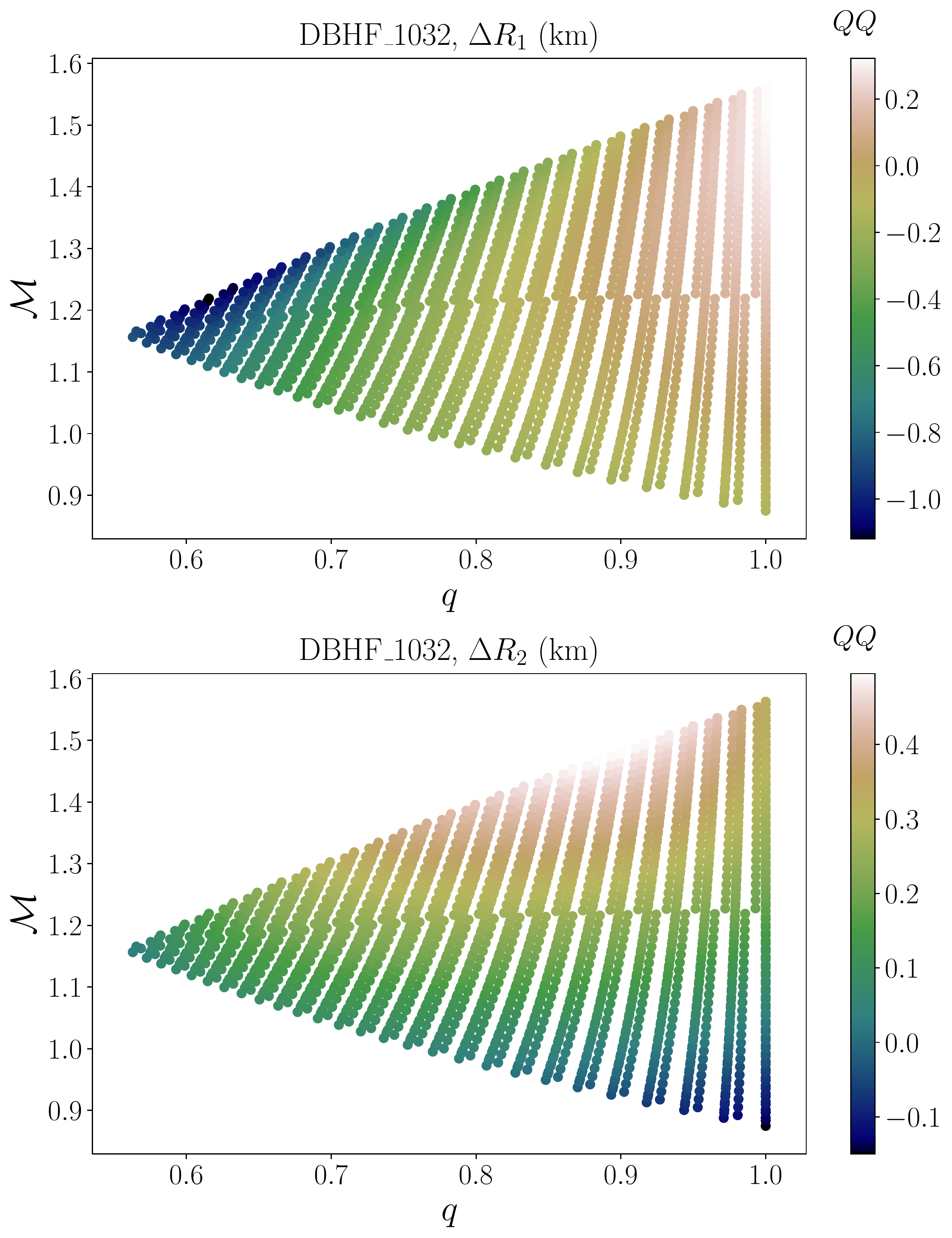}
\includegraphics[width=.65\columnwidth,clip=true]{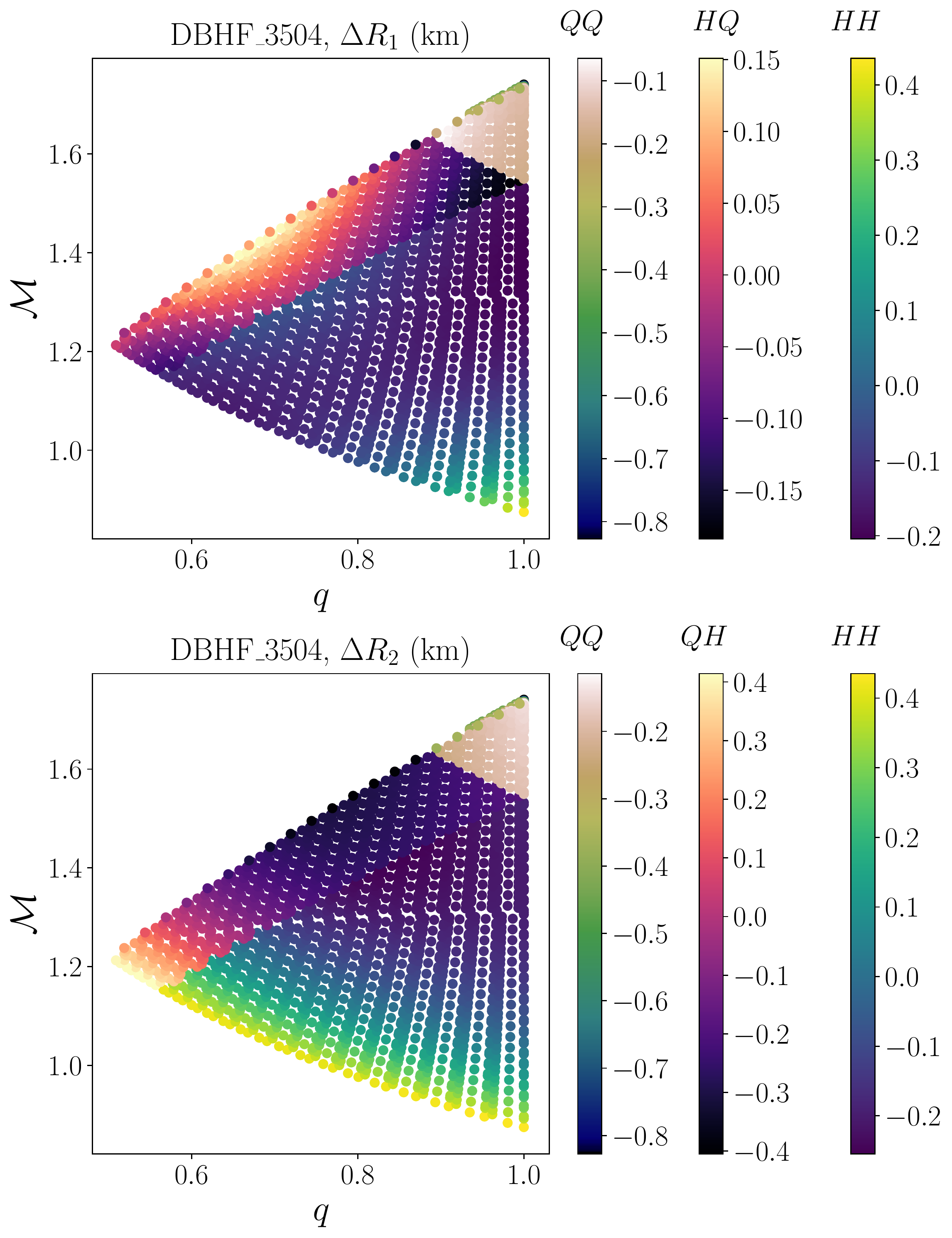}
\includegraphics[width=.65\columnwidth,clip=true]{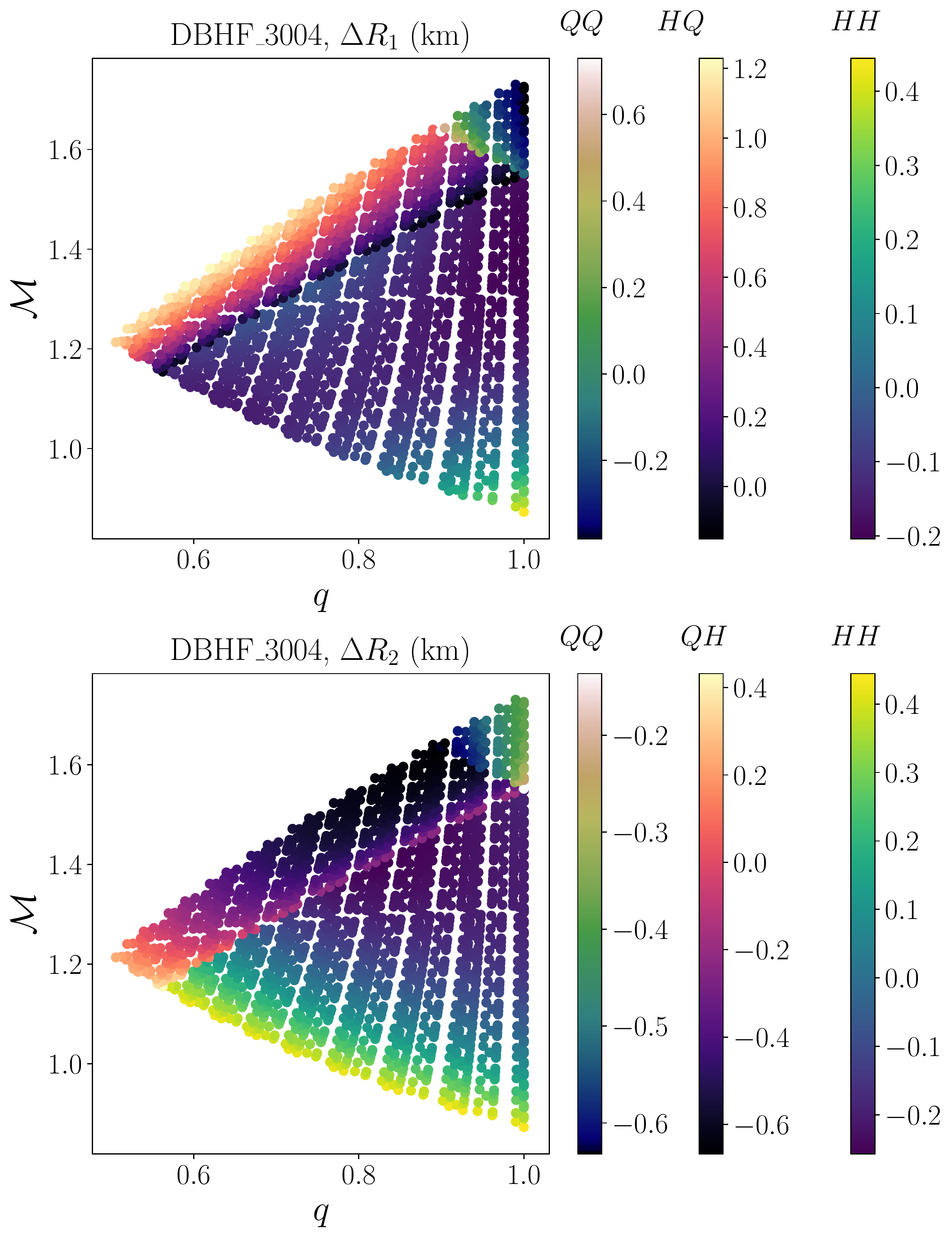}\\
\includegraphics[width=.65\columnwidth,clip=true]{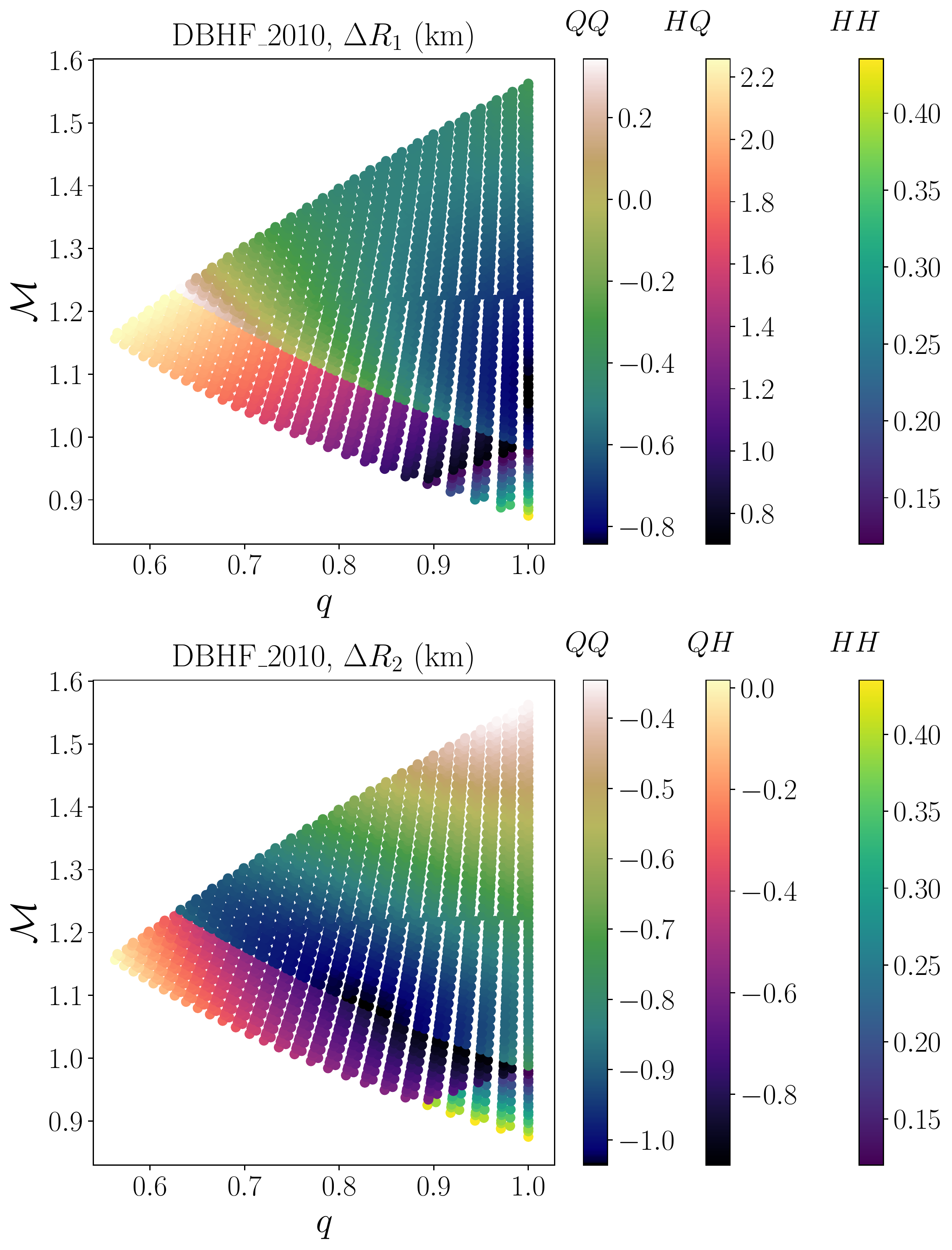}
\includegraphics[width=.65\columnwidth,clip=true]{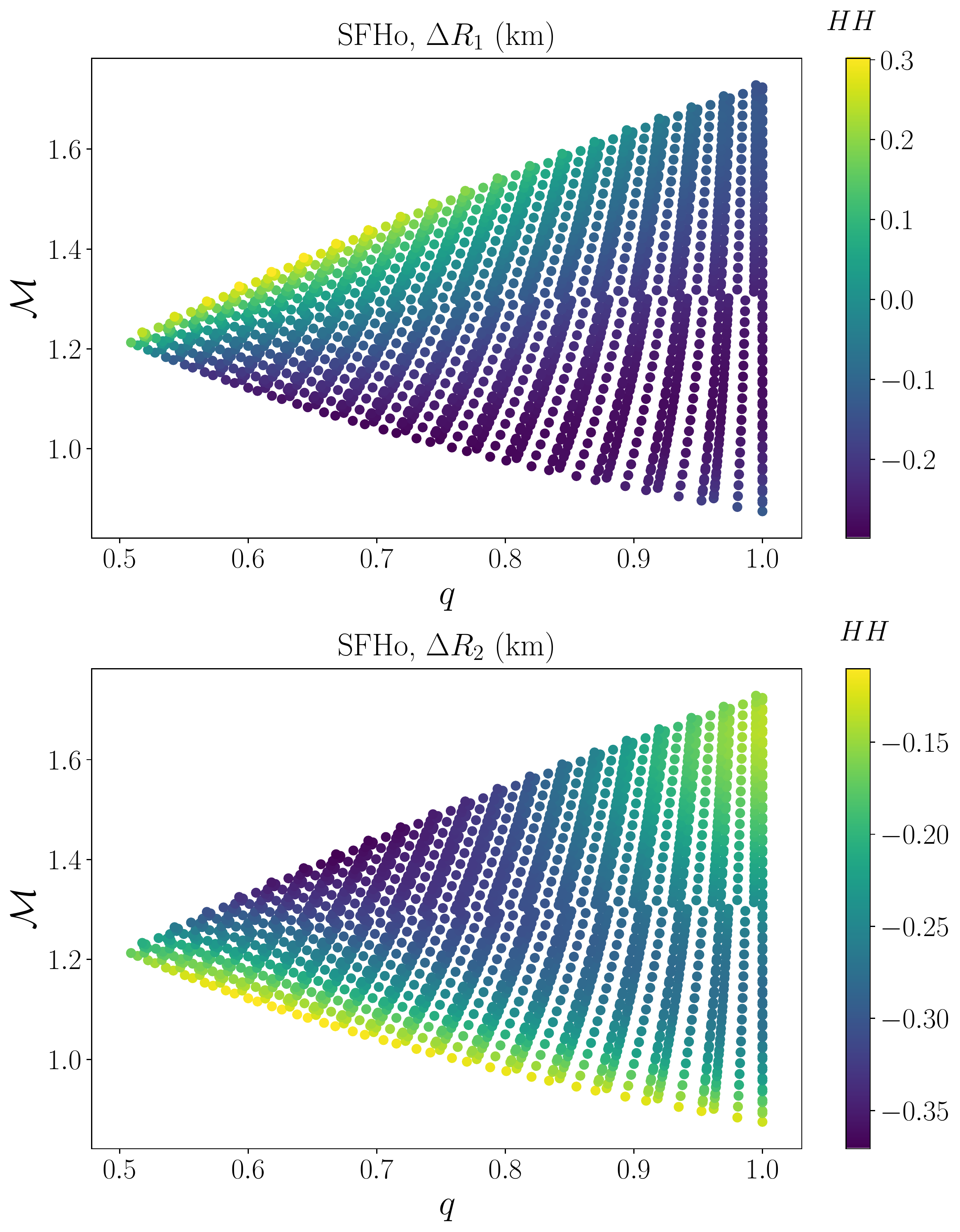}
\includegraphics[width=.65\columnwidth,clip=true]{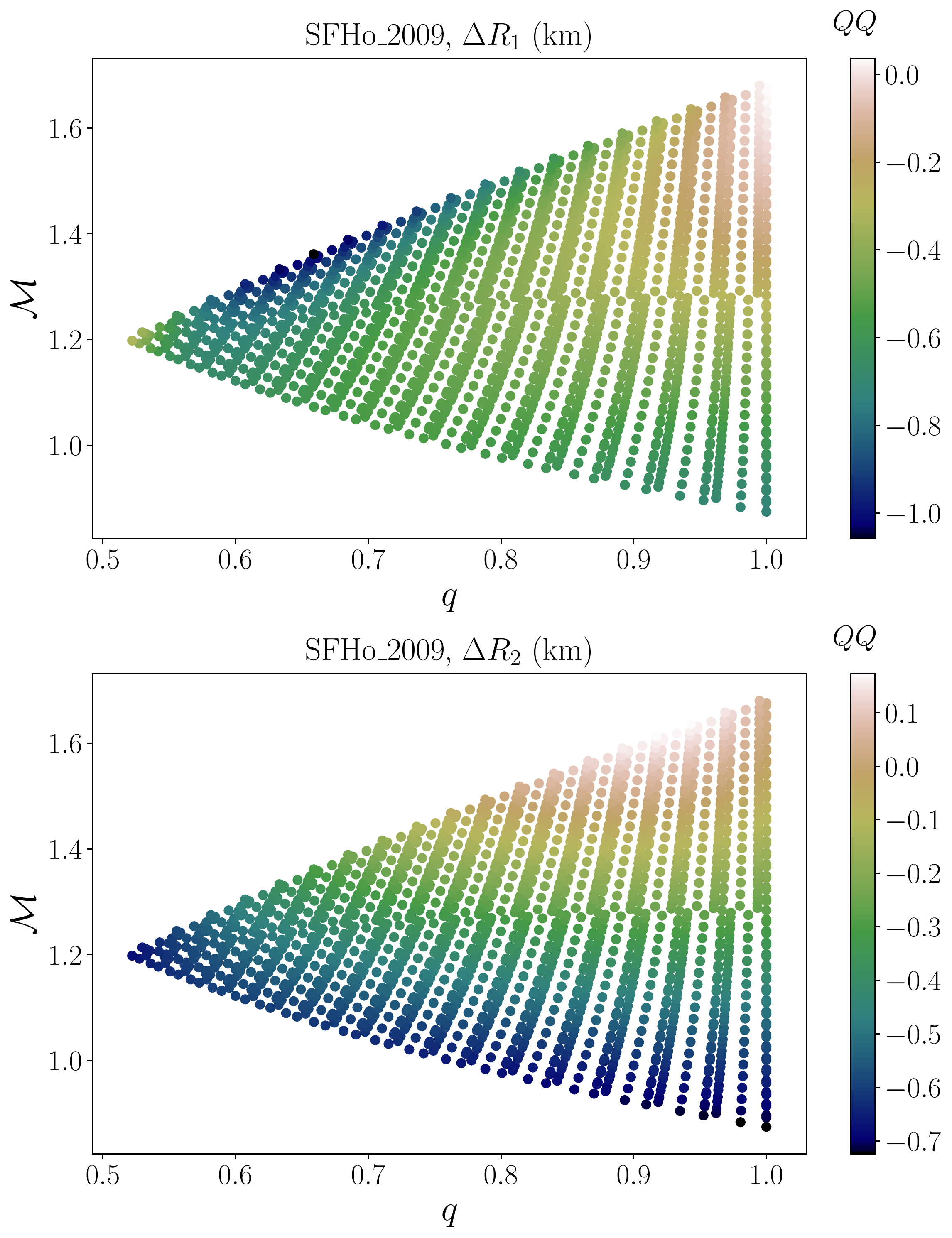}\\
\includegraphics[width=.65\columnwidth,clip=true]{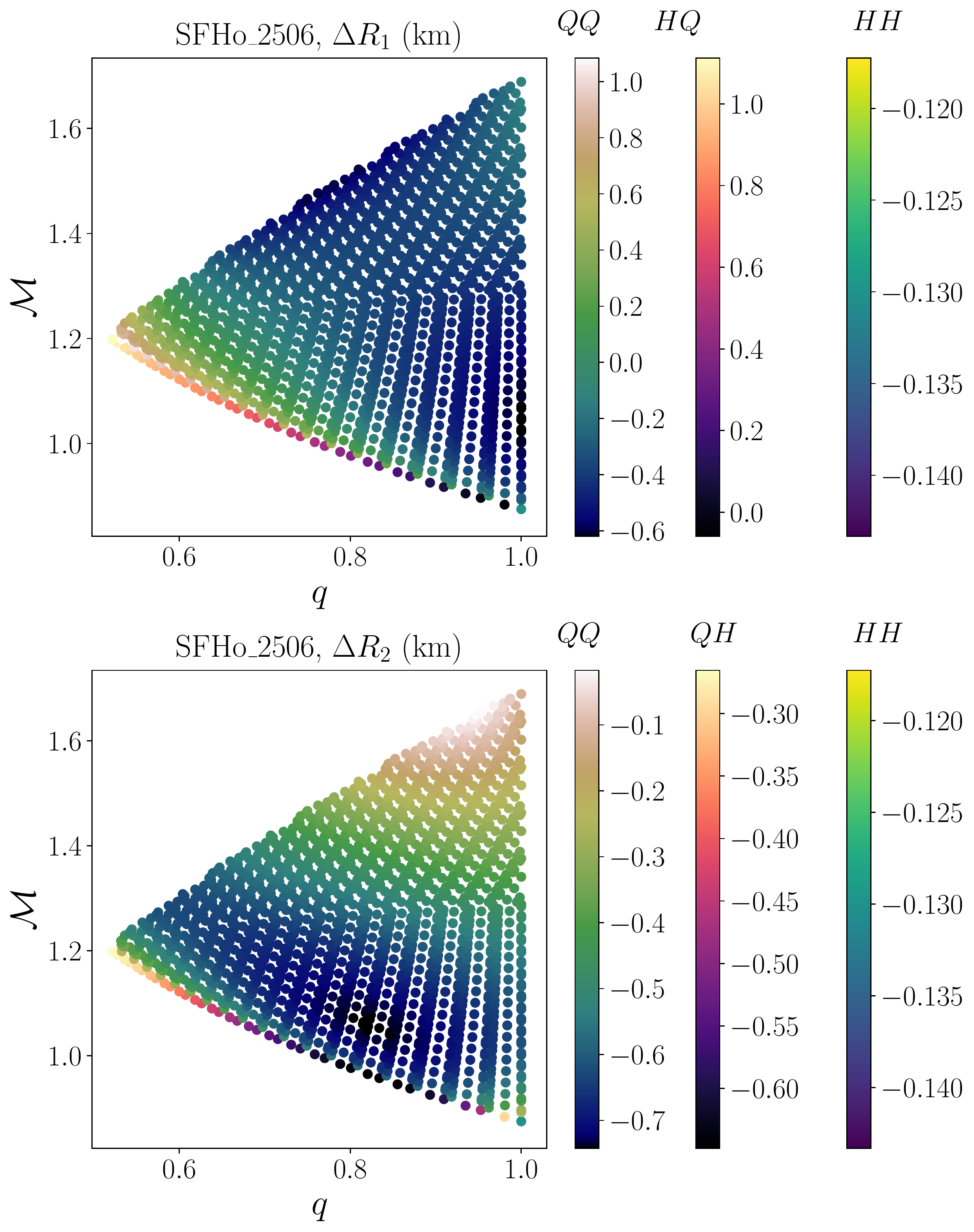}
\includegraphics[width=.65\columnwidth,clip=true]{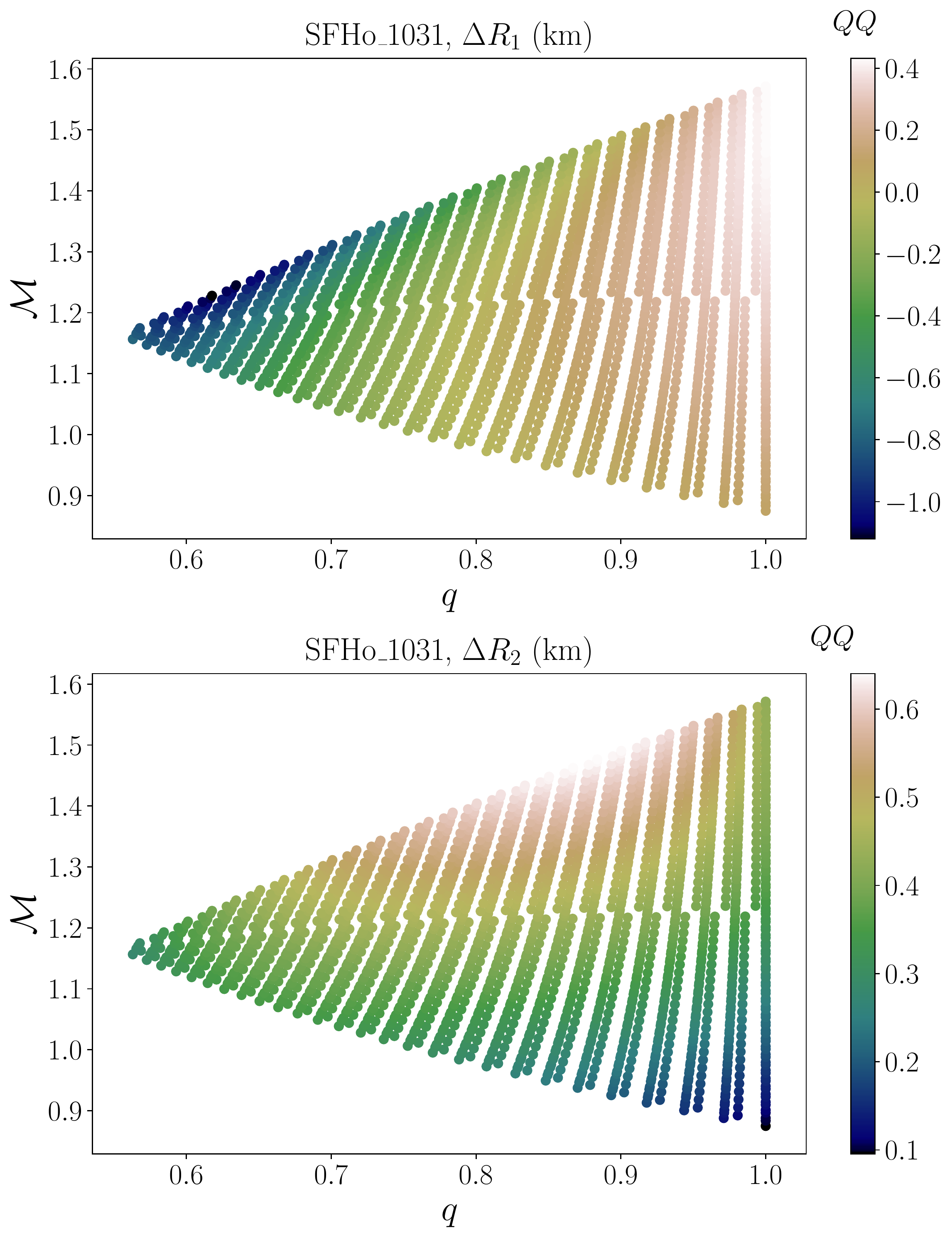}
\includegraphics[width=.66\columnwidth,clip=true]{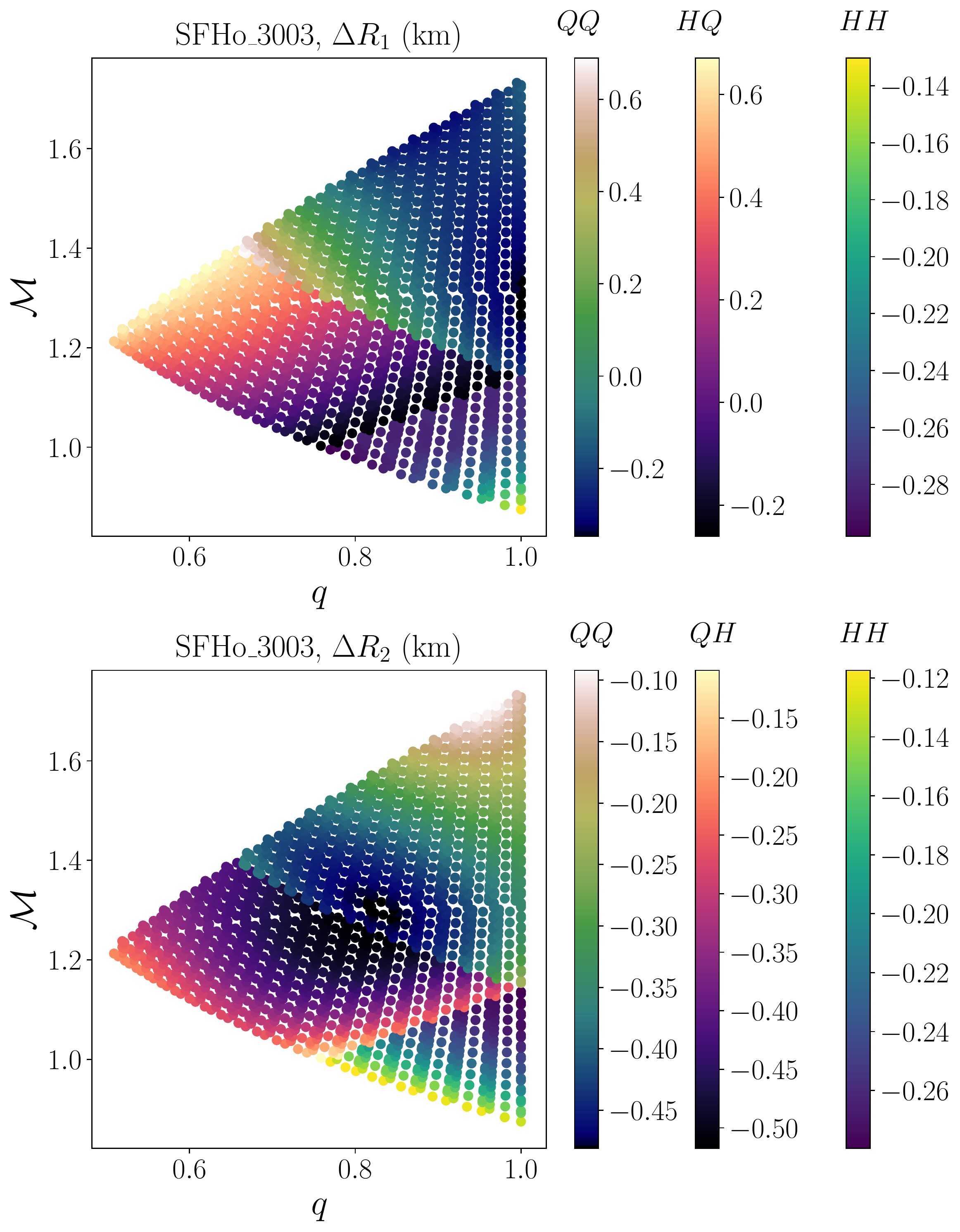}
\caption{Same as Fig.~\ref{fig:DeltaR} for the remaining EoSs of Fig.~\ref{fig:MR}.}
\label{fig:DeltaR2}
\end{figure*}

In this appendix we present in Fig.~\ref{fig:DeltaR2} $\De R_i$ plots similar to Fig.~\ref{fig:DeltaR} for the remaining EoSs from Fig.~\ref{fig:MR}. In all cases the values of $\De R_i$ are smaller than $\sim 2$~km.

\section{Effects of softer quark matter}
\label{AppB}

\begin{figure*}[htb]
\parbox{0.5\hsize}{
\includegraphics[width=3.5in]{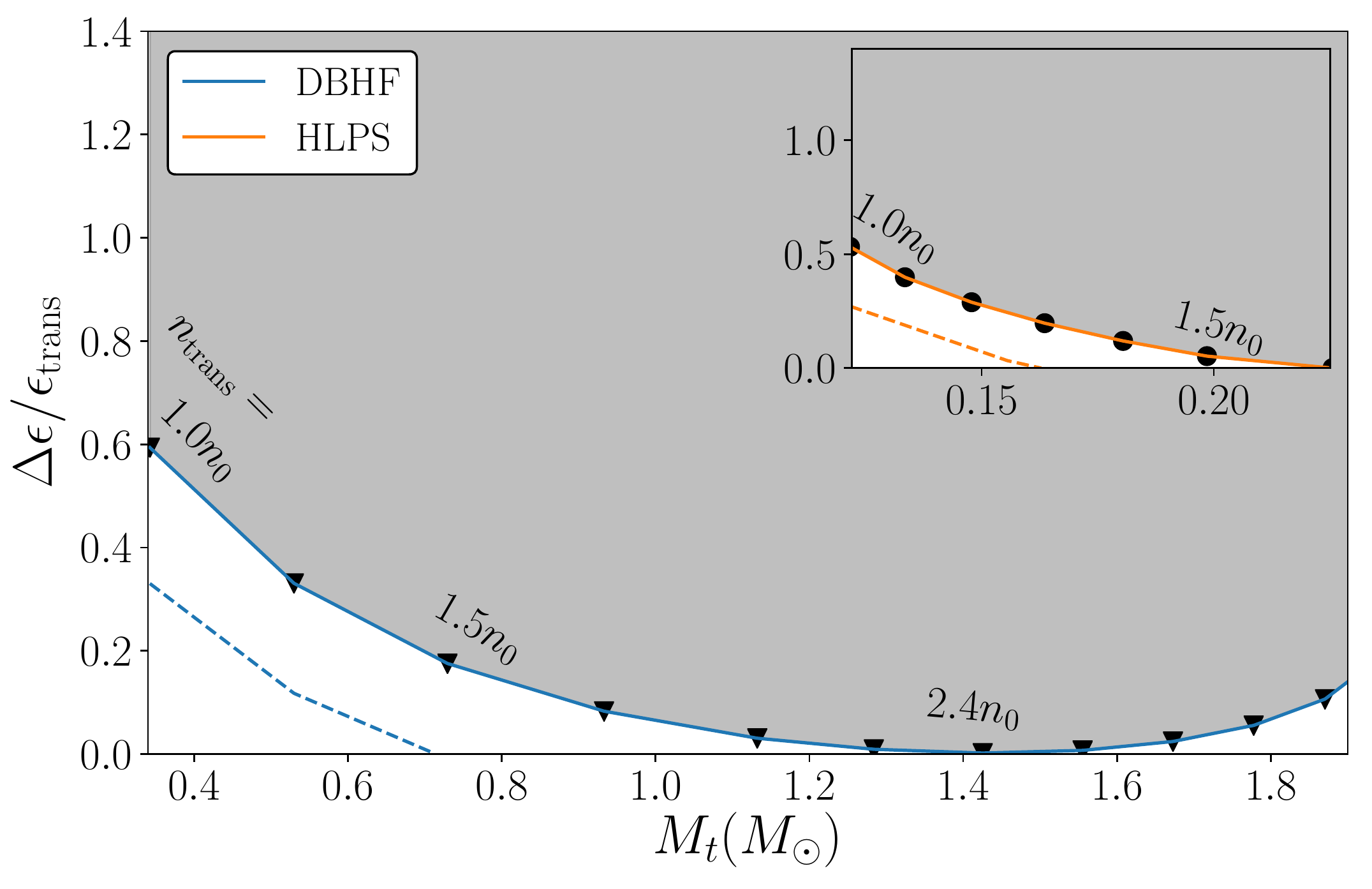}\\[-2ex]
}\parbox{0.5\hsize}{
\includegraphics[width=3.5in]{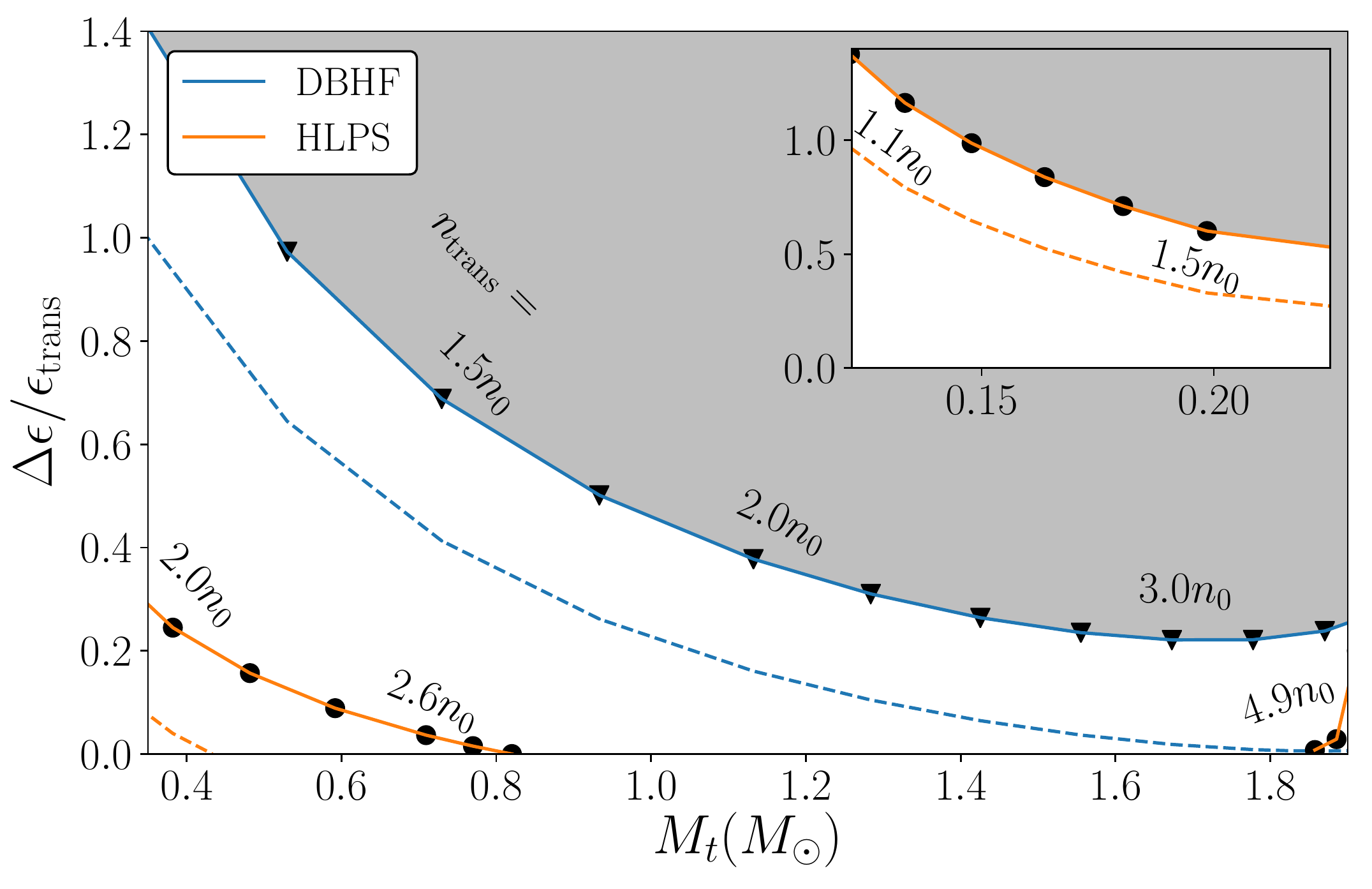}\\[-2ex]
}
\caption{$\Mmax=2.0\,\Msolar$ (solid) and $\Mmax=2.2\,\Msolar$ (dotted) contours for the DBHF and HLPS hadronic EoSs with soft ($\cQMsq=0.33$; left panel) and relatively stiff ($\cQMsq=0.5$; right panel) quark matter. For the softest quark EoS, 
the allowed values of $\De\ep/\etrans$ are so small that only very weak transitions can occur, 
leading to indistinguishable features compared to purely hadronic stars. 
By increasing the stiffness in quark matter (right panel), with the stiff DBHF hadronic baseline EoS it is 
now possible to have strong transitions ($\De\ep/\etrans\gtrsim 0.5$).
}
\label{fig:Mt_De-softer}
\end{figure*}

Similar to Fig.~\ref{fig:Mt_De_fit}, in Fig.~\ref{fig:Mt_De-softer} we show exemplary plots with $\Mmax$ contours for the DBHF and HLPS hadronic baselines, 
but with a softer quark EoS. Shaded regions again correspond to the region of the parameter space that is excluded due to the lower
limit on $\Mtrans$. The plots confirm that the choice of maximal stiffness $\cQMsq=1$ in the paper corresponds to the largest parameter space on the ($\De\ep/\etrans$, $\Mtrans$) plane available and is therefore conservative.

\bibliography{paper_GW-fin}

\end{document}